\documentclass[]{aastex631}
\usepackage{graphicx}
\usepackage{txfonts}
\usepackage{multirow}
\usepackage{array}
\usepackage[toc,page]{appendix}
\usepackage{amssymb}
\usepackage{pdfrender}
\newcommand*{\boldcheckmark}{%
  \textpdfrender{
    TextRenderingMode=FillStroke,
    LineWidth=.5pt, 
  }{\checkmark}%
}

\newcommand{\kps}{km\,s$^{-1}$}

\begin{document}

\title{Probing progression of heating through the lower flare atmosphere via high-cadence IRIS spectroscopy}

\correspondingauthor{J. L\"{o}rin\v{c}\'{i}k, V. Polito}
\email{lorincik@baeri.org, polito@lmsal.com}

\author[0000-0002-9690-8456]{Juraj L\"{o}rin\v{c}\'{i}k}
\affil{Bay Area Environmental Research Institute, NASA Research Park, Moffett Field, CA 94035, USA}
\affil{Lockheed Martin Solar \& Astrophysics Laboratory, Org. A021S, Bldg. 203, 3251 Hanover St., Palo Alto, CA 94304, USA}
\author[0000-0002-4980-7126]{Vanessa Polito}
\affil{Lockheed Martin Solar \& Astrophysics Laboratory, Org. A021S, Bldg. 203, 3251 Hanover St., Palo Alto, CA 94304, USA}
\affil{Department of Physics, Oregon State University, 301 Weniger Hall, Corvallis, OR 97331}
\author[0000-0001-5316-914X]{Graham S. Kerr}
\affil{Department of Physics, Catholic University of America, 620 Michigan Avenue Northeast, Washington, DC 20064, USA}
\affil{NASA Goddard Space Flight Center, Heliophysics Science Division, Code 671, 8800 Greenbelt Road, Greenbelt, MD 20771, USA}
\author[0000-0002-6835-2390]{Laura A. Hayes}
\affil{European Space Agency (ESA), European Space Research and Technology Centre (ESTEC), Keplerlaan 1, 2201 AZ Noordwĳk, The Netherlands}
\affil{Astronomy $\&$ Astrophysics Section, School of Cosmic Physics, Dublin Institute for Advanced Studies, DIAS Dunsink Observatory, Dublin D15XR2R, Ireland}
\author[0000-0001-5690-2351]{Alexander J. B. Russell}
\affil{School of Mathematics \& Statistics, 
University of St Andrews,  
St Andrews, KY16 9SS, UK}

\begin{abstract}
Recent high-cadence flare campaigns by the Interface Region Imaging Spectrograph (IRIS) have offered new opportunities to study rapid processes characteristic of flare energy release, transport, and deposition. Here, we examine high-cadence chromospheric and transition region spectra acquired by IRIS during a C-class flare from 2022 September 25. Within the flare ribbon, the intensities of the \ion{Si}{4} 1402.77\,\AA, \ion{C}{2} 1334.53\,\AA~and Mg II k 2796.35\,\AA~lines peaked at different times, with the transition region \ion{Si}{4} typically peaking before the chromospheric \ion{Mg}{2} line by $1 - 6$ seconds. To understand the nature of these delays, we probed a grid of radiative hydrodynamic flare simulations heated by electron beams, thermal conduction-only, or Alfv\'{e}n waves. Electron beam parameters were constrained by hard X-ray observations from the Gamma-ray Burst Monitor (GBM) onboard the \textsl{Fermi} spacecraft. Reproducing lightcurves where \ion{Si}{4} peaks precede those in \ion{Mg}{2} proved to be a challenge, as only a subset of \textsl{Fermi}/GBM-constrained electron beam models were consistent with the observations. Lightcurves with relative timings consistent with the observations were found in simulations heated by either high-flux electron beams or by Alfv\'{e}n waves, while the thermal conduction heating does not replicate the observed delays. Our analysis shows how delays between chromospheric and transition region emission pose tight constraints on flare models and properties of energy transport, highlighting the importance of obtaining very high-cadence datasets with IRIS and other observatories.
\end{abstract}

\section{Introduction} 
\label{sec:intro}

Solar flares are characterized by a sudden increase of radiation across all wavelengths, from radio waves to gamma rays. The increase of radiation is a by-product of the release of energy stored in the magnetic field within the solar atmosphere, facilitated by magnetic reconnection. In the standard flare scenario \citep{carmichael64, sturrock66, hirayama74, kopp76} and its extensions, energy released in the corona manifests in bulk flows, particle acceleration \citep[e.g.][]{brown71, lin76, holman11}, \textsl{in-situ} heating \citep[e.g.][]{ricchiazzi83, czaykowska01, longcope14}, and excitation of magnetohydrodynamic (MHD) waves \citep[e.g.][]{emslie82, fletcher08, russell13, reep16, kerr16}.

The time scales at which the lower solar atmosphere responds to the energy transfer processes vary depending on the mechanism. Early flare observations demonstrated correlations between hard X-ray (HXR) and microwave radio emission generated by accelerated non-thermal electrons and the H$\alpha$ line emission probed by various ground-based observatories \citep[see e.g.][and references therein]{krivsky63, svestka70, kane74, lin76, zirin78}, suggesting that the accelerated electrons deposit their energy in the chromosphere. The chromosphere responds rapidly to electron bombardment, with more recent ground-based H$\alpha$ line observations acquired at time resolution of the order of seconds or less \citep[e.g.][]{trottet00, radziszewski07, radziszewski11}, revealing very short (down to $< 3$\,s) time lags relative to HXR lightcurve maxima. Observations from the Solar Maximum Mission showed that HXR signals also coincide in time with emission formed in the overlying transition region \citep[e.g.][]{cheng81, poland82, tandberg83}. Unlike the ground-based H$\alpha$ observations, the ultraviolet (UV) emission of ions forming therein (such as \ion{Si}{4}, \ion{C}{4}, \ion{O}{4}, \ion{O}{5}) was initially observed at low spatial resolution, limiting the identification of emission sources. Valuable insights \citep[see e.g. the review][and references therein]{fletcher11} into the spatial and temporal coherence between the lower-atmospheric emission and HXR sources were brought by the Transition Region and Coronal Explorer \citep[TRACE;][]{handy99}. This imaging telescope provided excellent 1\arcsec\ spatial resolution and cadence as high as 2\,s. However, this cadence was only available with a single passband at a time, with no multi-temperature coverage \citep[e.g.][]{hudson06, fletcher09, liu13}.

Longer time lags of the order of 10\,s, observed between H$\alpha$ and HXR lightcurve maxima, were attributed to heating via thermal conduction following \textsl{in-situ} coronal energy release \citep[e.g.][]{trottet00, radziszewski07, radziszewski11}. Relatively-slower chromospheric response to this heating mechanism is given by the propagation of conduction fronts along reconnected field lines at ion sound speeds of a few 100 -- 1000\,\kps. Hereafter we refer to this scenario as `thermal conduction' driven flares; thermal conduction of course plays a role in all flares, but here we refer to this as the dominant means by which energy is transported from the corona to the lower atmosphere. Another proposed energy transport mechanism that operates relatively slower compared to non-thermal particles are downward propagating Alfv\'{e}n waves generated from the reconnection site. Alfv\'{e}n waves travel from the corona at $c_\mathrm{A} = 10^3 - 10^4$\,km\,s$^{-1}$, reaching the chromosphere within a few seconds of their generation (assuming this occurs at the apex of loops). Most of the Poynting flux dissipates through the transition region and chromosphere, but the location of the deposition depends on wave parameters and the ionization fraction \citep[][]{haerendel09, reep16, reep18b}. The Alfv\'{e}n wave heating can effectively drive chromospheric emission \citep[][]{kerr16,reep18b}, however these models still require observational validation.

Building upon the extensive literature on the atmospheric response to flare energy deposition, relative timings of intensity enhancements of lines formed at different atmospheric depths can in principle trace the progression of heating throughout the solar atmosphere. Spectroscopic observations from the Coronal Diagnostic Spectrometer \citep[CDS;][]{harrison95} identified time lags at scales of minutes between emission formed at flare $\approx 8$\,MK and lower ($0.01 - 0.1$\,MK) temperatures, enabling the differentiation between energy transport mechanisms \citep[e.g.][]{brosius04, brosius10, brosius12a}. However, a detailed diagnostics of heating propagation focused on the lower atmosphere, where the bulk of the flare radiative response originates at rapid timescales of seconds, remains largely unexplored. This is due to the lack of simultaneous observations of the chromosphere and transition region at very high, ideally sub-second, cadence and with sufficient spatial resolution.

Since 2013, unprecedented simultaneous observations of the chromosphere and transition region have been made possible by the Interface Region Imaging Spectrograph \citep[IRIS;][]{depontieu14}. By observing spectral lines formed at different heights in the lower atmosphere, IRIS provides invaluable flare diagnostics  \citep[see e.g. the review][and references therein]{depontieu21}. For instance, the transition region \ion{Si}{4} 1402.77\,\AA~and the chromospheric \ion{Mg}{2} k (2796.35\,\AA) and \ion{Mg}{2} h (2803.52\,\AA) lines are among the most commonly studied spectra formed in flaring conditions, often in tandem with HXR or microwave radio signatures of energetic electrons \citep[e.g.][]{liu15, li15a, warren16, brosius18, polito23frass}. The newly-developed IRIS flare observing programs with cadence below 1\,s, are particularly well-suited to study the rapid response of the chromosphere and transition region to flare energy deposition. These invaluable datasets have already been utilized in detailed studies of chromospheric condensation \citep[][]{lorincik22, wang23}. High-cadence IRIS observations of a C4-class flare from 2022 September 25 recently provided key evidence of slip-running reconnection manifested in super-Alfv\'{e}nic apparent motions of flare ribbon kernels with velocities reaching thousands km\,s$^{-1}$ \citep{lorincik24}. Some of the kernels were also captured by the IRIS slit, providing spectral observations of different lines well suited to study the response of the chromosphere and the transition region to energy deposition simultaneously at very high cadence.

In recent years, IRIS observations are often used alongside flare simulations modeled with radiative hydrodynamic loop numerical models such as \texttt{RADYN}, \texttt{HYDRAD}, or \texttt{FLARIX} \citep[see e.g. recent reviews by][]{kerr22,kerr23}. To model flare emission in such codes, energy can be injected in several ways: thermalization of non-thermal electron beams \citep[e.g.][]{allred05,allred15,polito23}, thermal conduction-only from flare heating of the corona \citep[e.g.][]{testa14,polito18}, or an approximated form of dissipation of Alfv\'{e}n waves \citep[e.g.][]{kerr16}. Where and how fast the energy is deposited, the subsequent plasma response, and the resulting emission depends on the heating parameters and the initial physical conditions of the atmosphere. When modeled as a power-law energy distribution, the electron beam parameters include the power carried by non-thermal electrons, $P_{NTE}$, the electron spectral index, $\delta$, that describes its slope, and the minimum energy of the distribution, the low-energy cutoff $E_{c}$. For example, \citet{allred05} showed that in the atmosphere heated by strong non-thermal flux ($F = 10^{11}$\,erg\,cm$^{-2}$\,s$^{-1}$), both chromospheric and transition region line intensities increase rapidly after the heating onset. \ion{Ca}{2} k intensities however peak only after the chromospheric condensation fronts reach the lower chromosphere. \citet{testa14} noted on differences between the response of the transition region and chromosphere to electron beam parameters. Among other, they found that \ion{Mg}{2} intensities have much weaker dependency on small changes of $E_{c}$ than those of the \ion{Si}{4} line. The nanoflare model investigation of \citet{polito18} showed that electrons can penetrate deeper into the atmosphere if loops have lower initial densities, thus driving a stronger plasma response to the heating in chromospheric and transition region lines, compared to that observed in denser loops for the same amount of energy released. Lastly, in addition to the parameters discussed above, the penetration depth of non-thermal electron beams can depend on $\delta$. \citet{allred15} nicely illustrate this in their analysis of the stratification of heating rates in different solar/stellar atmospheres. Overall, the more power is carried by higher energy electrons, the more deeply a particle beam penetrates.

In this study, we revisit the observations of the 2022 September 25 flare to study, for the first time, the stratified response of the lower atmosphere to flare heating on short timescales. We found intriguing delays between \ion{Mg}{2} line intensity peaks relative to the \ion{Si}{4} and \ion{C}{2} emission observed in a flare ribbon. Physical mechanisms causing these delays are investigated using a grid of \texttt{RADYN} simulations heated by different energy transfer mechanisms. The manuscript is structured as follows. Section \ref{sec:data} introduces datasets analyzed in Sections \ref{sec:overview} and \ref{sec:observations}. Section \ref{sec:modeling} details the investigation of \texttt{RADYN} models and their comparison to the observations. The results of our analysis are discussed in Section \ref{sec:discussion} and summarized in Section \ref{sec:summary}. 

\section{Data}
\label{sec:data}

\subsection{Spectroscopic and imaging observations}

In this study we analyze spectroscopic observations of a C4-class flare  observed by IRIS on 2022 September 25. IRIS provides high-resolution spectroscopy in two far-ultraviolet (FUV, 1331.6\,\AA~ -- 1358.4\,\AA~and 1380.6\,\AA~ -- 1406.8\,\AA) and one near-ultraviolet (NUV, 2782.6\,\AA~ -- 2833.9\,\AA) bands. These bands contain a multitude of lines formed across a wide range of temperatures of log ($T$ [K]) = $3.7 - 7$, with spectral resolutions of 0.026\,\AA~and 0.053\,\AA~and spatial resolutions of $\approx0.33\arcsec-0.4\arcsec$, in the FUV and NUV channels, respectively. The Slit-Jaw Imager (SJI) of IRIS provides context imaging observations with a field of view (FOV) of up to 175$\arcsec$ $\times$ 175$\arcsec$. The event under study was observed in a high-cadence flare observing program in the sit-and-stare mode with the temporal resolution of $\approx$0.92\,s for spectra and $\approx$1.8\,s for the SJI observations. Onboard spatial and spectral binning of 2 was used. The spectral observations were carried-out with exposure times of 0.3\,s (FUV band) and 0.19\,s (NUV band) in four spectral windows centered on the \ion{C}{2} 1335 \& 1336\,\AA~lines, \ion{Si}{4} 1403\,\AA~line, \ion{Mg}{2} k 2796\,\AA~line, and NUV continuum at 2814\,\AA. In addition to the spectral observations, the dataset contains SJI 2796\,\AA~and 1330\,\AA~imagery of the chromospheric and transition region emission, respectively. The IRIS observations were processed using packages available within the \texttt{SolarSoft} and \texttt{irispy\_lmsal} software libraries. The level-2 science-ready IRIS observations were despiked and subsequently smoothed using a sliding boxcar to reduce the noise while preserving the overall trends (Section \ref{sec:obs_properties}). The boxcar size was two pixels in the time dimension and 4 and 2 wavelength pixels in the FUV and NUV channels, respectively, as to account for the difference in the wavelength resolution between these channels. The spectra were finally normalized by exposure times. Using the \texttt{iris\_orbitvar\_corr\_l2} routine we verified that no further wavelength correction due to IRIS orbital variations was needed.

\begin{figure}[!t]
\centering
\includegraphics[width=8.50cm, clip,   viewport=00 05 630 280]{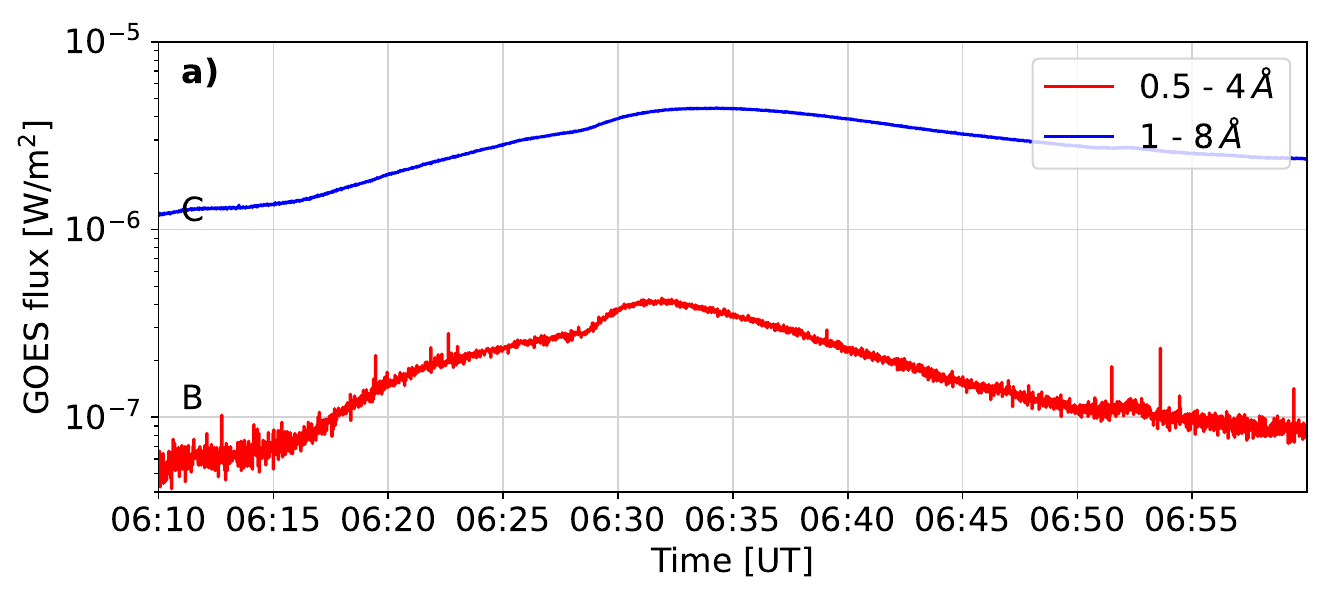} 
\includegraphics[width=8.50cm, clip,   viewport=00 05 630 280]{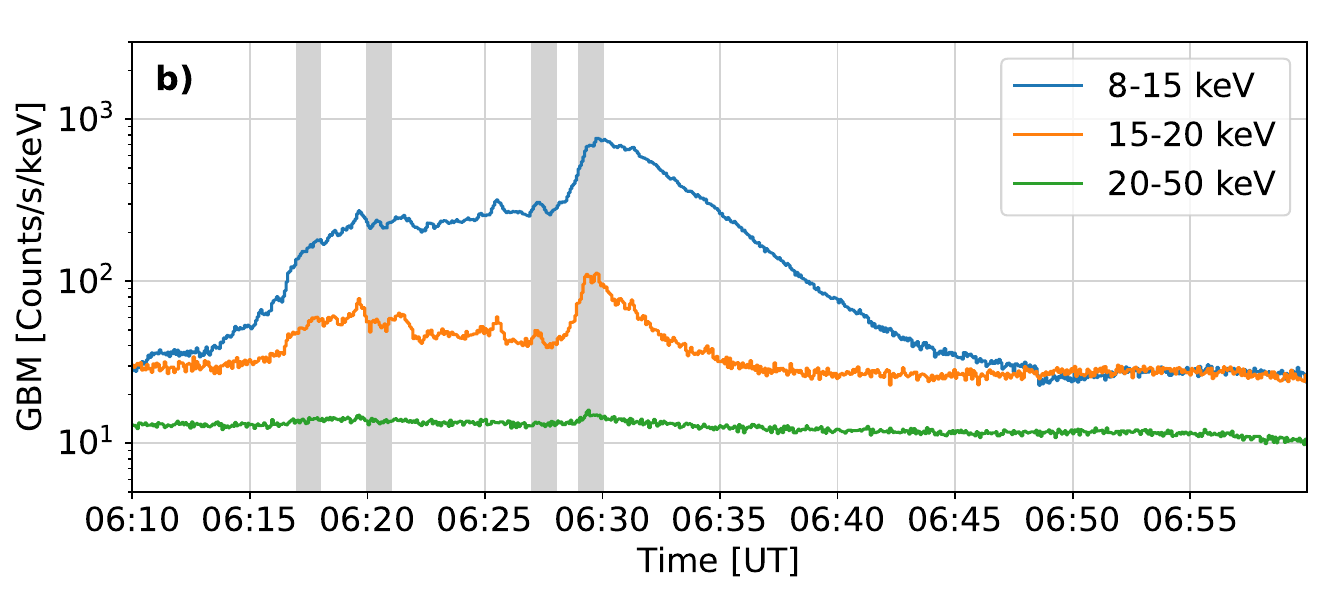} 
\\
\includegraphics[width=18.00cm, clip,   viewport=-15 05 1295 410]{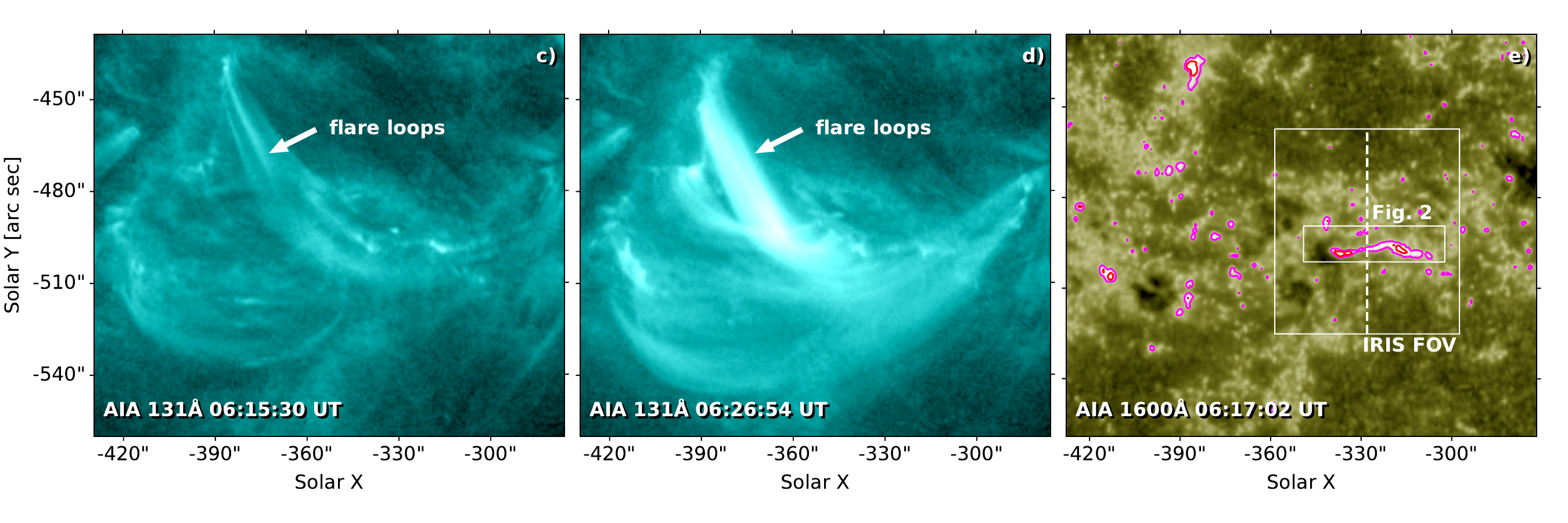} 
\caption{Context observations of the event. Panels a and b present flare SXR and HXR lightcurves measured in different wavelength and energy bands by GOES and \textsl{Fermi}/GBM, respectively. The grey stripes in panel b delineate time intervals of successful fits to the GBM spectra. Panels c -- d present AIA 131\,\AA~observations of flare loops developing during the flare. Observations of the flaring region in the 1600\,\AA~channel of AIA with the IRIS FOV and the FOV of Figure~\ref{fig:sji_ribbon} are shown in panel e. The magenta and red contours correspond to 30 and 50\% of the maximal intensity. \label{fig:overview}}
\end{figure}

We further investigated imaging observations of the event performed by the Atmospheric Imaging Assembly \citep[AIA;][]{lemen12} onboard the Solar Dynamics Observatory \citep[SDO;][]{pesnell12}. AIA captures full-disk images of the Sun at a spatial resolution of $\sim1\arcsec.5$ at a cadence of 12\,s and 24\,s, depending on the filter channel. Snapshots from the AIA 131\,\AA~channel, in flaring conditions dominated by the \ion{Fe}{21} line emission (log ($T$ [K]) = 7.05) \citep[e.g.][]{odwyer10}, were used to inspect flare loops forming during the event. Flare ribbons were studied in AIA 1600\,\AA~observations (log ($T$ [K]) = 5.0) with a major contribution from the transition region \ion{C}{4} lines, \ion{Si}{1} continua, as well as numerous chromospheric lines \citep[][]{simoes19}. The AIA observations were handled and visualized in the \texttt{SunPy} environment \citep{sunpy20}.  The full-disk AIA observations and SJI dataset were slightly misaligned. In order to co-align the two instruments, we determined the shift between the instrument coordinate systems based on comparing locations of bright ribbon emission observed in the SJI 1330\,\AA~and AIA 304\,\AA~as well as 1600\,\AA~snapshots, while taking the AIA observations for reference. The shift between the instruments determined via this method was $X$ = $-2\arcsec.2$, $Y$ = $-2\arcsec.4$.

\subsection{SXR observations and Fermi/GBM spectroscopy} 
\label{sec:data_fermi}

The soft X-ray (SXR) flux of the Sun was studied using data from the X-ray Sensor (XRS) onboard the GOES satellite. HXR observations from the Gamma-ray Burst Monitor \citep[GBM;][]{meegan_2009} onboard the \textsl{Fermi}/GBM space telescope were used to analyze the non-thermal electron sources during the flare. \textsl{Fermi}/GBM detects gamma rays and X-rays in the energy range between 8 keV and 40 MeV at the time resolution of 1\,s. While both GOES and \textsl{Fermi}/GBM observe the Sun-as-a-star emission, no other flare above A class occurred in the time period under consideration. The GBM spectra were analyzed via the \texttt{OSPEX} software. We employed the \texttt{v\_th} and \texttt{thick2} routines to fit spectra with thermal and thick-target bremsstrahlung models, respectively. {The \texttt{thick2} model represents thick-target bremsstrahlung emission from a population of non-thermal electrons following a single power-law energy distribution. Background emission was subtracted from the spectra first. This was determined by fitting by a polynomial to pre- and post-flare time intervals where no flare emission was observed.}  Spectra were integrated over 1-minute long periods centered at 06:17:30, 06:20:30, 06:27:30, and 06:29:30 UT (grey stripes in Figure~\ref{fig:overview}b). No traces of non-thermal emission were found before $\approx$06:16 UT, even using longer integration times. From this fitting we obtained $\delta$, $E_{c}$, and the power carried by non-thermal electrons, $P_{NTE}$. {Since \textsl{Fermi} does not image the X-ray sources, and therefore constrain the flare area $A$, which itself is necessary to obtain} the non-thermal energy flux densities $F = P_{NTE}/A$, we opted to constrain the footpoint area via UV ribbon imagery \citep[e.g.][]{rubiodacosta15, rubiodacosta16, graham20, polito23frass}. We first assumed that the energy released during the event was distributed to both ribbons defined by the 50\% intensity contours in the AIA 1600\,\AA~observations (red contours in Figure~\ref{fig:overview}e), providing the upper area limit $A_\mathrm{max}$ and thus the lower limit on $F$. The higher limit on $F$ was determined from the area $A_\mathrm{min}$ within bright kernels composing the southern ribbon observed in the 1330\,\AA~channel of SJI. The 50\% intensity level, designated by the red contours in Figure~\ref{fig:sji_ribbon}, was again used to delineate the pixels whose area was summed. {The lower limit on $A_\mathrm{min}$ was found to be $<10^{16}$ cm$^2$, roughly equivalent to 17 IRIS pixels. The broad range of $A$ listed in Table~\ref{tab:fermi} thus accounts for the possibility of non-thermal energy distributed to both small-scale IRIS kernels as well as extended ribbon structures.} 

\section{Overview of the event}
\label{sec:overview}

\begin{figure*}[!t]
\centering
\includegraphics[width=8.50cm, clip,   viewport=05 40 565 170]{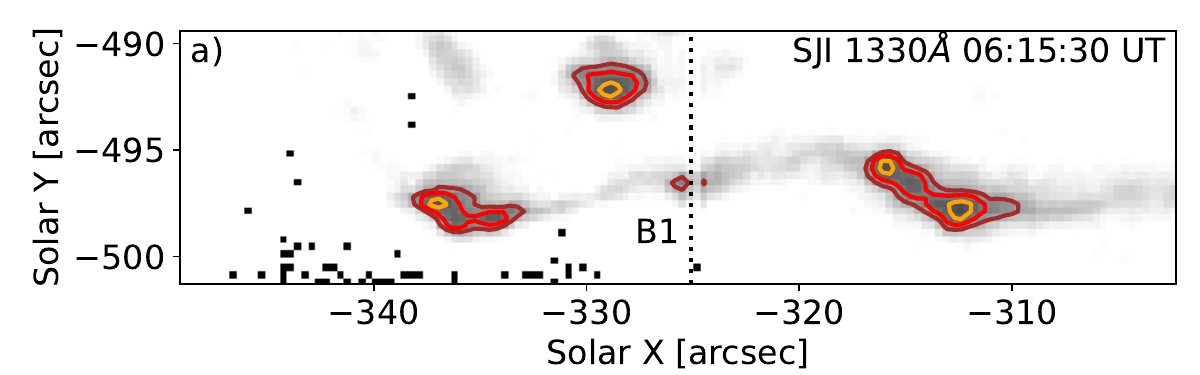} 
\\
\includegraphics[width=8.50cm, clip,   viewport=05 40 565 170]{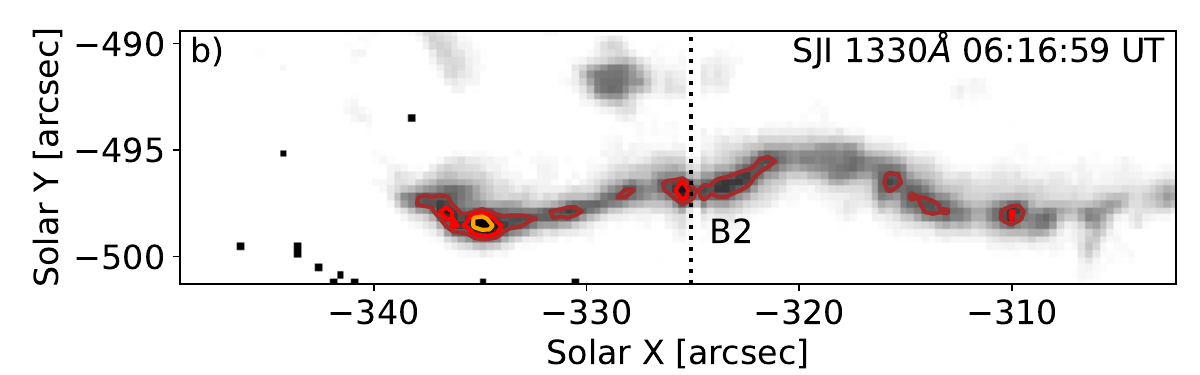} 
\\
\includegraphics[width=8.50cm, clip,   viewport=05 40 565 170]{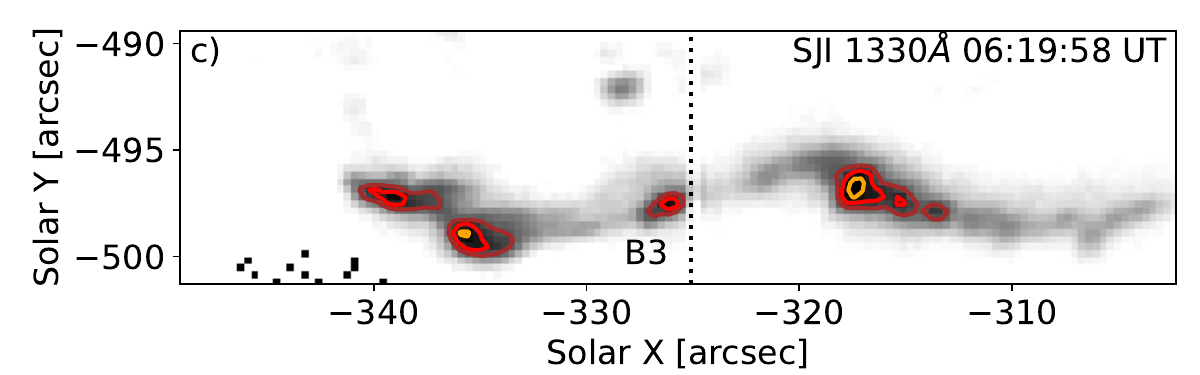} 
\\
\includegraphics[width=8.50cm, clip,   viewport=05 00 565 170]{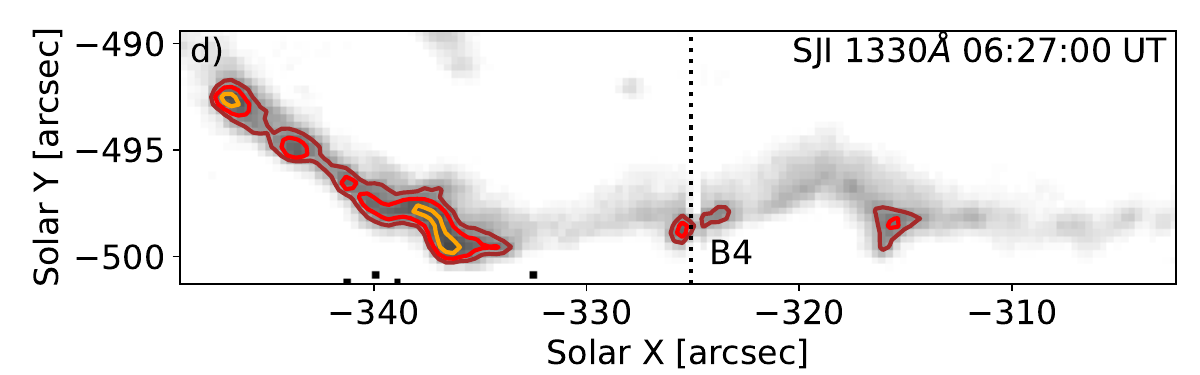} 
\caption{SJI 1330\,\AA~observations of the south-western ribbon. The brown, red, and orange contours correspond to 30, 50, and 80\% of the maximal intensity observed in each frame. Annotations B1 -- B4 outline flare kernel brightenings that crossed the slit of IRIS (vertical dotted line). {An animated version of this figure is available in the online journal. The animation covers the period between 06:15 -- 06:27:30 UT and the animation duration is 17\,s.}\label{fig:sji_ribbon}}
\end{figure*}

The C4-class flare under study occurred in the NOAA active region 13107 on 2022 September 25 and its context observations are presented in Figure~\ref{fig:overview}. The time evolution of the SXR flux detected in the $0.5 - 4$\,\AA~(red) and $1 - 8$\,\AA~(blue) channels of the GOES/XRS instrument is shown in panel a. The GOES/XRS data indicate that the flare started after 06:10 UT and peaked between 06:30 and 06:35 UT. A similar trend can be seen in the $8 - 15$\,keV HXR flux lightcurve (blue, panel b) detected by GBM. At relatively-higher energies ($15 - 20$\,keV; orange) there are multiple peaks between approximately 06:15 - 06:35 UT, corresponding to the impulsive and peak phases. Very little emission was detected at higher energies ($> 20 - 50$\,keV; green). Parameters of fits to the non-thermal emission in four time intervals during the impulsive phase of the flare are listed in Table~\ref{tab:fermi}/GBM. The energy flux density was moderate, ranging between $F=5 \times 10^8$ and $5 \times 10^{10}$\,ergs\,cm$^2$\,s$^{-1}$, depending on the intensity threshold used for the estimation of the footpoint area (see Section \ref{sec:data_fermi}). The highest energy flux density corresponds to the fit parameters obtained soon into the impulsive phase of the flare, assuming the smallest footpoint area. The lowest energy flux was measured towards the end of the impulsive phase. The other relevant fit parameters were $E_c = 12.8 - 18.2$\,keV and $\delta \approx 3.5 - 4.1$.

The second row of Figure~\ref{fig:overview} presents AIA 131\,\AA~observations (panels c, d) of the flare during its impulsive phase. Flare loops started to appear soon after the flare onset and the development of the flare loop arcade continued throughout the impulsive phase. The footpoints of these flare loops are found in a pair of flare ribbons, highlighted using the red and magenta contours in panel e as observed in AIA 1600\,\AA. The FOV of IRIS captured a flare ribbon located in the south-west. The evolution of this ribbon during the impulsive phase of the flare is detailed in Figure~\ref{fig:sji_ribbon} {and its animated version}. The background images correspond to SJI 1330\,\AA~filter snapshots at 06:15:30, 06:16:59, 06:19:58, and 06:27:00 UT (panels a -- d) {corresponding to the start of \textsl{Fermi} integration periods} during the impulsive phase. The brown, red, and orange contours outline intensity peaks corresponding to 30, 50, and 80\% of the maximal intensity in each frame. The contours outline flare kernel brightenings composing the ribbon and moving along it \citep{lorincik24} as a consequence of magnetic slipping reconnection \citep[][]{aulanier06, dudik14}. Some of the moving kernels crossed the IRIS slit, inducing enhancements visible across spectra formed in the chromosphere and the transition region. {In the remainder of the manuscript we focus on brightening events, referred to as B1 -- B4, characterized by the strongest spectral enhancements induced by the brightest kernels crossing the slit.} While {the event} B1 occurred before the onset of the non-thermal event, the B2 -- B4 correspond to the first three out of four periods of the \textsl{Fermi}/GBM fits.

\begin{deluxetable*}{ccccc}[h]
\tablecolumns{5}
\startdata
\tablehead{
Integration times {[UT]}    & 06:17 -- 06:18 & 06:20 -- 06:21 & 06:27 -- 06:28 & 06:29 -- 06:30 \\ 
IRIS brightening    							     & B2	&	B3   &	B4	 &		-		}
$P_{NTE}$ [10$^{26}$ erg\,s$^{-1}$]                 & 3.37  & 2.36  & 1.43  & 1.8      \\ 
$E_c$ [keV]                                         & 12.78 & 15.9  & 13.95 & 18.19     \\ 
$\delta$                                            & 3.95  & 4.05  & 3.48  & 4.06      \\ 
$A_\textrm{min}$ [10$^{16}$ cm$^2$]              & 0.76  & 2.56  & 6.34  & 1.63 \\
$A_\textrm{max}$ [10$^{16}$ cm$^2$]              & 20.63 & 12.3  & 27.06 & 31.43 \\
Min. energy flux [10$^9$\,ergs\,cm$^2$\,s$^{-1}$]   & 1.63  & 1.91  & 0.53  & 0.57      \\ 
Max. energy flux [10$^9$\,ergs\,cm$^2$\,s$^{-1}$]   & 44.52 & 9.2   & 2.25  & 11.07     \\
\enddata
\caption{Non-thermal fit parameters from \textsl{Fermi}/GBM at four instants during the impulsive phase of the flare. \label{tab:fermi}}
\end{deluxetable*}

\section{IRIS spectral analysis}
\label{sec:observations}

\subsection{Properties of line profiles}
\label{sec:obs_properties}

Our spectral analysis was primarily focused on the \ion{Si}{4} 1402.77\,\AA, \ion{Mg}{2} k 2796.35\,\AA, and \ion{C}{2} 1334.53\,\AA~lines. The properties of spectra were initially assessed via calculating the 0th (total intensity), 1st (centroid), and 2nd (variance) moments of the line profiles. The rest wavelengths of the \ion{Si}{4} and \ion{C}{2} lines were adapted from \citet[][]{sandlin86}. For the \ion{Mg}{2} line we used the reference value listed in the \texttt{irisspectobs\_define} routine in \texttt{SolarSoft}. The non-thermal velocity ($v_{\mathrm{nt}}$) of the \ion{Si}{4} line was computed using Equation (1) of \citet{lorincik22}. We assumed the square root of the variance to be equivalent to the standard deviation of the Gaussian used in the formula. The moments of the \ion{Si}{4} and \ion{C}{2} lines were obtained in the wavelength range of $\lambda_0 = \pm 0.6$\,\AA, while a slightly larger range of $\lambda_0 = \pm 1.0$\,\AA~was needed for the \ion{Mg}{2} line. These wavelength ranges were determined upon manual inspection of spectra under study (Appendix \ref{sec:appendix_a}). Prior to the calculation of the moments, average continua levels were subtracted from the spectra. The Si IV spectral window sometimes exhibits the presence of weak transition region lines of ions such as \ion{O}{4} and \ion{S}{4} \citep[e.g.][]{polito16}. As these lines were not present in our dataset (Appendix \ref{sec:appendix_a}), likely due to the low exposure times, they did not affect the average FUV continuum levels. On a similar note, by choosing the relatively-narrow spectral region around the \ion{Mg}{2} k line we avoided contributions from enhanced near-continuum emission typical for quiet Sun conditions. This approach allows us to estimate the NUV continuum levels in the same way as the FUV continuum for the \ion{Si}{4} and \ion{C}{2} lines. Analysis of \ion{Mg}{2} intensities corresponding to the line's k3 component is presented in Appendix \ref{sec:appendix_b}.

\begin{figure*}[b]
\centering
\includegraphics[width=8.80cm, clip,   viewport=00 05 620 510]{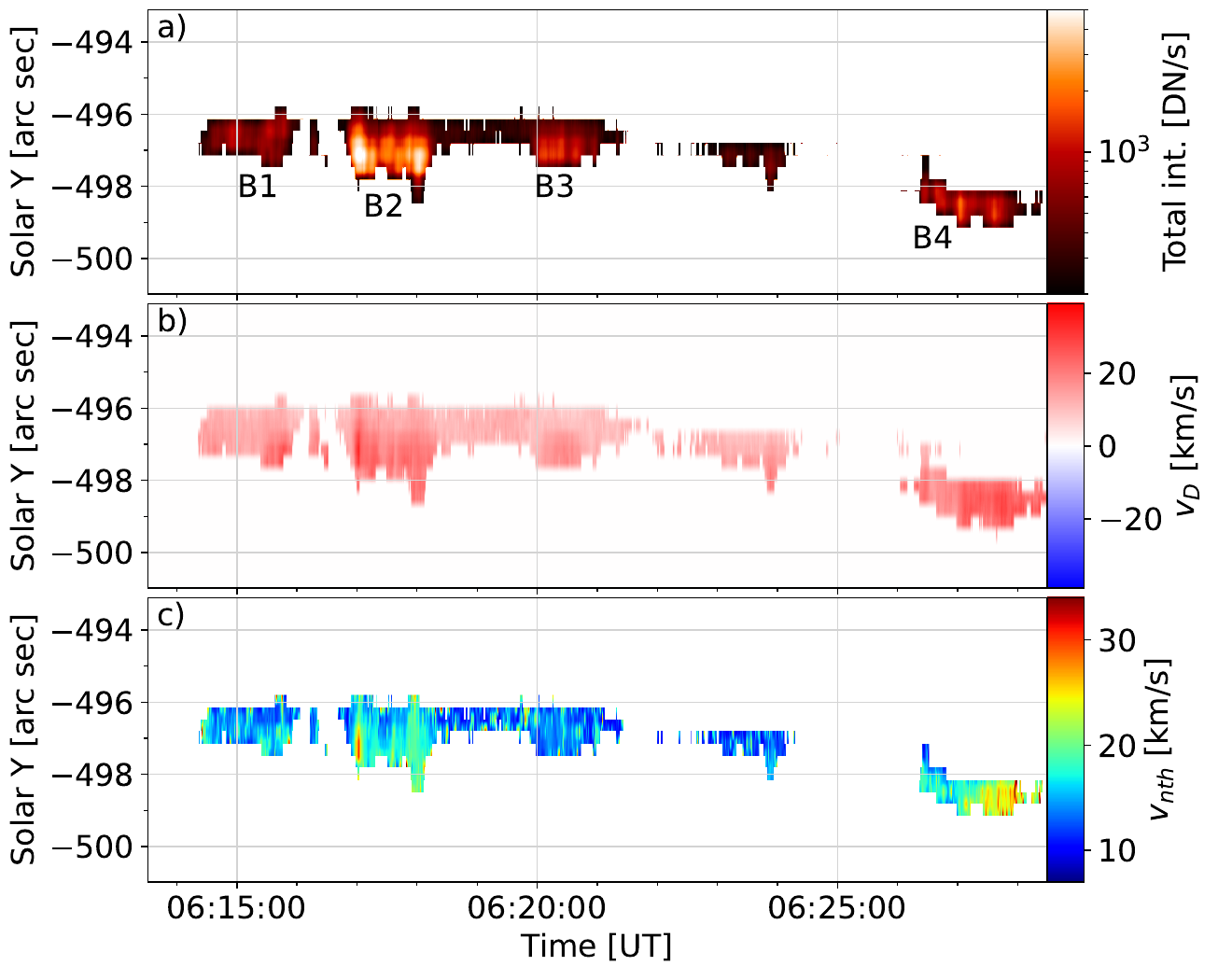} 
\\
\caption{Maps of the \ion{Si}{4} 1402.77\,\AA~line properties determined via the moment analysis along the slit of IRIS in the vicinity of the ribbon. Enhancements imprinted by the kernel brightenings B1 -- B4 are indicated. \label{fig:inspection}}
\end{figure*}

Figure~\ref{fig:inspection} shows space-time maps of \ion{Si}{4} line spectral properties determined via the moment analysis. Since the ribbon under study only exhibited negligible separation motion \citep{lorincik24}, the enhancements of spectral properties during B1 -- B4 were concentrated to a narrow region along the slit between $Y = -496\arcsec$ and $- 499\arcsec$. The \ion{Si}{4} line exhibited the highest intensities during the brightening B2 {imprinted by a particularly strong kernel present under the IRIS slit between $\approx$06:17 -- 06:18 UT (see the animation accompanying Figure~\ref{fig:sji_ribbon}).} The line was consistently redshifted (panel b, see also Appendix \ref{sec:appendix_a}), indicative of moderate downflows of $v_{\mathrm{D}} < 25$\,\kps, a behavior typically reported during flares \citep[e.g.][]{warren16, yu20, wang23}. The map of the non-thermal broadening $v_{\mathrm{nt}}$ plotted in panel c indicates a transient episode of increase to $v_{\mathrm{nt}} = 30$\,km\,s$^{-1}$ at 06:17 UT during the B2. Another increase of the non-thermal broadening was observed toward the end of the time period under study. During the B4, the line broadening typically ranged between $v_{\mathrm{nt}} = 15 - 35$\,km\,s$^{-1}$, compared e.g. to $v_{\mathrm{nt}} = 10 - 20$\,km\,s$^{-1}$ during the B3. Note that the \ion{Si}{4} line properties were also enhanced in the time period between 06:23 and 06:24 UT at the position $Y \approx -497\arcsec$, but the spectra were too noisy and we did not analyze them further. 

\begin{figure*}[!b]
\centering
\includegraphics[width=8.90cm, clip,   viewport=00 00 780 430]{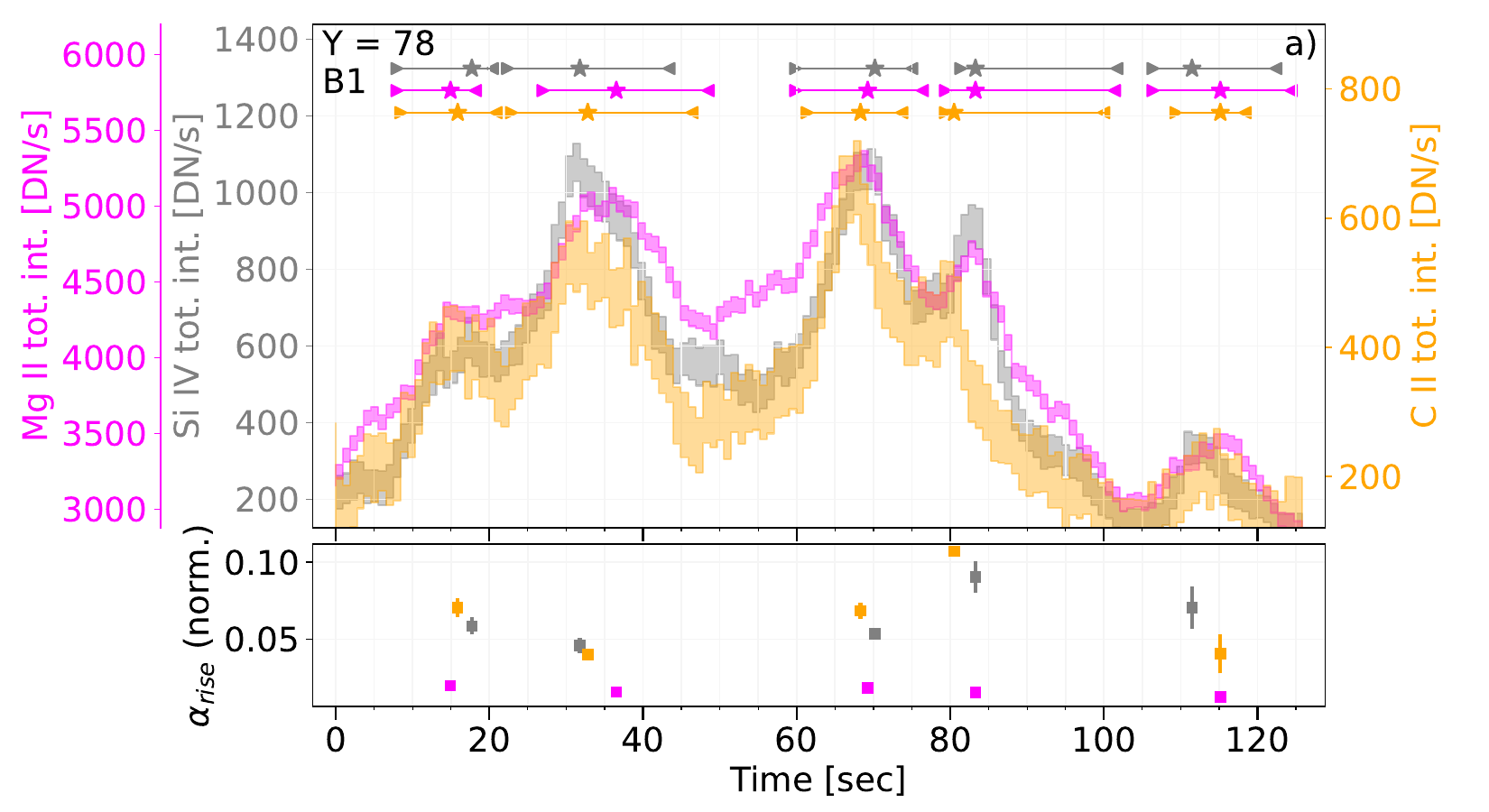} 
\includegraphics[width=8.90cm, clip,   viewport=00 00 780 430]{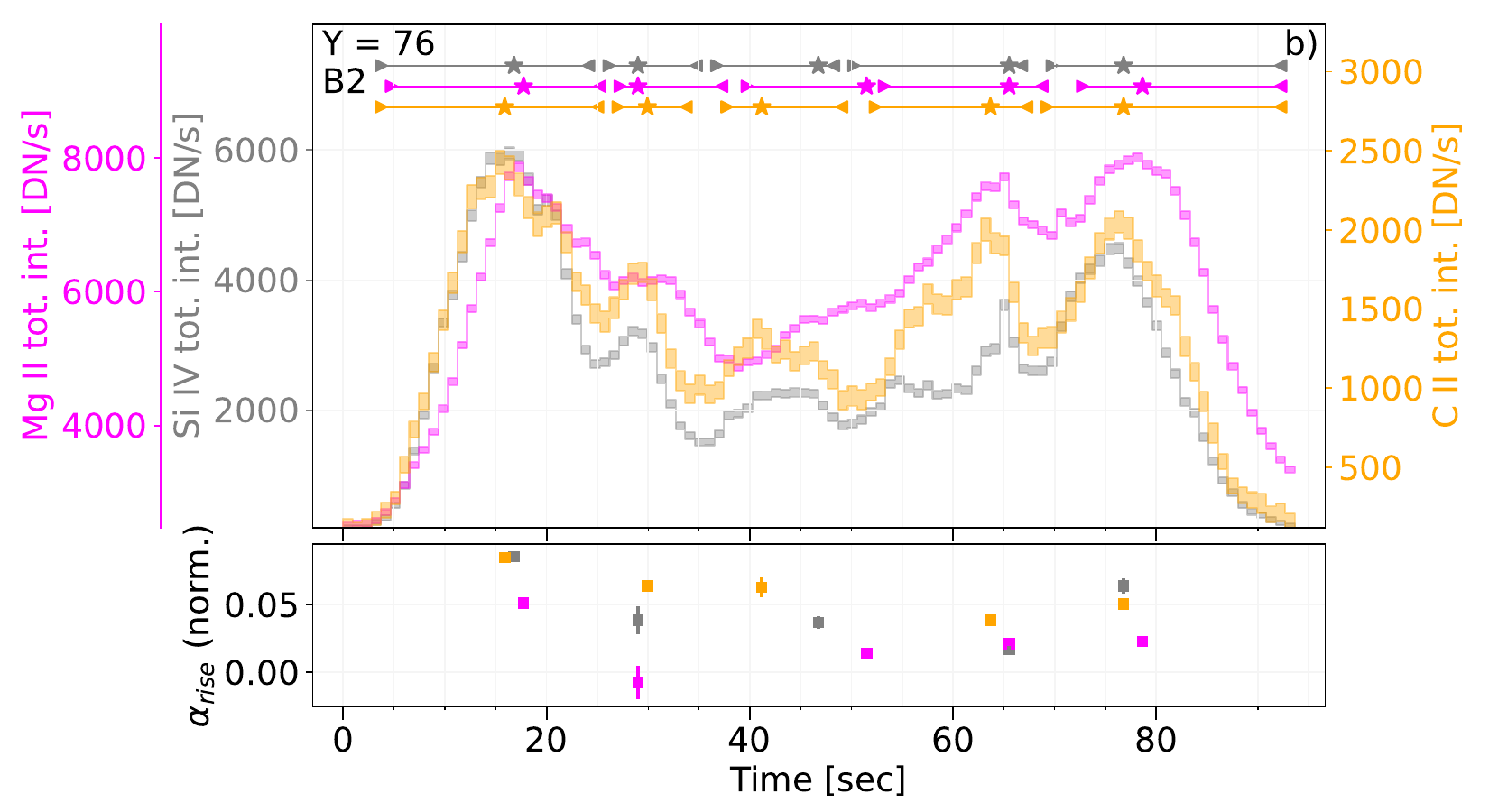} 
\\
\includegraphics[width=8.90cm, clip,   viewport=00 00 780 430]{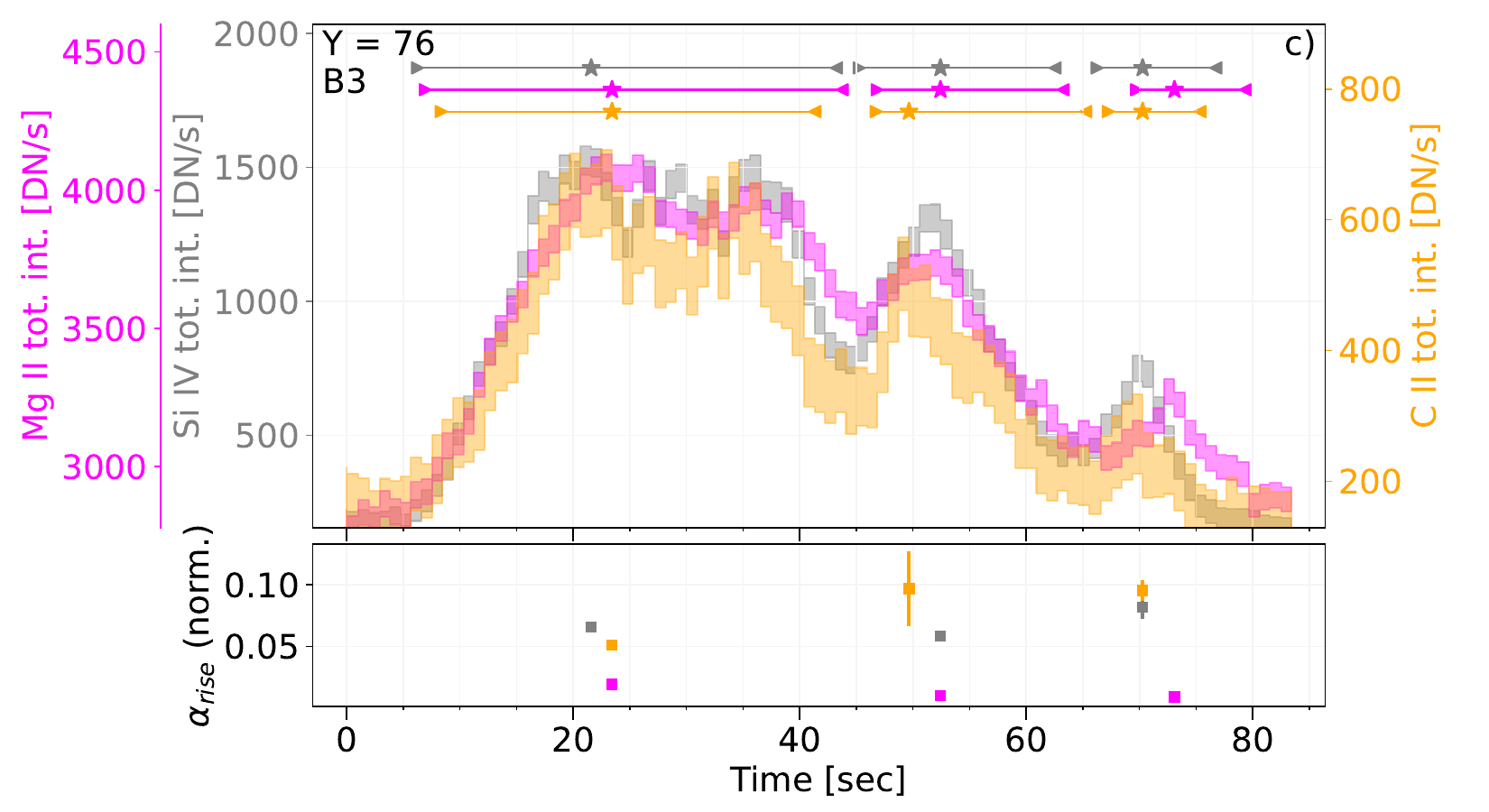} 
\includegraphics[width=8.90cm, clip,   viewport=00 00 780 430]{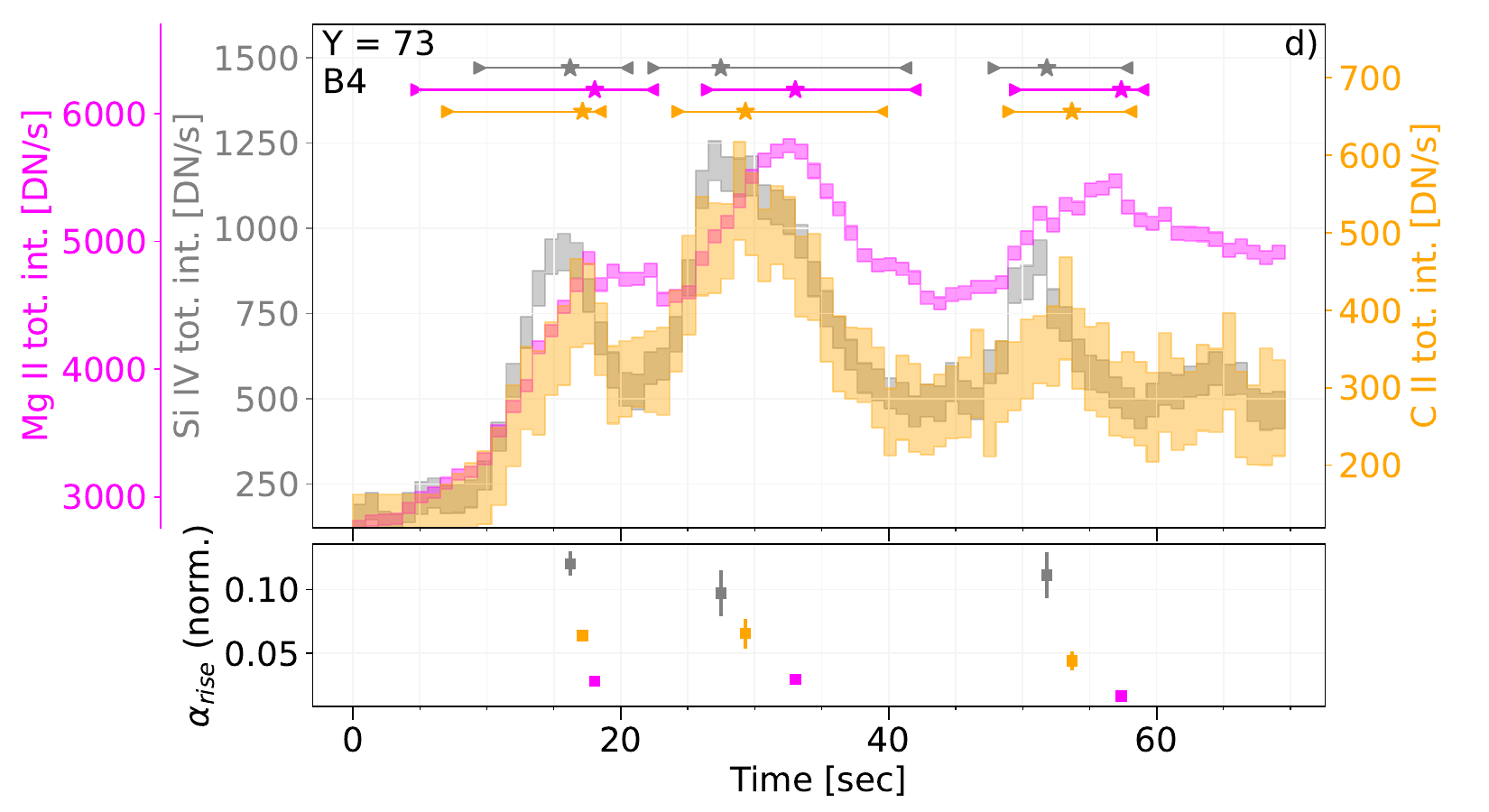} 
\caption{\ion{Si}{4} 1402\,\AA~(grey), \ion{Mg}{2} 2796.35\,\AA~(magenta), and \ion{C}{2} 1334.53\,\AA~(orange) line lightcurves during the B1 -- B4 brightening events in the ribbon observed by IRIS. The colored horizontal bars above the lightcurves indicate manually-identified peak intervals. The triangle symbols are the left and right bases (peak onset and end times), while the stars designate the maximum of each peak. The bottom part of each panel indicates the slopes of each peak after normalization. \label{fig:lightcurves}}
\end{figure*}

\subsection{Intensity lightcurves and peaks}
\label{sec:obs_lightcurves}

Figure~\ref{fig:lightcurves} presents \ion{Si}{4} 1402.77\,\AA~(grey), \ion{Mg}{2} k 2796.35\,\AA~(magenta) and \ion{C}{2} 1334.53\,\AA~(orange) lightcurves obtained via calculating the total intensity of the lines. Panels a -- d show lightcurves obtained at selected positions along the slit during the brightening events B1 -- B4. These lightcurves are characterized by an overall intensity increase from previous levels by a factor of few up to an order of magnitude. Several peaks can be distinguished along each lightcurve, with a duration between a few to roughly 20\,s. The peaks  appear to be quasi-periodic. Wavelet analysis \citep{torrence98} of the lightcurves revealed $16 - 18$\,s periods above $2 \sigma$ significance  occurring in short episodes. While some peaks appeared repeatedly one after another, the lightcurves also exhibited short periods where no peaks were present (see e.g. $t = 50$\,s in panel a or $t = 40$\,s in panel d). Note that intensity peaks with characteristics similar to those in the spectra were also observed in SJI 1330\,\AA~data in flare ribbon pixels adjacent to the slit of IRIS (not shown). This suggests that the intensity peaks were driven by the passage of slipping flare kernels under the slit (see Section \ref{sec:discussion}).

To study the timings and characteristics of the individual peaks, the lightcurves were segmented in order to isolate time periods when the peaks occurred in each ion separately. The peak time intervals are designated using the horizontal colored lines above the lightcurves in the same figure. Manually-identified onset time of a particular peak, referred to as the `left base', is indicated using the colored triangles ($\triangleright$) above the lightcurves. The time when the intensity drops before rising again (or the end of the lightcurve) is referred to as the `right base' ($\triangleleft$). The left and right bases were used to calculate the amplitude (in DN/s) of each peak as

\begin{equation}
I_{\text{max}} - \frac{|I_{\text{left base}} + I_{\text{right base}}|}{2}.
\end{equation}

When designating the peak time intervals, we only focused on peaks (and troughs separating them) common to all spectra. For instance, in Figure~\ref{fig:lightcurves}a we identified five major peaks across the three lightcurves. A distinct behavior is apparent in the time interval between 35 -- 70 seconds in Figure~\ref{fig:lightcurves}b. During this period,  the \ion{Si}{4} and \ion{C}{2} lines exhibited three clear peaks, while only one strong ($t$ = 60\,s) and one very weak ($t$ = 45\,s) peak are visible in the \ion{Mg}{2} line lightcurve. In such cases, the peak bases were adapted from the lightcurve exhibiting fewer peaks. Despite our efforts in the precise determination of the peak intervals, certain challenges arose, particularly in \ion{C}{2} lightcurves with higher noise level or \ion{Mg}{2} k3 lightcurves (Appendix \ref{sec:appendix_b}). Peaks visible in the lightcurves were also occasionally separated by flat plateau rather than deep troughs, complicating the selection of the interval boundaries. In order to mitigate the human bias in designating the peaks and their bases, the selection was also performed via an automatic routine. This process, described in Appendix \ref{sec:appendix_c}, confirmed the results of our manual analysis, however at the expense of using data with a higher grade of smoothing. In the remainder of the manuscript we thus focus on peaks determined manually. 

The most intriguing characteristic of the lightcurves is that the times when the individual peaks reach maxima differ across the three ions. These instants are for clarity indicated using the star symbols along the respective peak intervals above the lightcurves in Figure~\ref{fig:lightcurves}. The \ion{Si}{4} 1402.77\,\AA~line peaks (grey stars) and \ion{C}{2} line peaks (orange stars) were typically occurring before those visible along the \ion{Mg}{2} 2796.35\,\AA~lightcurves (magenta stars). The horizontal separation between the stars is a measure of the delay between the line peaks, and is typically of the order of a few seconds or less. To our knowledge, delays between the line emission formed in different regions of the lower solar atmosphere at these timescales have not yet been reported. This is the key result of our study and the observable we focus on in the remainder of this manuscript.

The delays between the peak maxima alone do not account for the overall time evolution of the peaks. While some of the peaks across the ions reach maxima simultaneously or close in time, they can exhibit time evolution characterized by distinct peak widths, asymmetry, or {steepness across} the three lightcurves. Here we determined the peak steepness by fitting the ascending segment of every peak, bounded by the left base and peak maximum, with a linear function weighted by intensity uncertainties via the \texttt{statsmodels} Python module. {Because the peak bases often correspond to different instants across the three ions due to the distinct peak time evolution, the endpoints of linear fits were chosen to be ion-specific.} To account for the differences of peak amplitudes across the lightcurves, we normalized the slopes of peak rise $\alpha_{\mathrm{rise}}$ by peak strengths. $\alpha_{\mathrm{rise}}$ of every peak is indicated at the bottom of each panel in Figure~\ref{fig:lightcurves}, with the same color-coding as used for the lightcurves above. The slopes of the \ion{Si}{4} and \ion{C}{2} intensity peaks are usually close one to another and higher than the ascending slopes of the \ion{Mg}{2} peaks. One exception occurs (at $t\sim65$~s in source B2), during the time period exhibiting a different number of peaks across the ions. The individual \ion{Si}{4} peaks at $t = 50 - 70$\,s are steeper than the single \ion{Mg}{2} peak in this time window, but when merged into a single peak interval (see above) the ascending slope appears less steep. The \ion{C}{2} intensity evolution is generally similar to that of the \ion{Si}{4} line, with the peaks being stronger and troughs separating them deeper compared to those along the \ion{Mg}{2} k lightcurves. Because of the similarity between the \ion{Si}{4} and \ion{C}{2} intensity evolution and the higher noise levels in the \ion{C}{2} lightcurves, hampering the peak determination therein, we opted to further analyze the delays between the transition region \ion{Si}{4} and chromospheric \ion{Mg}{2} k line emission only.

\subsection{Quantifying delays between \ion{Mg}{2} and \ion{Si}{4} line emission}
\label{sec:obs_delays}

\begin{figure}[h]
\centering
\includegraphics[width=8.80cm, clip,   viewport=00 00 710 350]{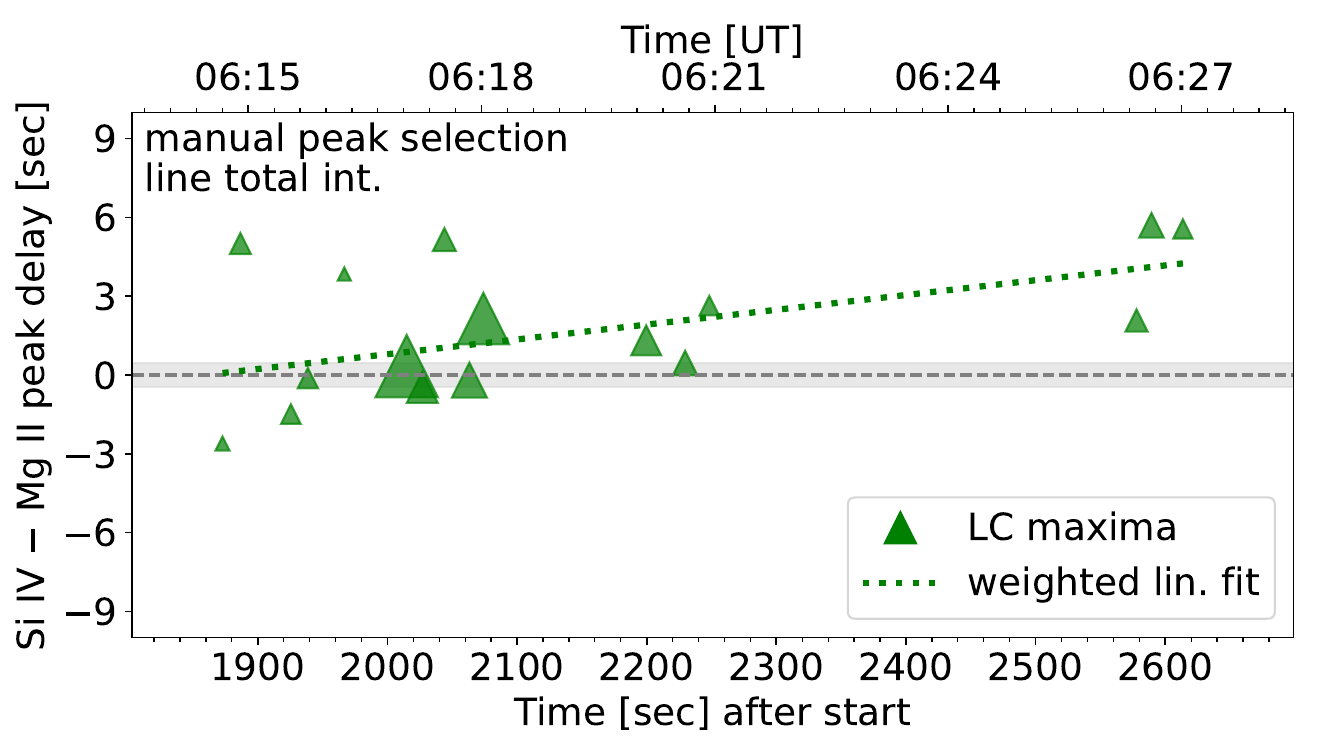}
\caption{Delays between the maxima of \ion{Si}{4} and \ion{Mg}{2} k line intensity peaks inferred from the total line intensities as a function of time during the flare. The peaks whose maxima were used for the delay calculation were selected manually. The size of the symbols is given by the amplitude of \ion{Si}{4} peaks. The dotted line represents linear fit to data weighted by the peak amplitude. The grey horizontal strip at $t = 0$\,s corresponds to $\pm \Delta_{t}/2$. \label{fig:delays}}
\end{figure}

The delays between the maxima of the individual \ion{Si}{4} and \ion{Mg}{2} intensity peaks are plotted in Figure~\ref{fig:delays} as a function of time. Panel a shows the  delays between the peaks in the lightcurves produced via the total line intensities (Figure~\ref{fig:lightcurves}). The time is indicated on the horizontal axis in seconds after the start of the observations and spans most of the impulsive phase of the flare. Positive delays (vertical axis) imply that the \ion{Si}{4} intensity peaks precede those visible in the \ion{Mg}{2} line and vice versa. The size of the symbols is given by the amplitude of the \ion{Si}{4} peaks whose delay the symbol indicates; the larger the symbol, the larger the peak amplitude. Finally, the grey strip at $y = 0$ corresponds to $\pm \Delta_{t}/2$, where $\Delta_{t} \approx 0.92$ is the cadence of the spectral observations. 

\ion{Si}{4} intensity peaks occurred before those in the \ion{Mg}{2} line the majority of the time, with 11 out of 16 measurements (69\%) exhibiting a positive delay. The longest delays are $\approx 6$\,s long, whereas 2 out of the 11 positive delays are below $\Delta_{t}$. The mean of the measured delays $\delta_{\mathrm{t}}$ was determined as $1.8 \pm 0.6$\,s with a standard deviation $\sigma$ of 2.5\,s. The mean positive delay is $\delta_{\mathrm{t}} = 3.1 \pm 0.6$\,s with $\sigma = 1.9$\,s. The green dotted line plotted in Figure~\ref{fig:delays} represents linear fit to the delays, with \ion{Si}{4} peak amplitudes used as weights. Despite the low number of the datapoints, the trend indicated by the linear fit is suggestive of an overall increase of the emission delays as the flare was progressing {in the region under study}. The \ion{Si}{4} peak maxima were occurring consistently before their \ion{Mg}{2} k counterparts after $t \approx 2100$\,s in Figure~\ref{fig:delays}. The last three datapoints plotted in Figure~\ref{fig:delays} correspond to the B4 at the end of the impulsive phase of the flare. 

Delays between the \ion{Si}{4} and \ion{Mg}{2} k3 intensity maxima are briefly discussed in Appendix \ref{sec:appendix_b}. Analysis of line emission delays inferred via peak cross-correlations and peak centroids is presented in Appendix \ref{sec:appendix_d}. We find that, while the relative delays between the line emission can vary when the peak time evolution is accounted for, the results are generally consistent with the delays inferred from peak maxima. \\

The IRIS spectral analysis of the newly-discovered delays between the chromospheric and transition region emission can be summarized in few key points:

\begin{enumerate}
    \item{The transition region emission intensity peaks typically precede the emission originating in the chromosphere (\ion{Si}{4} precedes \ion{Mg}{2});}
    \item{{The typical delay time is up} to $\approx 6$ seconds;}
    \item{There is seemingly an increase of the delay toward the end of the impulsive phase.}
\end{enumerate}

{Even though the peak delays can also be discerned in slit positions adjacent to those presented in Figure \ref{fig:lightcurves}, the high-cadence sit-and-stare IRIS observations cannot address the spatial distribution of the delays in the full length of the ribbon structure. These outcomes are only inferred from the four brightening events B1 -- B4 in the ribbon portion coincident with the IRIS slit, and they might not be representative of the general flare behavior (see Section \ref{sec:discussion}). On the other hand, these results} hold irrespective of the method used for detection of intensity peaks and peak intervals (Appendix \ref{sec:appendix_c}) and determination of delays (Appendix \ref{sec:appendix_d}). The robustness of this result is underlined by the fact that the delays are evident in both level-2 data free of any post-processing (not shown), as well as lightcurves with various degrees of smoothing (Figure~\ref{fig:lightcurves}, Appendix Figure~\ref{fig:lightcurves_app}). The short {delay times are} the likely reason why they have not been reported in previous spectroscopic observations, often limited by lower time resolution. 

\section{Interpreting the observed delays using RADYN simulations}  \label{sec:modeling}

\subsection{\texttt{RADYN} flare models} \label{sec:mod_method}

Motivated by the new observations of delays between transition region \ion{Si}{4} and chromospheric  \ion{Mg}{2} emission, one-dimensional (1D) field-aligned flare simulations were performed, in which the agent of energy transport through the solar atmosphere varied. Our aim was to determine what these observations tell us about flare energetics and about the energy propagation through {flaring atmospheres, that leads to the formation of the observed delays}. For this we utilized the \texttt{RADYN} code \citep{carlsson92, carlsson95, carlsson97, allred05, allred15}, which is a powerful tool to study the radiative and hydrodynamic response of the solar atmosphere to flares, from the photosphere through the corona. \texttt{RADYN} solves the coupled, non-linear equations of hydrodynamics, non-LTE radiative transfer and non-equilibrium atomic level populations, on a one-dimensional (1D) adaptive grid, allowing for resolution of shocks and gradients in the chromosphere and the transition region. Optically thick radiation transfer is included for H, Ca, and He. We subjected different pre-flare atmospheres to electron beam (EB), thermal conduction (TC), TC + EB (hybrid; H), and Alfv\'{e}n wave (AW) heating.

Upon estimating the lengths of flare loops connecting the two ribbons (Figure~\ref{fig:overview}c -- d) we assumed a half-loop length of 50\,Mm. Flare energy was injected at a constant rate for $\tau_{\mathrm{inj}}= 10$\,s. This duration was estimated from the intensity peaks along the IRIS intensity lightcurves (Section \ref{sec:obs_lightcurves}). We assumed four different initial loop atmospheres:  (1) an atmosphere in radiative equilibrium \citep[`RE', e.g.][]{allred15} with no ad-hoc chromospheric heating at coronal apex $T=1$~MK, (2) a similar setup but with coronal apex $T=3$~MK, (3) an atmosphere with initial ad-hoc chromospheric heating to create an extended chromospheric plateau \citep[referred to as `plage'-like atmosphere, following][]{carlsson15,polito18}, with coronal apex $T=1$~MK, and (4) that same plage chromosphere but with coronal apex $T=3$~MK. {Initial loop apex temperatures of $T=1$, 3~MK are based on active region observations and commonly used to model flare emission \citep[e.g.][]{kasparova09, allred15, carlsson23}.}
 
The properties of the EB heating were constrained by parameters of fits to non-thermal components of \textsl{Fermi}/GBM spectra (Table~\ref{tab:fermi}).{ As discussed in Section \ref{sec:data_fermi}, the inferred energy flux densities injected to the \texttt{RADYN} simulations were constrained via two sets of footpoint area measurements, $A_\mathrm{min}$ \& $A_\mathrm{max}$. The corresponding flux densities $F$ were: [$1 \times 10^9$, $5 \times 10^9$, $1 \times 10^{10}$, $5 \times 10^{10}$]\,erg\,cm$^{-2}$\,s$^{-1}$.} The low-energy cutoff was fixed at $E_c = 15$ keV, and the spectral index at $\delta = 4$, to approximate the values found in the \textsl{Fermi}/GBM analysis. The lowest observationally-constrained value of $F=5 \times 10^8$\,erg\,cm$^{-2}$\,s$^{-1}$ (see $A_\mathrm{max}$ during the B4, Table~\ref{tab:fermi}) produced EB model flare atmospheres that did not contain measurable \ion{Si}{4} 1402.77\,\AA~emission, so they were excluded from the analysis. Indeed, according to the models of \cite{polito23, kerr24b}, such low values of energy flux densities are closer to what would be expected to appear in the ribbon fronts, rather than the bright kernels that we study here. The \textsl{Fermi}/GBM inspired simulations were supplemented by experiments with $F = 1 \times 10^9$\,erg\,cm$^{-2}$\,s$^{-1}$, $E_c = 5$\,keV, and $\delta = 4$, for each of the four pre-flare atmospheres. These models were tested to account for the possibility that the non-thermal electron distribution had an $E_c = 5$\,keV, below the sensitivity threshold of \textsl{Fermi}/GBM.

Further, we ran flare models assuming purely \textsl{in-situ} heating in the corona and consequent energy propagation to the lower atmosphere through a thermal conduction front (here referred to as ``TC models"). Similar to previous studies \citep[e.g.][]{testa14, polito18}, we assume a volumetric heating rate (ergs s$^{-1}$ cm$^{-3}$) distributed over a certain region along the loop. We use the same values of total energy flux density energy as that of the EB simulations, but spread over the coronal portion of the loops. For all TC models, the heating is distributed between the loop apex and $3\times10^8$ cm (at the boundary between the corona and the top of the TR). To aid comparison with the EB models, we list the TC models based on energy flux rather than volumetric heating rate. Some of the simulations with the largest energy fluxes ($F=$ [$5 \times 10^9$, $5 \times 10^{10}$]\,erg\,cm$^{-2}$\,s$^{-1}$) were excluded from our analysis, as they caused the atmosphere to reach temperatures above 100\,MK that are unrealistic for the relatively small C-class flare analyzed here. The excluded runs were replaced with simulations with slightly lower flux ($F=$ [$2.5 \times 10^9$, $2.5 \times 10^{10}$]\,erg\,cm$^{-2}$\,s$^{-1}$).

We have also ran two test cases where the heating was equally distributed between the TC \textsl{in-situ} heating and EB energy models. We have done so for two energy fluxes, $F=$ [$5 \times 10^9$, $5 \times 10^{10}$]\,erg\,cm$^{-2}$\,s$^{-1}$, and called these models ``Hybrid (H)". 

The last energy transport mechanism that we explored was downward propagating Alfv\'en waves \citep{emslie82,fletcher08,haerendel09,russell13,reep16,kerr16,reep18b}. The implementation of AW in \texttt{RADYN} flare simulations of \citet{kerr16} was updated to include the wave travel time in the computation. This closely follows the approach of \citet{reep18b}, such that the wave moves at the local Alfv\'en speed as it travels down the loop. A complete description of updates to AW energy transport in \texttt{RADYN} will be presented in a forthcoming manuscript \citep{kerr_awnanoflares}. To model energy transport via AW we impose a magnetic field stratification for the purpose of calculating the Alfv\'en speed. It does not vary in time. This was set to $B=1500$~G in the photosphere, and $B\sim500$~G in the low corona, evolving with pressure according to Eq. 10 of \cite{kerr16}. {Photospheric fields of similar strengths can be found in the strong-field regions of the NOAA AR13107. While \citet{lorincik24} suggested that the environment where the flare occurred was disperse with potentially weaker coronal fields, these field strengths were chosen to provide a reasonable upper constraint on $B_\mathrm{LOS}$ consistent with the HMI observations.} Injected AWs had frequency = $5$~Hz and perpendicular wave number at loop apex $k_{x,a} = 5\times10^{-4}$~cm$^{-1}$ (varying linearly with magnetic field through the loop). The Poynting flux corresponds to that of the EB models constrained by the HXR fitting, although here we only selected three example models $S = [5\times10^8, 1\times10^9, 1\times10^{10}]$~erg~s$^{-1}$~cm$^{-2}$ for a detailed study. The particular values of the frequency and wavenumber were chosen because they produce a significant increase of \ion{Si}{4} intensity, which is needed to explain the peaks in intensity. A broader parameter study of the \ion{Si}{4} response to wave heating has been conducted by \citet{kerr_awnanoflares}. As with the EB and TC experiments, energy was injected for a duration of 10~s. Unlike the EB experiments, in these AW experiments the energy injected in the corona takes time to reach the lower atmosphere, where energy losses takes place due to resistive damping, including ambipolar effects.

\begin{deluxetable*}{lcccc}[h]
\tablecolumns{4}
\startdata
\tablehead{
Model	&	$F$ [\,ergs\,cm$^2$\,s$^{-1}$]		&	$T_{\mathrm{init}}$ [MK]	& Init. atm. &   Delays reproduced?} 
EB1    	&	1F09	& 1  & RE     & $\times$   \\ 
EB1p    	&	1F09	& 1  & plage     & $\times$   \\ 
EB2 	&	1F09	& 3  & RE     & $\times$   \\ 
EB2p 	&	1F09	& 3  & plage     & $\times$   \\ 
EB3		&	5F09	& 1  & RE     & $\times$   \\ 
EB3p     &	5F09	& 1  & plage     & $\times$   \\ 
EB4    	&	5F09	& 3  & RE     & $\times$   \\ 
EB4p 	&	5F09	& 3  & plage     & $\times$   \\ 
EB5 	&	1F10	& 1  & RE     & $\times$   \\ 
EB5p	&	1F10	& 1  & plage     & $\times$    \\ 
EB6   	&	1F10	& 3  & RE     & $\times$    \\ 
EB6p   	&	1F10	& 3  & plage     & $\times$    \\ 
EB7 	&	5F10	& 1  & RE     & $\boldcheckmark$   \\ 
EB7p 	&	5F10	& 1  & plage     & $\boldcheckmark$    \\ 
EB8	&	5F10	& 3  & RE     & $\boldcheckmark$    \\ 
EB8p	&	5F10	& 3  & plage     & $\boldcheckmark$    \\   
EB9$^{*}$	&	1F09	& 1  & RE  & $\boldcheckmark$     \\ 
EB9p$^{*}$	&	1F09	& 1  & plage  & $\boldcheckmark$     \\  \hline
H1		&	5F09	& 3  & RE      & $\times$   \\ 
H2		&	5F10	& 3  & RE      & $\boldcheckmark$   \\ \hline 
TC1		&	1F09	& 1  & RE      & $\times$   \\ 
TC2	 	&	2.5F09	& 1  & RE      & $\times$   \\ 
TC3		&	5F09	& 3  & RE      & $\times$   \\ 
TC4		&	1F10	& 3  & RE      & $\times$   \\ 
TC5		&	2.5F10	& 3  & RE      & $\times$   \\ \hline
AW1		&	5F08	& 3  & RE      & $\boldcheckmark$   \\ 
AW2		&	1F09	& 3  & RE      & $\boldcheckmark$   \\ 
AW3		&	1F10	& 3  & RE      & $\times$   \\ 
\enddata
\caption{List of electron beam (`EB'), hybrid (`H'), thermal conduction (`TC'), and Alfv\'{e}n wave (`AW') heating models analyzed in this study. The columns of the table list the varying model parameters. The rightmost column indicates whether the model reproduces the observed delays between the \ion{Si}{4} and \ion{Mg}{2} peaks. EB simulations assume $\delta = 4$ and $E_c = 15$\,keV. \\ $^{*} E_c = 5$\,keV models.  \label{tab:models}}
\end{deluxetable*}

\begin{figure}[!h]
\centering
\includegraphics[width=.99\textwidth, clip,   viewport = 00 10 860 930]{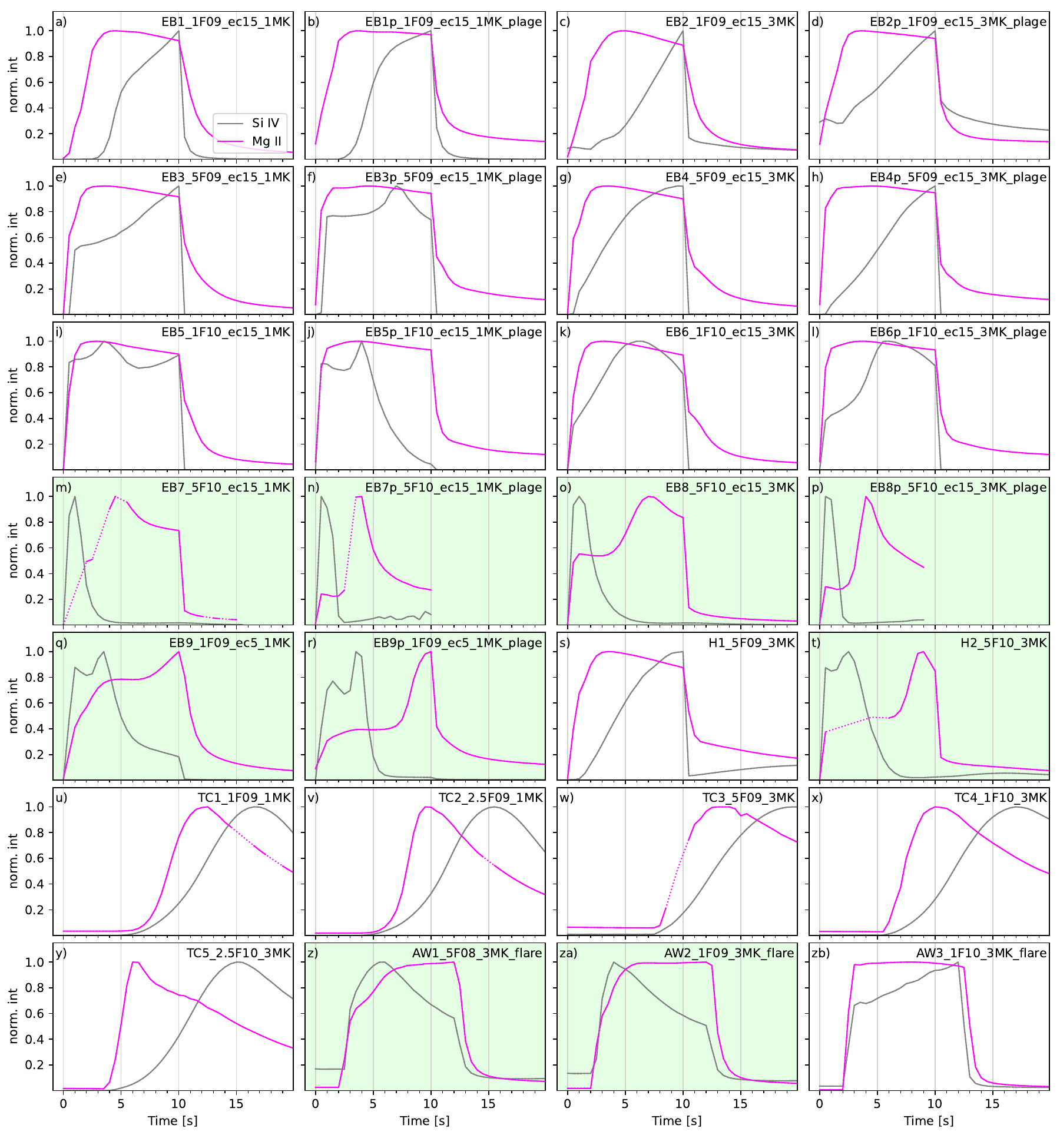}
\caption{Lightcurves of  \ion{Si}{4} 1402.77\,\AA~(grey) and \ion{Mg}{2} 2796.35\,\AA~(magenta) synthesized from our grid of \texttt{RADYN} simulations. Panels a -- r present curves in electron-beam heated atmospheres, panels s, t were obtained in hybrid simulations, panels u -- y correspond to thermal conduction-heated atmospheres, and panels z -- zb to simulations with the Alfv\'{e}n wave heating. Dotted sections along the lightcurves represent interpolated values (where applicable). Green semi-transparent background was used to highlight lightcurves that replicate the observed delays. \label{fig:model_lc}}
\end{figure}

\subsection{Synthetic lightcurves} \label{sec:mod_lc}

Synthetic \ion{Si}{4} 1402.77\,\AA~emission was produced assuming equilibrium conditions and atomic data from the CHIANTI version 10 model \citep{dere97, delzanna21} at a timestep of 0.5\,s \citep[see][for details]{polito18}. To produce synthetic \ion{Mg}{2} 2796.35\,\AA~emission, the radiation transfer code \texttt{RH15D} \citep{uitenbroek01,pereira15} was used, taking account of partial redistribution effects. The spectra were synthesized assuming microturbulence of $v_\mathrm{turb} = 7$\,km\,s$^{-1}$ and a filling factor of one. For further details on the experiment setup, including conversion to IRIS count rates, we refer the reader to \citet{polito23} and \citet{kerr24a}. To facilitate the comparison between the observations and models, the properties of the synthetic profiles were obtained via the moment analysis using the same wavelength intervals as those used for the observations (Section \ref{sec:obs_properties}). 

Figure~\ref{fig:model_lc} presents \ion{Si}{4} 1402.77\,\AA~(grey) and \ion{Mg}{2} 2796.35\,\AA~(magenta) synthetic lightcurves corresponding to each line's total intensity. The dotted sections along certain \ion{Mg}{2} lightcurves are interpolated values, where the \texttt{RH15D} code failed to converge to a solution. The lightcurves were normalized to facilitate comparison with the observations. Observed and modeled spectral characteristics such as total intensities, Doppler shifts, and line broadening are compared in Section \ref{sec:comparison}.  

In contrast to the observations, the modeled \ion{Si}{4} and \ion{Mg}{2} lightcurves usually consisted of a single peak only, a key difference caused by the properties of the heating profile. The lightcurves obtained in the EB (panels a -- r) and H (panels s \& t) models are characterized by a steep increase of the intensity of both or either of the two lines immediately after the start of the heating. More gentle intensity increases are characteristic for the TC model lightcurves (Figure~\ref{fig:model_lc} u -- y), while the AW heating (z -- zb) again results in a steep intensity increase (for the chosen wave parameters), albeit delayed from the simulation start by 2 -- 3\,s due to the wave's travel at the local Alfv\'en speed. The intensity decrease begins at $t = 10$\,s, right after the heating is switched off, in the EB and H models (panels a -- t) and again slightly later in the AW models (z -- zb). In the EB and H models, the \ion{Si}{4} emission is only at a level that would be detectable by IRIS during the heating period. \ion{Mg}{2} k lightcurves decline more gradually compared to \ion{Si}{4}.

\begin{figure}[h]
\centering
\includegraphics[width=8.80cm, clip, viewport = 0 0 250 195]{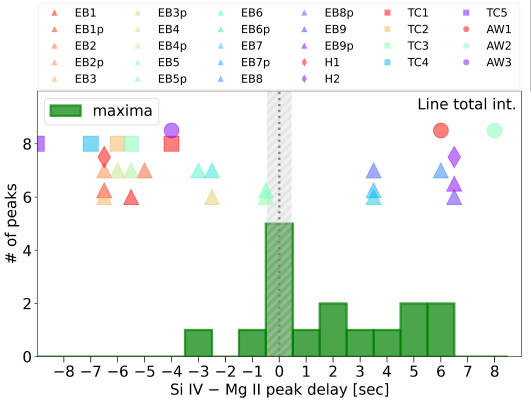}
\caption{Histograms depicting the distribution of the observed delays inferred from the total \ion{Si}{4} and \ion{Mg}{2} line intensities. The symbols above the histograms correspond to the delays between synthetic lightcurves, and are the same for both panels. Different symbol styles and colors were used to distinguish between the models (see the legend above the figure). The hatched grey strip at $t = 0$\,s corresponds to $\pm \Delta_{t}/2$. \label{fig:histo}}
\end{figure}

The delays between the modeled \ion{Si}{4} and \ion{Mg}{2} emission and their comparison to the observations are visualized in Figure~\ref{fig:histo}. The histogram presents the distribution of the observed delays corresponding to the total intensities. The grey strip at $0$\,s corresponds to $\pm \Delta_{t}/2$. The symbols above the histogram indicate the delays between the maxima of the synthetic lightcurves. Different symbol styles were used to distinguish between the models (see the legend of Figure~\ref{fig:histo} for further reference). Of the 28 runs studied here, 9 simulations (6 different heating models, with some that worked for both pre-flare atmospheres) produced a delay of \ion{Mg}{2} lightcurve maxima relative to \ion{Si}{4}, also highlighted using the green semi-transparent background in Figure~\ref{fig:model_lc}. This occurred in:

\begin{itemize}
    \item the high-flux ($F = 5 \times 10^{10}$\,ergs\,cm$^2$\,s$^{-1}$) EB7 -- EB8p models (Figure~\ref{fig:model_lc} m -- p);
    \item the low flux, small $E_{c}$ ($F = 1 \times 10^{9}$\,ergs\,cm$^2$\,s$^{-1}$, $E_c = 5$\,keV) EB9 and EB9p models (Figure~\ref{fig:model_lc} q \& r); 
    \item the high-flux H2 model (Figure~\ref{fig:model_lc} t); and 
    \item the AW1 \& AW2 models (Figure~\ref{fig:model_lc} z \& za).
\end{itemize}

The delays obtained via cross-correlating the lightcurves and calculating peak centroids are discussed in Appendix \ref{sec:appendix_d}.

\subsection{Comparison of observed and modeled spectral properties}  \label{sec:comparison}

\begin{figure}[!h]
\centering
\includegraphics[width=18.00cm, clip, viewport=670 50 1725 140]{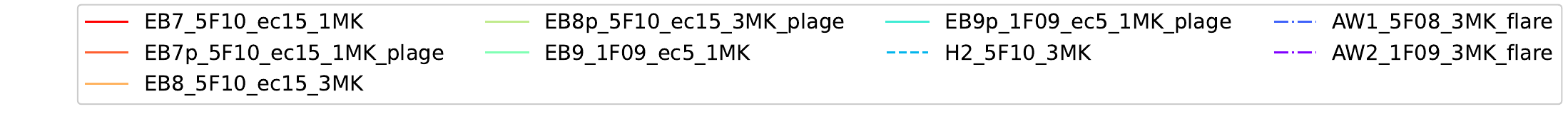}
\includegraphics[width=5.90cm, clip, viewport=00 00 430 510]{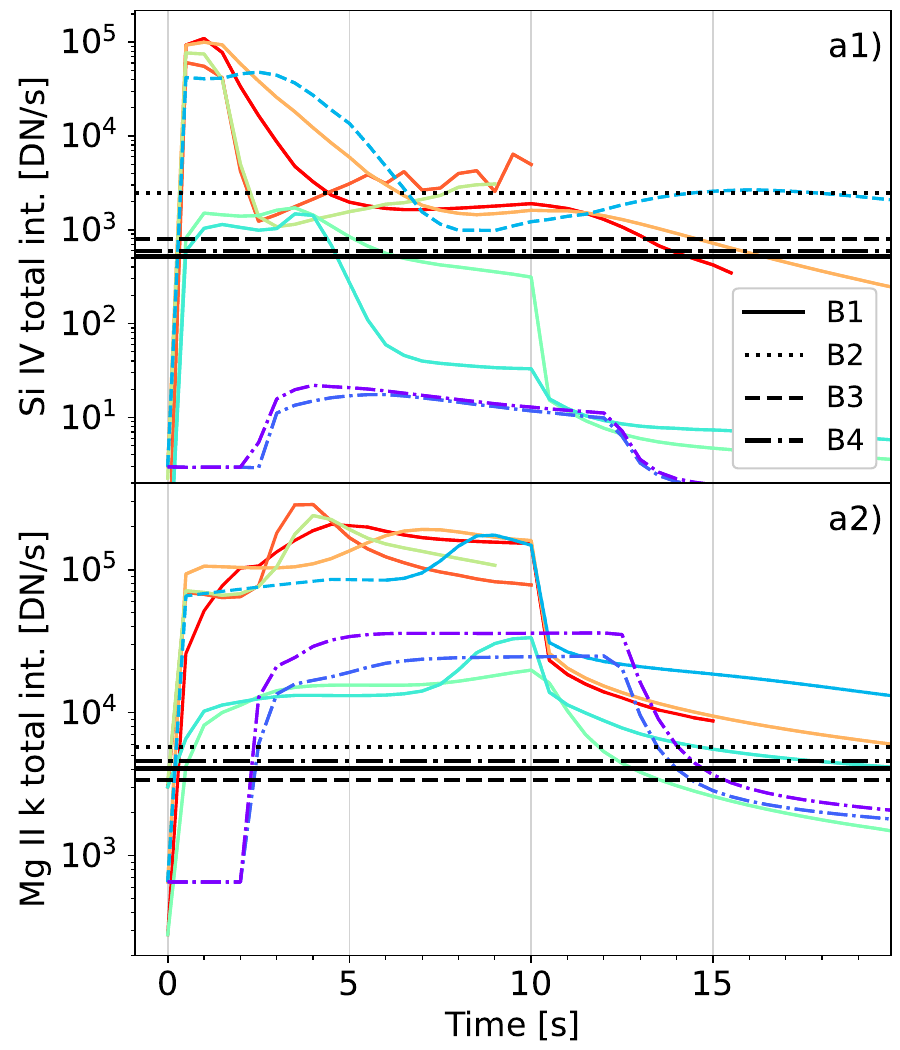}
\includegraphics[width=5.90cm, clip, viewport=00 00 430 510]{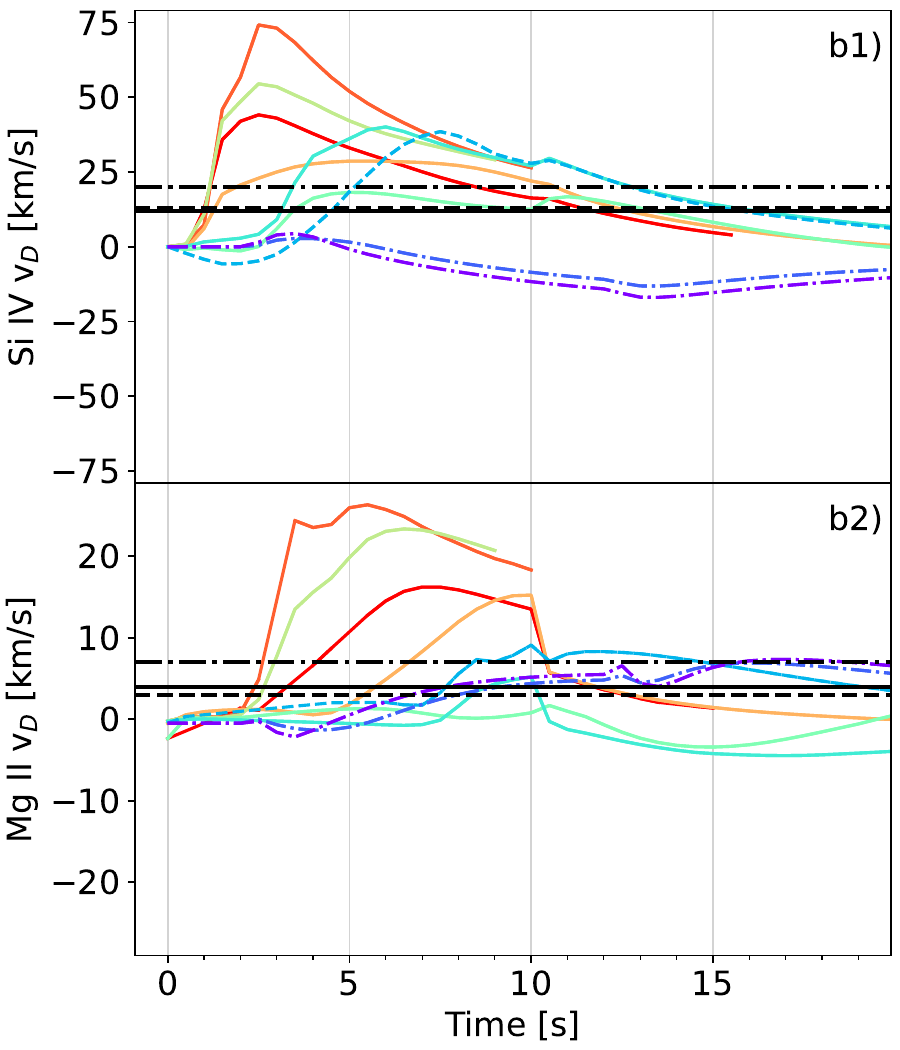}
\includegraphics[width=5.90cm, clip, viewport=00 00 430 510]{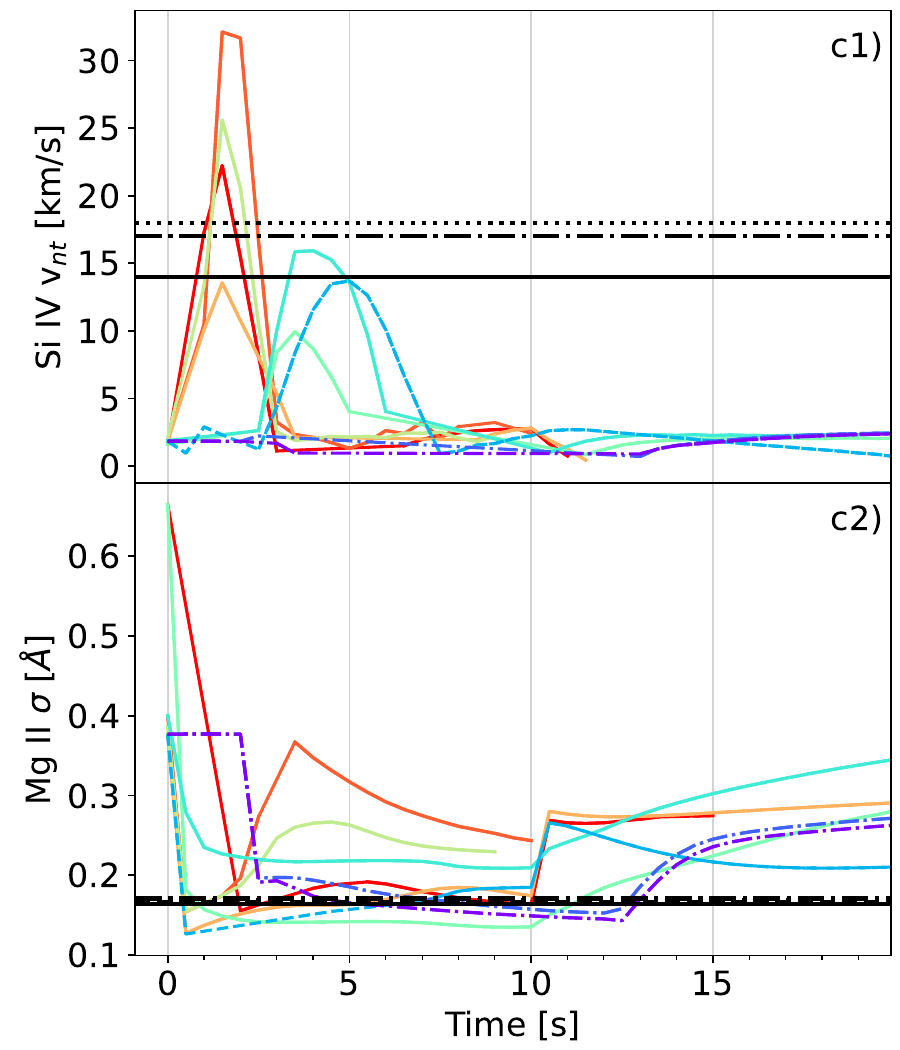}
\caption{Evolution of the total intensity (panel a), the Doppler velocity (panel b), and the broadening ($v_{\mathrm{nt}}$, $\sigma$; panel c) of the \ion{Si}{4} 1402.77\,\AA~(top) and \ion{Mg}{2} 2796.35\,\AA~(bottom) spectra synthesized in the models which reproduce the observed delays. The color-coding and different linestyles distinguish between the models. The black horizontal lines correspond to the observed values averaged during the B1 -- B4 events. \label{fig:synth_q}}
\end{figure}

The time evolution of the spectral properties of the \ion{Si}{4} 1402.77\,\AA~and \ion{Mg}{2} 2796.35\,\AA~lines synthesized in the models which reproduce the observed delays is presented in Figure~\ref{fig:synth_q}. Panels a1 \& a2 show the total line intensities, panels b1 \& b2 depict the Doppler velocities $v_{\mathrm{D}}$, and panels c1 \& c2 detail the evolution of the broadening, expressed in terms of the non-thermal velocity $v_{\mathrm{nt}}$ for \ion{Si}{4}. The black horizontal lines correspond to the average of the quantities observed during the brightening events B1 -- B4. Since reproducing some of the observed \ion{Mg}{2} line properties in flare models can be difficult \citep[see e.g.][and references therein]{rubiodacosta17, polito23, kerr23, kerr24a}, the following analysis is mostly focused on the \ion{Si}{4} line.

%Since reproducing the large observed \ion{Mg}{2} line widths in flare models can be difficult \citep[see e.g.][and references therein]{rubiodacosta17, polito23, kerr23, kerr24a}, the following analysis is mostly focused on the \ion{Si}{4} line properties.

The line intensities observed during the B1 -- B4 were typically of the order of $10^2 - 10^3$\,DN\,s$^{-1}$ for the \ion{Si}{4} line and $10^3$\,DN\,s$^{-1}$ for \ion{Mg}{2}. As seen in Figure~\ref{fig:synth_q}a1, a2, the synthetic intensities rarely align with the observations, in some cases exceeding the observed ones by up to two orders of magnitude, particularly for the \ion{Mg}{2} line. The difference between the modeled and observed intensities can be attributed to using an ad-hoc filling factor and possible multithreaded structure of loops. These factors, on the other hand, should not affect the ratios of the synthetic intensities. The observed ratios between the \ion{Mg}{2} and \ion{Si}{4} line intensities averaged during the B1 -- B4 range roughly between 2 -- 8. Similar synthetic ratios can be found in the EB7 \& 7p, EB8 \& 8p (red $\rightarrow$ gold solid lines) and EB9 (green) models in the first $\approx 4$\,s after the start of the heating. These ratios later grow to $\approx 10^{1} - 10^{2}$, values also found in the H2 model (blue dashed line). The ratios obtained in the EB9p model (turquoise) are also fairly close, around 10 in the same time period. The \mbox{\ion{Mg}{2} vs. \ion{Si}{4}} ratio in the AW1 \& 2 runs is of the order of $10^{3}$, far from the observed values. 

The \ion{Si}{4} profiles synthesized in the EB and H models are redshifted and the magnitude of the Doppler shifts is comparable to the observations (Figure~\ref{fig:synth_q}b1). The high-flux plage EB7p and EB8p models (light-red and gold solid lines) show redshifts in the excess of 50\,km\,s$^{-1}$, roughly factor of 2 higher than the observations. The AW models reproduce mild ($< 20$\,km\,s$^{-1}$) \ion{Si}{4} blueshifts, particularly towards the end of the heating period, which do not align with the observations (c.f. blue curves and black horizontal lines). These blueshifts are preceded by very weak ($v_{\mathrm{D}} < 5$\,km\,s$^{-1}$) and brief redshifts $\sim 2 - 4$\,s after the start of the simulation during the rise of the \ion{Si}{4} intensity. All of the models under consideration reproduce mild \ion{Mg}{2} redshifts consistent with the observations (Figure~\ref{fig:synth_q}b2).

The last spectral property constraining our models is the line broadening. The \ion{Si}{4} non-thermal velocity (Figure~\ref{fig:synth_q}c1) exhibits a transient increase to $ v_{\mathrm{nt}} \sim 30$\,km\,s$^{-1}$ after the start of the heating in the high-flux EB7 \& 8 models, whereas the $E_c = 5$\,keV EB9, 9p models as well as the H2 model exhibit a milder increase to  $\sim 15$\,km\,s$^{-1}$  $\sim$3\,s into the heating. The modeled $v_{\mathrm{nt}}$ generally correspond well to the observations, indicative of $v_{\mathrm{nt}}$ ranging between $10 - 20$\,km\,s$^{-1}$. The AW models do not exhibit $v_{\mathrm{nt}}$ increase in the \ion{Si}{4} line, although we note that we have not included any potential non-thermal broadening due to the Alfv\'en wave itself. The proxy on the \ion{Mg}{2} line broadening obtained from the 2nd moment is comparable to the observations. The standard deviation $\sigma$ drops from the pre-flare levels to $\sigma \sim 0.15 - 0.2$\,\AA~during the heating period in all but the EB7p and EB8p models. Converted to the Gaussian FWHM of $\sim 0.35 - 0.47$\,\AA, this range is consistent with the widths of the observed profiles (Appendix \ref{sec:appendix_a}). We emphasize that, in contrast to the observed typically single-peaked \ion{Mg}{2} profiles, the synthetic spectra were usually double-peaked and therefore this should be taken as a quantitative, and not fully accurate, comparison. \\

To summarize this comparison, the high-flux ($F = 5 \times 10^{10}$\,ergs\,cm$^2$\,s$^{-1}$) EB7 \& EB8 simulations produce synthetic spectra that are generally consistent with the observations. In addition to reproducing the observed delays, the properties of spectra synthesized in the $E_c = 5$\,keV EB9, 9p models show relative intensities, widths, and Doppler shifts comparable to the observations. Two out of three AW models included in our study reproduce the observed delays, but the synthetic \ion{Si}{4} spectra are blueshifted and the relative intensities do not align with the observations (see also Section \ref{sec:discussion}).

\subsection{Comparing the chromospheric and transition region electron density evolution}   \label{sec:mod_tne}

Next, we aim to investigate the differences between physical conditions in example models which do and do not reproduce the observed delays between the \ion{Si}{4} 1402.77\,\AA~and \ion{Mg}{2} 2796.35\,\AA~line emission. In the dataset under study, the \ion{Si}{4} line spectra were only observed in the flare ribbon during the relatively-short brightening events B1 -- B4. \ion{Mg}{2} line spectra, on the other hand, were visible even outside of the ribbon as well as before the flare onset. The visibility and the increase of intensity of the \ion{Si}{4} line at such short (0.3\,s) exposure times is thus contingent upon ongoing heating, which is confirmed by the modeling (Figure~\ref{fig:model_lc}). If we consider the \ion{Si}{4} line to be optically thin, its intensity is proportional to the square of the electron number density $N_{\mathrm{e}}$. This is a reasonable assumption given the magnitude of the non-thermal electron flux detected by \textsl{Fermi}/GBM and the Gaussian-like line profiles with no traces of absorption features (see Appendix Figure~\ref{fig:profiles_app}, left), though possible opacity effects cannot be entirely excluded \citep[see details in][]{kerr19}.

\begin{figure}[h]
\centering
\includegraphics[width=6.20cm, clip, viewport=00 68 698 570]{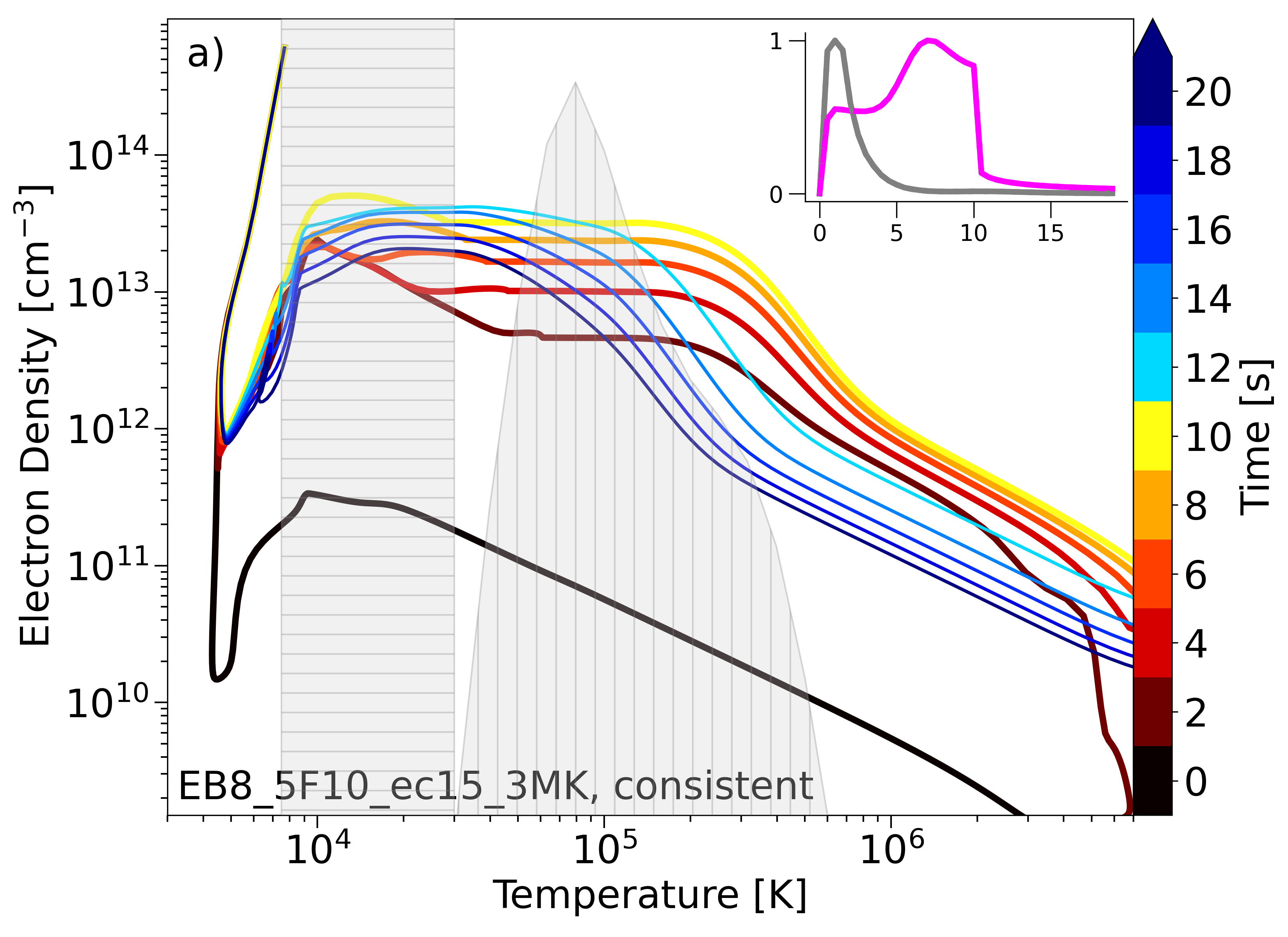}
\includegraphics[width=5.356cm, clip, viewport=95 68 698 570]{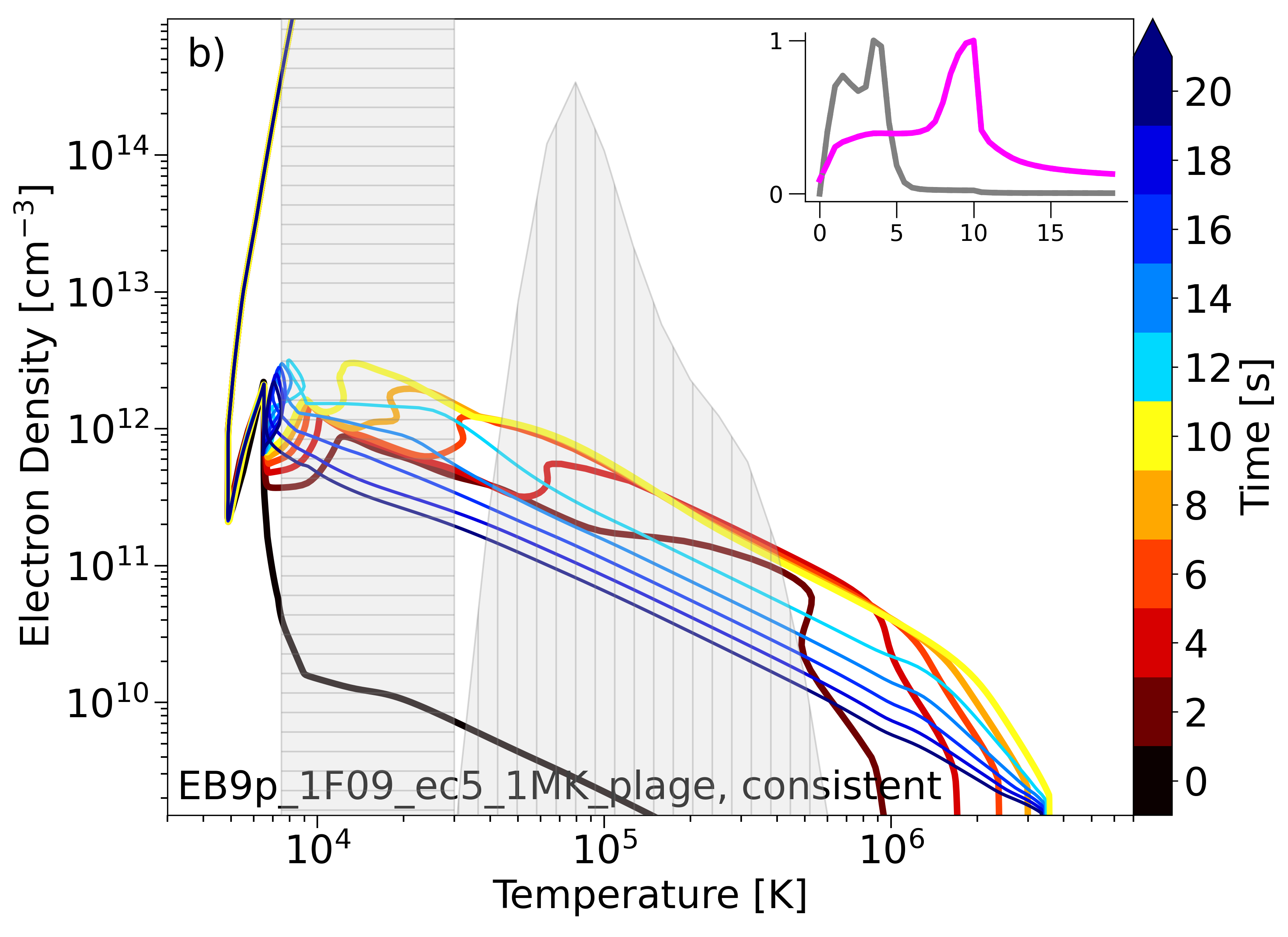}
\includegraphics[width=6.218cm, clip, viewport=95 68 795 570]{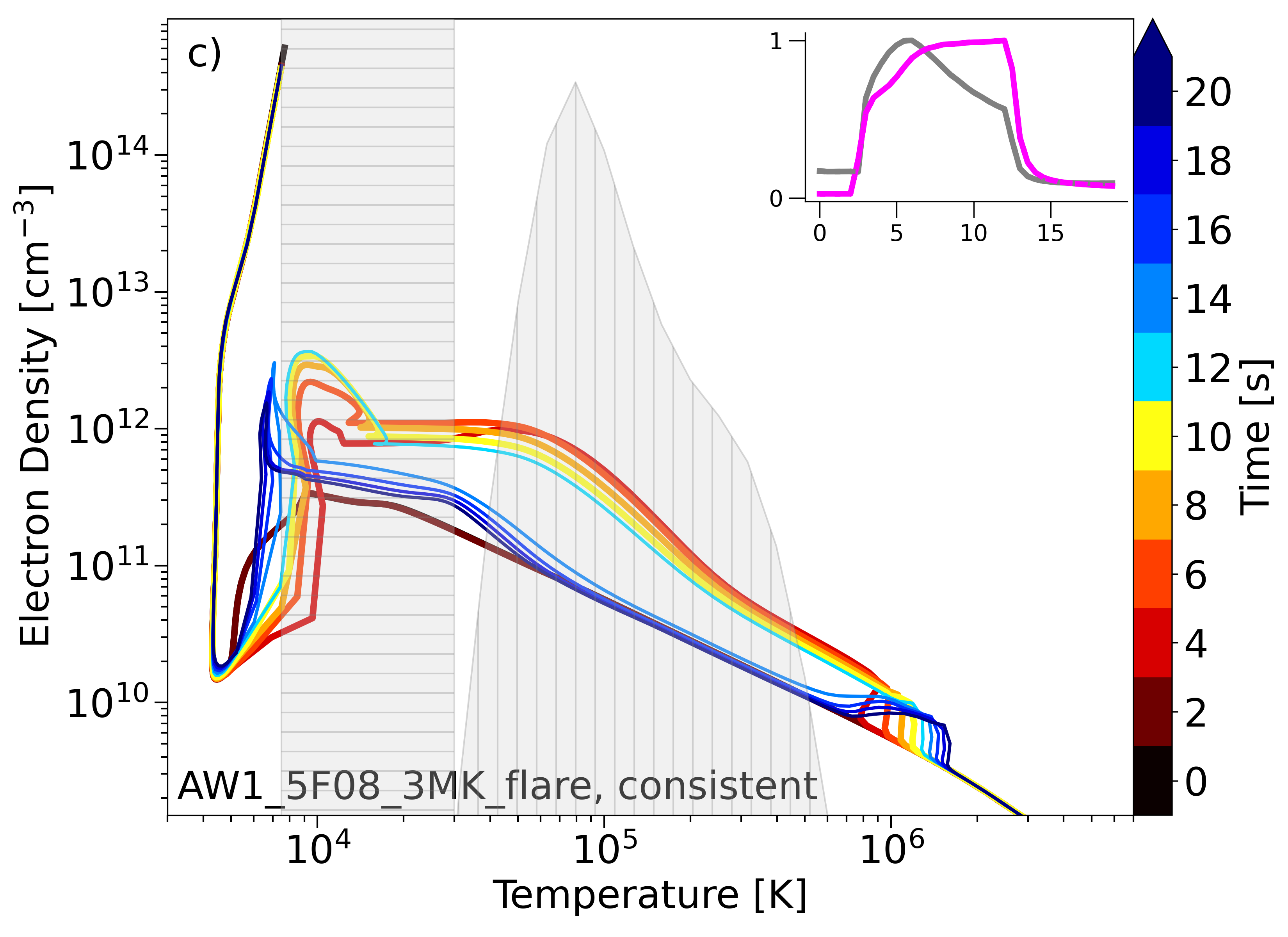}
\\
\includegraphics[width=6.20cm, clip, viewport=00 00 698 570]{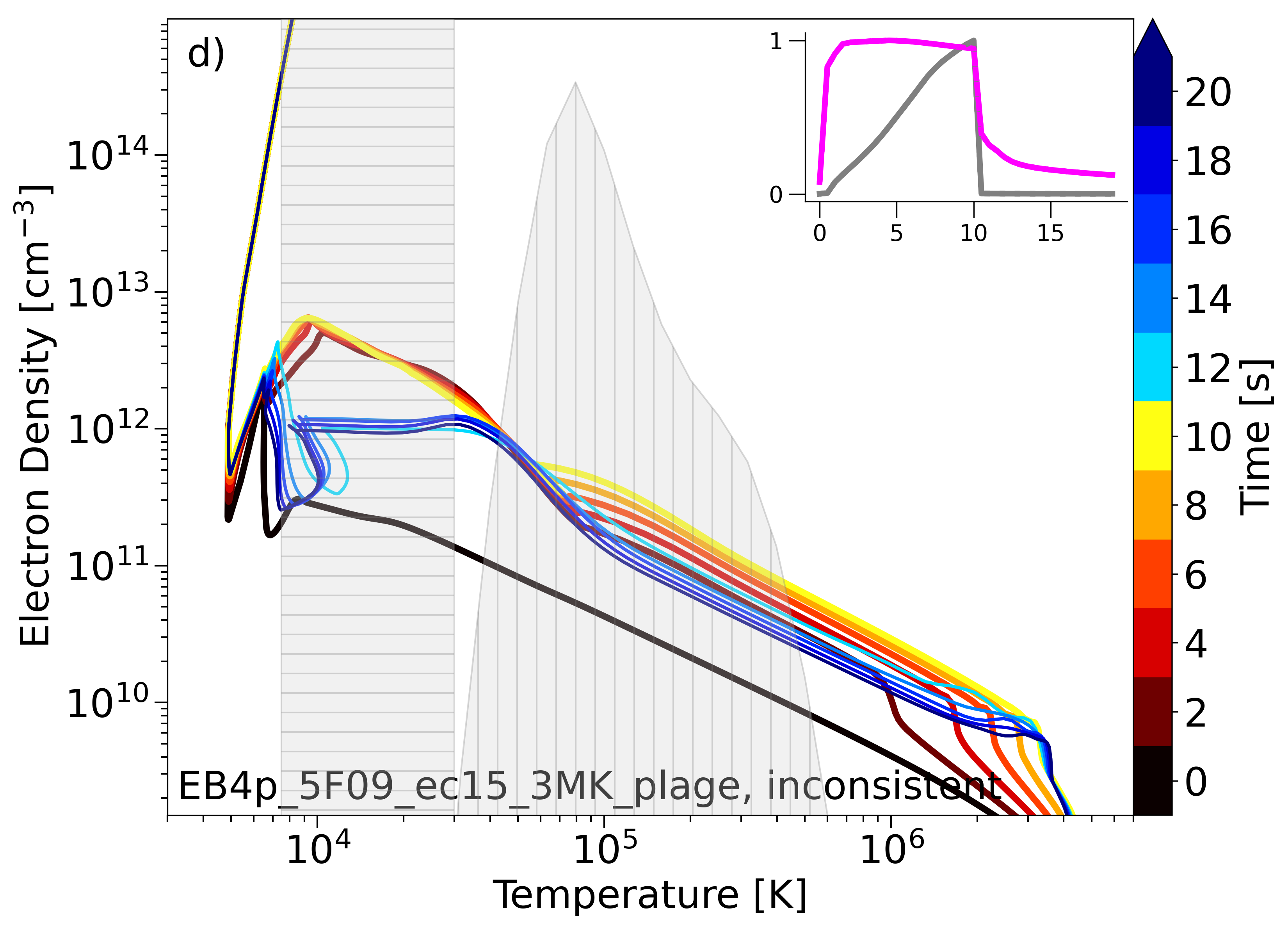}
\includegraphics[width=5.356cm, clip, viewport=95 00 698 570]{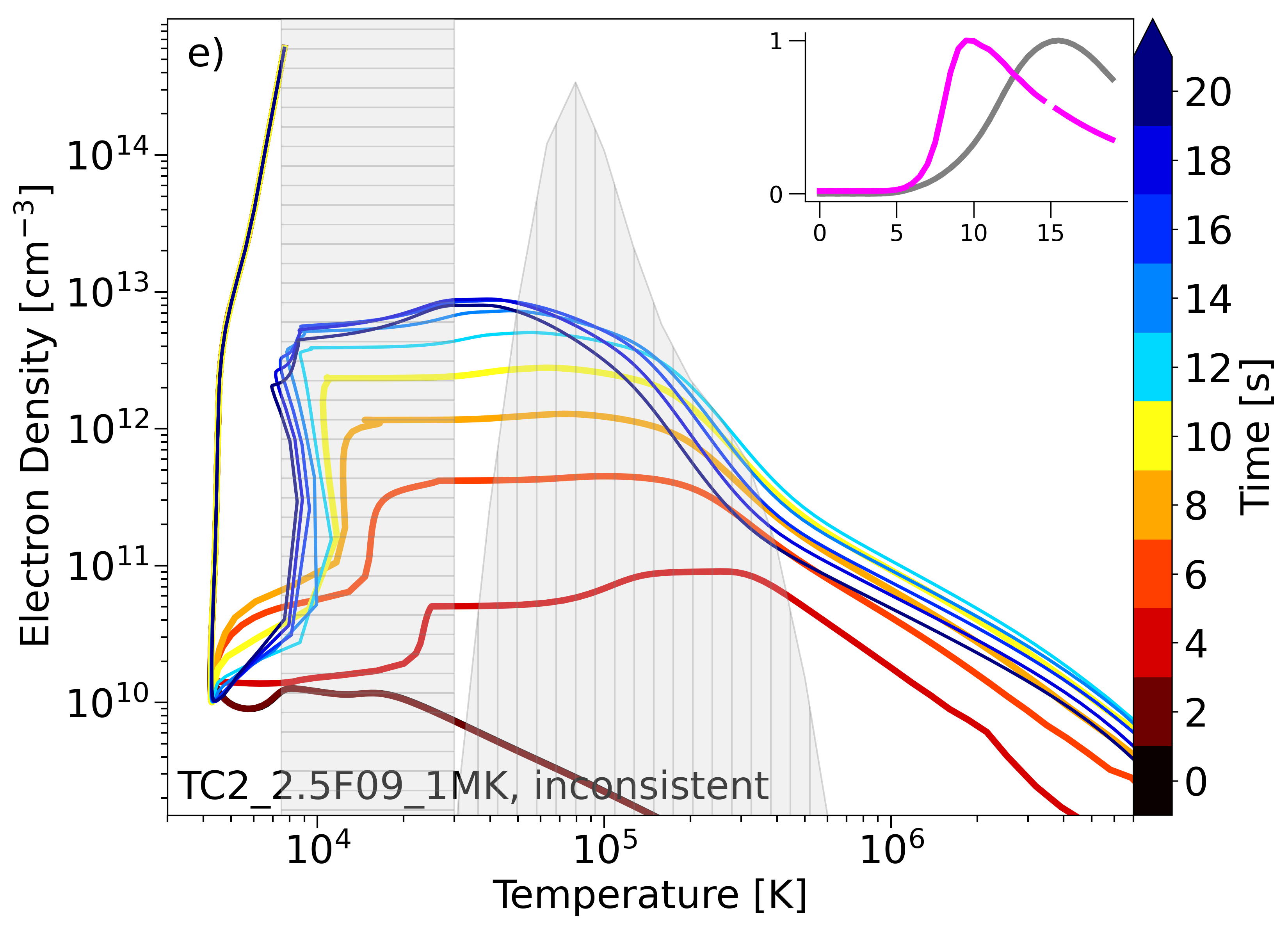}
\includegraphics[width=6.218cm, clip, viewport=95 00 795 570]{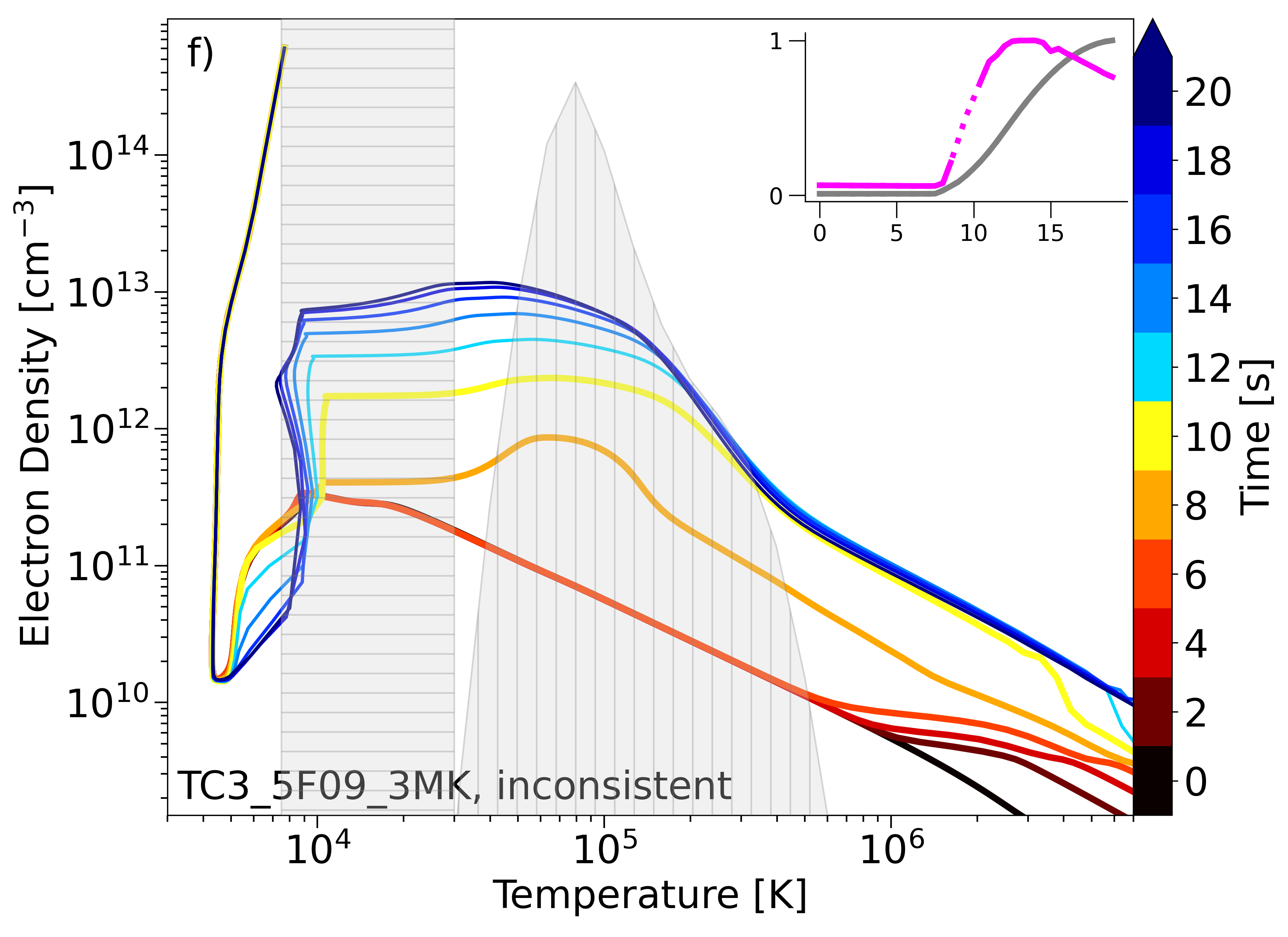}
\caption{Electron density as a function of temperature in selected models that are consistent (top row) and inconsistent (bottom row) with observed delays. Color represents simulation time, with curves shown at 2\,s increments. The pre-flare is the black curve. The heating period (brown $\rightarrow$ yellow thick curves) and the cooling period (cyan $\rightarrow$ dark blue thin curves) are both included. The hatched gray semi-transparent areas in the background indicate the rough formation region of the \ion{Mg}{2} 2796.35\,\AA~line (horizontal hatches) and the normalized contribution function of the \ion{Si}{4} 1402.77\,\AA~line (vertical hatches). Inserts in the top-right corner show the \ion{Si}{4} (gray) and \ion{Mg}{2} (magenta) lightcurves synthesized in each model. Details of heating parameters are provided in the lower left corners. \label{fig:model_tne}}
\end{figure}

As a proxy for probing $N_{\mathrm{e}}$ in the formation region of each line we explored the temporal variation of $N_{\mathrm{e}}$ as a function of the gas temperature $T$. The resulting curves are shown in Figure~\ref{fig:model_tne} at a 2\,s cadence, where color represents time during the simulation (black is the pre-flare). The `warm' color palette (brown $\rightarrow$ yellow thick curves) corresponds to the heating period, whereas the `cold' colors (cyan $\rightarrow$ dark blue thin curves) designate the cooling. On each panel, the gray semi-transparent area with vertical hatches is the normalized contribution function of the \ion{Si}{4}, assuming log($N_{\mathrm{e}}$ [cm$^{-3}$]) = 12. The formation region of the \ion{Mg}{2} k line cannot simply be outlined in the $N_{\mathrm{e}}$ vs. $T$ parameter space as it varies with time due to NLTE radiation transfer effects. For a crude estimation of the line's formation region we plot a gray semi-transparent area with horizontal hatches, corresponding to $7.5 \times 10^3 - 30 \times 10^3$\,K. This area outlines the range of radiation temperatures where most of the \ion{Mg}{2} k line emission is formed \citep[][]{kerr24a}. The heating parameters as well as whether the model is consistent with the observations are indicated in the lower-left corner of each panel. The inserts in the upper-right corner show the normalized lightcurves synthesized in the selected atmospheres, the same as those in Figure~\ref{fig:model_lc}.

The top row of Figure~\ref{fig:model_tne} (panels a -- c) contains selected models consistent with the observations. The EB8 and EB9p models (panel a, b) are characterized by a rapid increase of $N_{\mathrm{e}}$ over more than 2 orders of magnitude in both the chromospheric and transition region line formation regions. This rapid $N_{\mathrm{e}}$ increase causes a commensurately rapid intensity increase of both lines. In the EB8 model, the transition region $N_{\mathrm{e}}$ continues to increase over the following $\sim 8$\,s (brown $\rightarrow$ yellow curves, panel a), although the \ion{Si}{4} lightcurve peaks within $\approx 2$\,s after the heating onset. The gradual $N_{\mathrm{e}}$ increase in the transition region occurs at $\sim 4 - 6$\,s (red curves, panel b) and the \ion{Si}{4} lightcurve reaches maximum at $t = 4$\,s in the EB9p model. The $N_{\mathrm{e}}$ increase subsequently progresses into the upper chromosphere, reached at $t = 6$ and 8\,s in the EB8 and EB9p models, respectively. These instants are close in time to the \ion{Mg}{2} lightcurve maxima. The $N_{\mathrm{e}}$ increase in the AW1 model (panel c) is delayed by $t \sim 4$\,s from the injection of the wave into the loop. The transition region $N_{\mathrm{e}}$ growth in that model is more significant and abrupt compared to the more gradual growth in the chromosphere, resulting in the \ion{Mg}{2} lightcurve trailing the \ion{Si}{4} one. 

Model atmospheres that do not reproduce the observed delays are plotted in the bottom row of Figure~\ref{fig:model_tne}. The $N_{\mathrm{e}}$ evolution in the EB4p model (panel d), is characterized by rapid and significant density increase in the chromosphere immediately after the onset of the heating, remaining nearly constant for the first 10~s, of the simulation. A consequence of this density increase is the near-instantaneous growth of the \ion{Mg}{2} line intensity, reaching maximum within the first two seconds after the heating onset. This \ion{Mg}{2} behaviour occurs in all of the $F = [5 \times 10^{9}$, $1 \times 10^{10}]$\,ergs\,cm$^{-2}$\,s$^{-1}$ EB models as well as the $F = 5 \times 10^{9}$ H1 model. The $N_{\mathrm{e}}$ increase in the transition region, and the subsequent \ion{Si}{4} intensity growth is more gradual in those models, peaking at $t = $10\,s. 

The TC2 model (panel e) shows a jump in the transition region $N_{\mathrm{e}}$ from the pre-flare $N_{\mathrm{e}} \lesssim 10^9$ cm$^{-3}$ by roughly two orders of magnitude at $t \sim 4$\,s. This initial jump is also evident in the upper chromosphere, albeit less pronounced. The $N_{\mathrm{e}}$ increase propagates from the upper to the lower chromosphere as the time goes by. After $t = 6$\,s, $N_{\mathrm{e}}$ increases in both the chromosphere and the transition region at roughly the same rate, at which the intensities of both lines start to rise. Non-normalized lightcurves (not shown) indicate that the \ion{Si}{4} line becomes observable only when the $N_{\mathrm{e}}$ in the transition region exceeds the order of $10^{11}$\,cm$^{-3}$. On the other hand, the continuing chromospheric $N_{\mathrm{e}}$ increase contributes to the \ion{Mg}{2} line intensity observed at $\approx 200$\,DN\,s$^{-1}$ as soon as the heating sets on. The \ion{Mg}{2} emission shows a steeper rise compared to \ion{Si}{4} and peaks when the $N_{\mathrm{e}}$ in the chromosphere exceeds the order of $10^{12}$\,cm$^{-3}$ at $t \sim 10$\,s (yellow). The \ion{Si}{4} emission continues to progressively increase together with the transition region $N_{\mathrm{e}}$, peaking roughly 6\,s after \ion{Mg}{2}. The onset of the $N_{\mathrm{e}}$ rise in the chromosphere and the transition region in the TC3 model is delayed roughly by 8\,s after simulation start (orange curve in panel f). The atmosphere with $T_{\mathrm{init}} = 3$\,MK is significantly denser than the 1\,MK one (compare the black curves in panels e\& f), leading to more efficient energy dissipation in the corona in the first $\sim 6$\,s \citep[e.g. see Section 4.1 in][]{polito18}. Just like in the TC2 model, the intensities of both lines start to rise simultaneously, but the \ion{Mg}{2} lightcurve shows a steeper increase and peaks first. \\

In summary, the analysis presented above suggests that the observed increase of \ion{Si}{4} emission before \ion{Mg}{2} is reproduced in models with rapid density (and hence emission measure) enhancements in the transition region that occur prior to the equivalent increases in the chromosphere. This condition is rarely met in most of the EB models apart from those with a large energy flux density, or a very small $E_{c}$. The latter is however inconsistent with (or below the sensitivity of) \textsl{Fermi}/GBM observations (Table~\ref{tab:fermi}). The rapid density increase in the transition region is reproduced in the AW simulations, where the chromospheric increase is more gradual. None of the TC heating models in our parameter survey satisfy this requirement, because it takes longer for the density in the transition region to rise via the evaporation process to values high enough for the \ion{Si}{4} line to become visible and eventually peak.

%None of the TC heating models in our parameter survey satisfy this requirement, because it takes longer for the density in the transition region to rise to high enough values for the \ion{Si}{4} line to become visible and eventually peak.

\begin{figure}[h]
\centering
\includegraphics[width=18.00cm, clip, viewport=00 15 510 480]{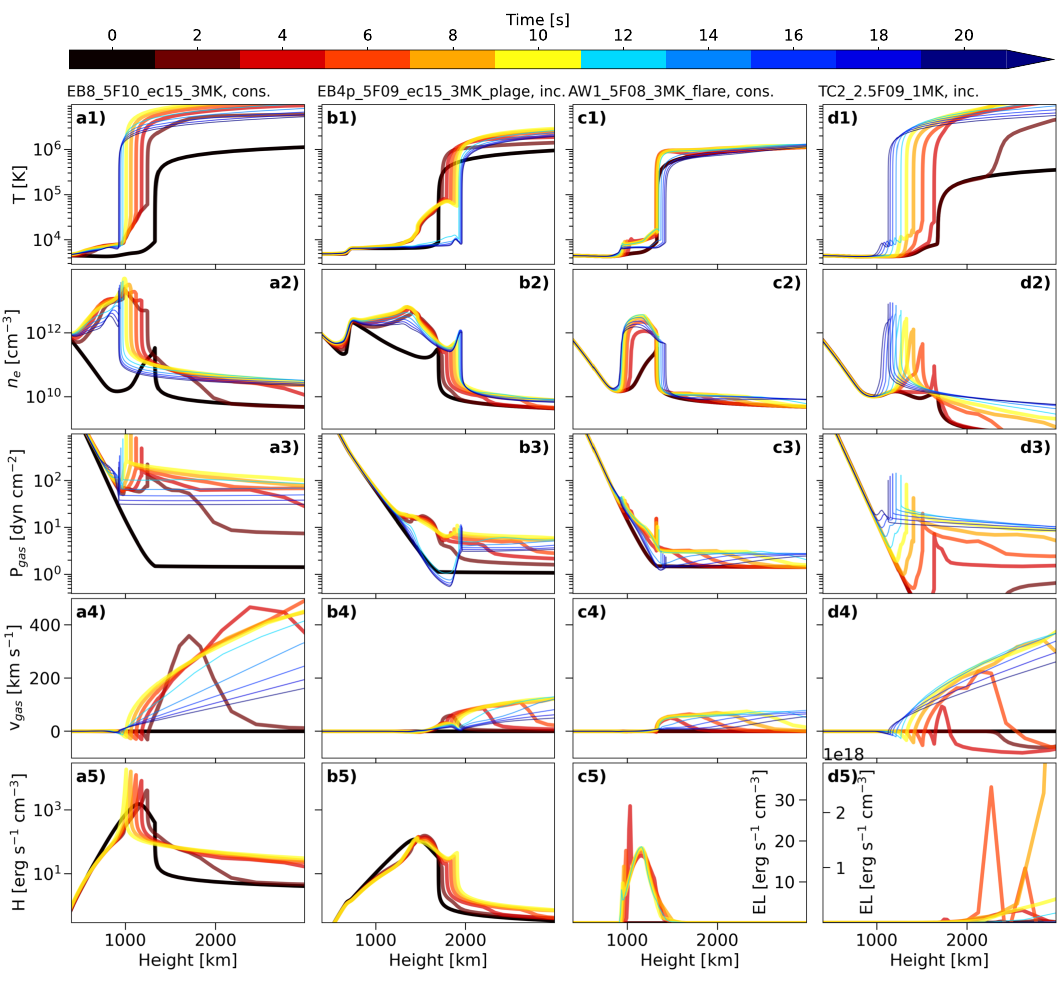}
\caption{Spatio-temporal evolution of atmospheric parameters in example \texttt{RADYN} atmospheres. The color-coding was used to mark the time in the simulation with 2\,s increments, starting from the initial state (black), the heating period (brown $\rightarrow$ yellow thick curves), followed by the cooling period (cyan $\rightarrow$ dark blue thin curves). Panels a and c present the models EB8 and AW1 that reproduce the observed positive delays between the \ion{Si}{4} and \ion{Mg}{2} emission. Panels b and d show the evolution in the EB4p and TC2 models which are not consistent with the observations. From top to bottom are:  gas temperature $T$, electron number density $N_{\mathrm{e}}$ gas pressure, bulk velocity, and energy terms. This is expressed in terms of the EB heating $H$ (in [erg\,s$^{-1}$\,cm$^{-3}$], panels a5 and b5) and the energy losses due to heating (normalized, in [erg\,s$^{-1}$\,cm$^{-3}$]) plotted in panels c5, d5.\label{fig:model_evol}}
\end{figure}

\subsection{Comparing flare dynamics from different energy transport mechanisms}   \label{sec:mod_evol}

Comparing the atmospheric evolution from simulations that are more consistent with the observed \ion{Si}{4} and \ion{Mg}{2} delays can further guide us as to why these delays exist. Figure~\ref{fig:model_evol} presents a comparison from sample models that do (columns of panels a \& c) and do not (columns of panels b \& d) reproduce the observed delays. The parameter evolution is plotted between the heights $z = 400 - 3000$\,km, providing a detailed view of the chromosphere, transition region, and the lower corona. The figure uses the same color-coding as Figure~\ref{fig:model_tne}. Top rows indexed `1', `2', and `3' detail the evolution of gas temperature, electron number density ($N_{\mathrm{e}}$), and pressure, respectively. The macroscopic plasma velocity is plotted in row `4', where positive values correspond to upflows. The last row (`5') presents energy terms expressed in terms of the EB heating (panels a\& b) and energy losses (`EL') due to heating (panels c \& d).

\begin{itemize}
\item{Consistent Model 1: The energy deposition rate (panel a5) in the EB8 model exhibits a narrow maximum in the transition region, whose tails extend to the upper chromosphere as well as the corona. As a result of the heating, the temperature, density, and pressure exhibit a prompt (at $t = 2$\,s, brown color) rise in the entire region under study (panels a1 -- a3). The region where the transition region forms shifts from $\sim$1300\,km at the simulation onset to $\sim$900\,km at the end of the investigated period. Strong density gradients and a pressure pulse are formed at the transition region at $t = 4$\,s, both propagating downwards. The increase of pressure drives strong ($v_\mathrm{g} \sim 400$\,\kps, panel a4) evaporative upflows above the transition region immediately after the onset of the heating.}

\item{Inconsistent Model 1: The energy deposition in the EB4p simulation exhibits a broader peak spanning from the chromosphere to the transition region, peaking in the upper chromosphere at $z \sim 1500$\,km (panel b5). The order-of-magnitude EB flux decrease, compared to the EB8 model, translates to an order of magnitude drop in the peak heating. The temperature, density, and pressure increase at chromospheric altitudes (panels b1 -- b3; $z \sim 1700$\,km) are evident, while the position of the transition region shifts slightly upwards by  $\sim$200\,km. The pressure exhibits an increase in the chromosphere and, after $t = 6$\,s, a weak pulse in the transition region, driving modest evaporative flows with $v_\mathrm{g} \lessapprox 100$\,\kps above it.}

\item{Consistent Model 2: The Alfv\'e{n} wave heating in the AW1 model causes energy losses over a broad region below the transition region, with a narrow peak in the lower chromosphere at $t = 4 - 8$\,s. The $\sim 4$\,s delay of the energy losses after the start of the heating (panel c5) is caused by the wave travel time after injection. Unlike the rest of the models discussed here, the simulation shows only a modest temperature increase, both in the chromosphere and the lower corona ($T_\mathrm{g} \sim 10^6$\,K, panel c1). The density in the transition region peaks at $t = 4 - 6$\,s after which it starts to slowly decrease as the location of the wave heating ``bores'' deeper into the chromosphere, as discovered by \citet{reep18b}. The fact that the wave propagates as opposed to traveling effectively instantaneously like the electron beams means that the chromospheric density peak is delayed until the end of the heating period (orange \& yellow curves in panel c2). The weak pressure pulse in the transition region drives slow evaporation at $v_\mathrm{g} \lessapprox 80$\,\kps\ in the corona (panels c3 \& c4). We also found that at $t = 6$\,s, the transition region position shifts slightly downward by $\sim$3\,km, followed by the formation at progressively higher altitudes after $t = 8$\,s.}

\item{Inconsistent Model 2: The \textsl{in-situ} direct heating in the model TC2 at altitudes $z > 2000$\,km causes significant (by more than an order of magnitude) temperature increase in the corona (panel d1) and steepens the temperature gradient in the transition region. The heating leads to a pressure gradient in the transition region which shifts down (to higher column masses) to balance the incoming conduction flux. The pressure gradient causes the temperature and density in the underlying chromosphere to rise at $t \sim 6$\,s (red curve in panels d1 \& d2). The temperature continues to grow during the rest of the heating period (orange $\rightarrow$ yellow curves), although gradual increase of the density and pressure progresses even beyond (panels d2 \& d3). Pronounced pressure pulse drives evaporative upflows above the transition region, whose strength rises from $v_\mathrm{g} \sim 100$\,\kps at $t = 4$\,s to $v_\mathrm{g} > 200$\,\kps at $t = 6$\,s (panel d4).} 
\end{itemize}

The formation of \ion{Si}{4} line emission is contingent upon electron number density increase in the transition region. The plasma that fills the transition region originates in the underlying chromosphere and is carried upward by the evaporative flows driven by pressure imbalance due to heating. The location of energy deposition has a major effect on the evolution of the atmospheric parameters, including the gas velocity. The peak of the heating rate coincides with the transition region for example in the EB8 (Figure~\ref{fig:model_evol}a) and EB9p (not shown) models that reproduce the observed delays. In both of those types of non-thermal electron distributions (larger $F$ or small $E_{c}$), there are a greater number of low-energy electrons that are easily thermalized in the transition region. Atmospheres where the bulk of the heating is released in the chromosphere, such as the EB4p (Figure~\ref{fig:model_evol}b) as well as most of the EB models exhibit a steep \ion{Mg}{2} emission increase preceding that of \ion{Si}{4}, not consistent with the observations. Our analysis suggests that whichever of the two lines in the EB as well as the AW models peaks first is given by the location of energy dissipation. Such interpretation is not applicable for the TC simulations. Since the downward directed heat flux from the the thermal conduction models must travel through the transition region prior to the chromosphere, one might expect the transition region emission increase prior to the chromospheric one. Our analysis (Figure~\ref{fig:model_tne}e, Figure~\ref{fig:model_evol}d) however challenges this view, as the density increase sets on soon after the heating onset in both the chromosphere and the transition region. While the TC heating effectively dissipates in the corona (even more so for the 3\,MK model), the lower atmosphere responds to the heating by increasing radiative losses from the chromosphere, driving \ion{Mg}{2} emission before \ion{Si}{4}. In conclusion, both the density and the location of the heating deposition are key to explain why the \ion{Si}{4} peaks before \ion{Mg}{2} or vice-versa.

\section{Discussion}
\label{sec:discussion}

\subsection{Intensity peaks and kernel motions}

The observations in Section \ref{sec:obs_lightcurves} showed that peaks along the \ion{Mg}{2} k (2796.35\,\AA), \ion{C}{2} 1334.53\,\AA, and \ion{Si}{4} 1402.77\,\AA~intensity lightcurves appeared quasi-periodically, at periods between $16 - 18$\,s. {Solar flare emission often exhibits time variations or pulsations that can display a quasi-periodic pattern over a broad range of timescales, commonly referred to as quasi-periodic pulsations (QPPs) \citep[see e.g.][and references therein]{nakariakov09, mclaughlin18, zimovets21}. For examples of studies reporting on QPP periods close to those reported here, see e.g. \citet{hayes20, kou22}. While a broad range of studies have investigated QPPs, their underlying physical drivers remain uncertain. This is in part because the term `QPP' encompasses a wide variety of observational signatures with different characteristics and likely different origins. Commonly proposed physical mechanisms include magnetohydrodynamic (MHD) waves, and `bursty', time-dependent or oscillatory reconnection. QPPs may also originate in the upper reconnection region via instabilities or MHD wave-driven modulation of the reconnection rate \citep[see discussions in][]{hayes19, kou22}. Oscillations of \ion{Si}{4} non-thermal broadening \citep[][]{jeffrey18} and intensity \citep[][]{chitta20} with periods of about 10\,s have also been previously interpreted as signatures of turbulence. However, with the exception of the event B2 when signal was strong, we did not find evidence for broadening increases associated to the intensity peaks in the event under study, suggesting that turbulence is unlikely to be the primary driver here. By investigating SJI 1330\,\AA~intensities in ribbon portions adjacent to the IRIS slit, we found that the signal enhancements were caused by kernel brightenings either appearing or moving across the IRIS slit as a consequence of slipping reconnection,  as further investigated in \citet{lorincik24}. QPPs driven by or associated to apparent slipping motions have been reported in the past at periods of a few minutes \citep{li15b} down to $\approx 10$\,s \citep{zhang25}, close to the timescales we report on here. While a detailed investigation into the origin of these QPP signatures is beyond the scope of this work, it represents a valuable direction for future research.} 

{Since the dataset under study was acquired in the sit-and-stare mode, our analysis was limited to several IRIS pixels coincident with the ribbon portion where these kernels induced spectral enhancements. Given the orientation of the IRIS slit with respect to the ribbon, it is difficult to state how general these results are across the entirety of the flare. However, comparable behavior was observed across four distinct brightening events (B1 -- B4), increasing credibility of our results. It ought to be emphasized that the cadence of conventional IRIS raster observations would not be high enough to resolve the delays between the chromospheric and the transition region emission. This highlights the importance of spectral measurements covering a large spatial region simultaneously and at a high cadence by instruments such as the upcoming Multi-slit Solar Explorer \citep[MUSE;][]{depontieu22, cheung22} or the proposed Spectral Imager of the Solar Atmosphere \citep[SISA;][]{calcines24}.}

\subsection{Delays and HXR signatures of particle acceleration}

Despite the ubiquitous reconnection manifestations in the form of kernel motions (Section \ref{sec:overview}), the HXR observations from \textsl{Fermi}/GBM showed evidence of a weak non-thermal component only (Figure~\ref{fig:overview}b). The non-thermal event set on $\approx$2 minutes after the first brightening event B1 had already occurred. Therefore, our interpretation of processes causing the observed delays (Section \ref{sec:mod_tne}, \ref{sec:mod_evol}) is likely only relevant to the time period coincident with the B2 -- B4.

Examples of simulations reproducing the observed delays between the \ion{Si}{4} and \ion{Mg}{2} peaks include the EB7 -- EB8p atmospheres heated by high non-thermal EB flux. The $F = 5 \times 10^{10}$\,ergs\,cm$^2$\,s$^{-1}$ used in those models is the upper constraint on the flux {obtained from $A_\mathrm{min}$ during B2. This result implies that, for the EB models to reproduce the observed delays, the non-thermal energy needs to be deposited to a small-scale ($\sim
15$ IRIS pixels$^2$) flare kernel (see also Section \ref{sec:data_fermi}).} The elevated flux levels were measured soon after the onset of the non-thermal event (Table~\ref{tab:fermi}), and therefore the high-flux models are relevant in the period coincident with the brightening event B2. Considering that the non-thermal flux decreased by a factor of $\approx 20$ during the B2 and B4, but the delays were still mostly positive in the same period (Figure~\ref{fig:delays}), indicates that the driver of these delays was likely changing during the flare.

Unlike the high-flux EB and Alfv\'{e}n wave models, the EB9 and EB9p models reproduce both the observed delays and most of the observed spectral properties (Section \ref{sec:comparison}). Although this is an interesting result, it has to be taken with caution as $E_c = 5$\,keV used therein is much lower than the $E_c = 15$\,keV adopted from the HXR spectral fitting. These models have been included in the analysis to explore the possibility of the delays being a consequence of heating by less energetic electrons with $E_c$ below the detection threshold of \textsl{Fermi}/GBM (8\,keV). 

\subsection{Open questions in delay simulations} 

Only a fraction of the \texttt{RADYN} models reproduce the observations (Section \ref{sec:mod_lc}). The delays thus pose tight, previously unexplored constraints on RHD simulations of flare emission. {In Figure~\ref{fig:histo}, the positive delays reproduced in some of the electron-beam heated models (symbols) are clustered at $\delta_{\mathrm{t}} = 3.5$, 6.5\,s, whereas the observed delays are distributed more uniformly. This clustering can be attributed to similar time evolution of these model atmospheres, where the electron density increase in the line formation regions sets on close in time. Differences of peak times of emission synthesized therein, if any, are below the 0.5\,s temporal resolution of the synthetic lightcurves. None of the models reproduce the observed delays of $ \Delta_{t}/2 < \delta_{\mathrm{t}} \lessapprox 3$ and $\approx 5$\,s. A broader heating parameter survey would possibly fill this gap, but reproducing the exact properties of the observed delays was not the goal of this work.} 

Most of the lightcurves synthesized in the EB models, including those that reproduce the observed delays between the \ion{Si}{4} and \ion{Mg}{2} emission, are characterized by steep intensity increase right after the heating onset (Figure~\ref{fig:model_lc}). This behavior does not align with the observed lightcurves where the peaks exhibit gradual intensity increase and decrease, albeit with varying slopes, widths, and amplitudes (see Section \ref{sec:obs_lightcurves} and Appendix \ref{sec:appendix_d}). Time evolution aligning with the observations is exhibited by the smoother lightcurves synthesized in the TC models (Figure~\ref{fig:model_lc}u -- y) which however do not reproduce the observed delays. An interesting result of its own, even though conduction fronts pass the transition region before reaching the chromosphere, the chromospheric emission is the first to rise and peak. This is due to the fact that the increase of the electron density in the transition region is driven by evaporation from the chromosphere which is the first to respond to \textit{in-situ} heating in the corona by increasing radiative losses therein \citep[see also][]{polito18}. Because of the {efficient heating dissipation in the corona, especially in the 3\,MK loops, the density increase in the transition region is delayed and the synthetic \ion{Si}{4} intensities in some cases peak as long as 20\,s after the heating onset (Figure \ref{fig:model_lc}w). These lightcurves in addition last for} tens of seconds, and the up to 15\,s long delays between the \ion{Si}{4} and \ion{Mg}{2} intensity peaks synthesized therein are much longer than the observed ones. {We therefore conclude that it is unlikely for TC heating to reproduce the observed delays. Note that using an even higher initial loop apex temperature of 5\,MK results in more effective dissipation in the corona and, as a consequence, reduced heating of the transition region, as shown in simulations of \citet{polito18}.} Relatively-shorter ($< 3$\,s) delays between the \ion{Mg}{2} and trailing \ion{Si}{4} emission were found in TC-heated nanoflare loop models of \citet{polito18} that we revisited in our investigation (not shown). The 15\,Mm loop length used in those simulations likely facilitates quicker post-evaporation density increase in the transition region compared to the 50\,Mm loop length in the model grid constrained by our observations. 

The last heating mechanism analyzed in our study are the Alfv\'{e}n waves (AW). The lightcurves produced via two out of three AW-heated simulations under study were found to replicate the observed delays, with the \ion{Si}{4} line peaking first. The \ion{Si}{4} intensity lightcurve synthesized in the AW1 model also exhibited a relatively-gentle increase to the maximum (Figure~\ref{fig:model_lc}z), in line with the observations. The AW models that reproduced the delays, on the contrary, did not replicate \ion{Si}{4} redshifts and broadening characteristic for the flare (although broadening due to the wave itself was not synthesized). Mild ($v_\mathrm{D} < 10$\,km\,s$^{-1}$) \ion{Si}{4} redshifts were found in the higher-flux AW3 model, which however does not reproduce the observed delays (Figure~\ref{fig:model_lc}zb). When the atmospheric evolution therein (not shown) is compared to the lower-flux AW1 model (Figure~\ref{fig:model_evol}c), the peak of the energy losses shifts to the chromosphere which also exhibits a strong increase of the electron number density. It ought to be noted that none of the simulations analyzed here account for persistent mild ($v_\mathrm{D} \lessapprox 15$\,km\,s$^{-1}$) redshifts ubiquitous in transition region spectra \citep[see e.g.][and references therein]{hardi99a, hardi99b, feldmann11}. Had these `reference' redshifts been added to the synthetic spectra (Figure~\ref{fig:synth_q}), the resulting Doppler shifts in the AW runs would align with the observations. Interestingly, the Alfv\'{e}n wave models replicated the observations fairly well with no bespoke adjusting in otherwise broad parameter space of wave frequency $f$, perpendicular wavenumber $k_x$ and magnetic field stratification. The chosen values of $f$ and $k_x$ were selected based on previous experience, as realistic values likely to produce damping near the transition region and in the upper chromosphere \citep{emslie82,russell13,reep16,kerr16,russell24}. Meanwhile, the magnetic field stratification had photospheric $B_0 = 1500$\,G and coronal $B_\mathrm{cor} = 500$\,G (c.f. Eq. (10) of \citet{kerr16}) {representing the upper constraint on the magnetic field strength in the active region (Section \ref{sec:mod_method}).} The lack of precise measurements of coronal magnetic fields during flares poses a limitation on constraining the AW heating parameters. This has an impact on the Alfv\'en speed and wave damping properties. {Spectral synthesis in \texttt{RADYN} simulations subjected to updated AW heating with different heating parameters will be presented in the follow-up study by \citet{kerr_awnanoflares}.}

We finally have to reiterate numerous assumptions and limitations of our experiment. First, \ion{Mg}{2} line profiles synthesized in most \texttt{RADYN+RH} models are typically double-peaked, in contrast to single-peaked profiles commonly observed during flares, making comparisons between modeling and observations difficult. Next, even though the observed intensity peaks appeared quasi-periodically, all models considered in our study for simplicity assume a single instance of constant heating as we are here focused on simulating individual impulsive peak episodes in a single loop. Whether repeated heating of a single loop can reproduce the quasi-periodic intensity peaks delayed across different ions remains to be simulated in the future. In addition to these assumptions, we can not rule-out the possibility of coexistence of different non-thermal electron populations heating the same or adjacent volumes of plasma \citep{polito23}, or simultaneous occurrence of different energy transport mechanisms. Last, the simulations under study are neither multithreaded \citep{falewicz15, polito19, reep20}, nor account for cross-sectional expansion of flux tubes \citep{reep24}, properties important for replicating flare lightcurves or description of mass flows from the chromosphere into the corona, respectively.

\section{Summary}
\label{sec:summary}

This manuscript presents spectral analysis of the \ion{Si}{4} 1402.77\,\AA, \ion{C}{2} 1334.53\,\AA, and \ion{Mg}{2} 2796.35\,\AA~lines observed by the IRIS satellite during a C-class flare from 2022 September 25. Line intensity lightcurves observed in periods coincident with the appearance of flare ribbon kernels beneath the IRIS slit exhibited multiple peaks. The most notable outcome of this study is the discovery of delays between the peaks along the intensity lightcurves. The delays between the transition region \ion{Si}{4} and the chromospheric \ion{Mg}{2} line, we focused on primarily, were on average $1.8$\,s and up to $\approx 6$\,s long. Data indicate that the positive delays (\ion{Si}{4} peaks before equivalent \ion{Mg}{2} peaks) were becoming more consistent as the time went by in the impulsive phase of the flare. These delays present an intriguing observable indicative of sequential excitation of different regions of the lower atmosphere of the Sun subjected to flare energization. 

In order to {investigate physical conditions leading to} the delays revealed by the high-cadence IRIS spectroscopy, we analyzed a grid of \texttt{RADYN} flare atmospheres heated by electron beams, thermal conduction following \textsl{in-situ} heating in the corona, and Alfv\'{e}n waves. Parameters of electron beam heating were constrained from HXR observations of \textsl{Fermi}/GBM, which detected a weak ($E_c \approx 15$\,keV) non-thermal event during the flare impulsive phase. Synthetic lightcurves consistent with the observations were reproduced in the high-flux ($F = 5 \times 10^{10}$\,ergs\,cm$^2$\,s$^{-1}$) electron beam and hybrid models, low flux ($F = 1 \times 10^{9}$\,ergs\,cm$^2$\,s$^{-1}$) but small $E_c = 5$\,keV models, and certain models heated by Alfv\'{e}n waves. Simulation analysis revealed that the peaks in the \ion{Si}{4} emission precede their \ion{Mg}{2} counterparts when the flare heating causes rapid and significant density increase from pre-flare levels in the transition region where \ion{Si}{4} forms. On the contrary, heating causing strong density increase in the chromosphere prior to the transition region causes \ion{Mg}{2} intensity to increase before \ion{Si}{4}, which does not align with the observations. Model atmospheres heated by thermal conduction were found to be inconsistent with the observations as the \ion{Mg}{2} emission precedes that of \ion{Si}{4} due to increased radiative losses in the chromosphere in the wake of \textsl{in-situ} coronal heating. Alfv\'{e}n wave heating presents a promising candidate mechanism reproducing the observed delays, but the properties of synthetic spectra do not align with the observations and the wave parameters can be difficult to constrain. We have however just started to explore the Alfv\'{e}n wave heating using the \texttt{RADYN} code, and further investigation of this mechanism is needed. It ought to be noted that majority of the observed delays, especially later in the flare impulsive phase, occurred when the \textsl{Fermi}/GBM flux was low ($F = 5 \times 10^{8} - 10^{9}$\,ergs\,cm$^2$\,s$^{-1}$), supporting the notion that the delays are due to smaller energetic heating events or the Alfv\'{e}n waves. It is also likely that the heating mechanism causing the delays could have been changing as the time went by during the flare.

To conclude, the newly-observed delays between the chromospheric and transition region emission have a rich diagnostic potential to probe the response of the lower atmosphere to flare heating and provide valuable constraints on models of flare emission. Future studies leveraging now extensive database of IRIS flare observations at very high, sub-second cadence will address the prevalence of these delays and their implications for flare energetics.

\section*{Acknowledgments}

JL and VP acknowledge support from NASA under the contract NNG09FA40C (IRIS). JL also acknowledges support from the Heliophysics Guest Investigator (H-GI) Open grant 80NSSC24K0553. V.P. also acknowledges support from NASA H-GI grant No. 80NSSC20K0716. GSK acknowledges support from a NASA Early Career Investigator Program award (grant 80NSSC21K0460). L.A.H was supported by an ESA Research Fellowship during the main part of this research. She is now supported by a Royal Society-Research Ireland University Research Fellowship. This manuscript benefited from discussions held at a meeting of the International Space Science Institute team: “Interrogating Field-Aligned Solar Flare Models: Comparing, Contrasting and Improving,” led by Dr. G. S. Kerr and Dr. V. Polito. Resources supporting this work were provided by the NASA High-End Computing (HEC) Program through the NASA Advanced Supercomputing (NAS) Division at Ames Research Center GID s2306. IRIS is a NASA small explorer mission developed and operated by LMSAL with mission operations executed at NASA Ames Research Center and major contributions to downlink communications funded by ESA and the Norwegian Space Agency. SDO data were obtained courtesy of NASA/SDO and the AIA and HMI science teams. This research has made use of the Astrophysics Data System, funded by NASA under Cooperative Agreement 80NSSC21M00561. CHIANTI is a collaborative project involving George Mason University, the University of Michigan (USA), University of Cambridge (UK) and NASA Goddard Space Flight Center (USA).

\appendix 
\newcounter{appendix}
\renewcommand{\theappendix}{Appendix}
\refstepcounter{appendix}

\setcounter{figure}{0}    
\renewcommand{\figurename}{Appendix Figure}

\section{Typical IRIS line profiles}
\label{sec:appendix_a}

\begin{figure*}[h]
\centering
\includegraphics[width=6.900cm, clip,   viewport=000 00 635 380]{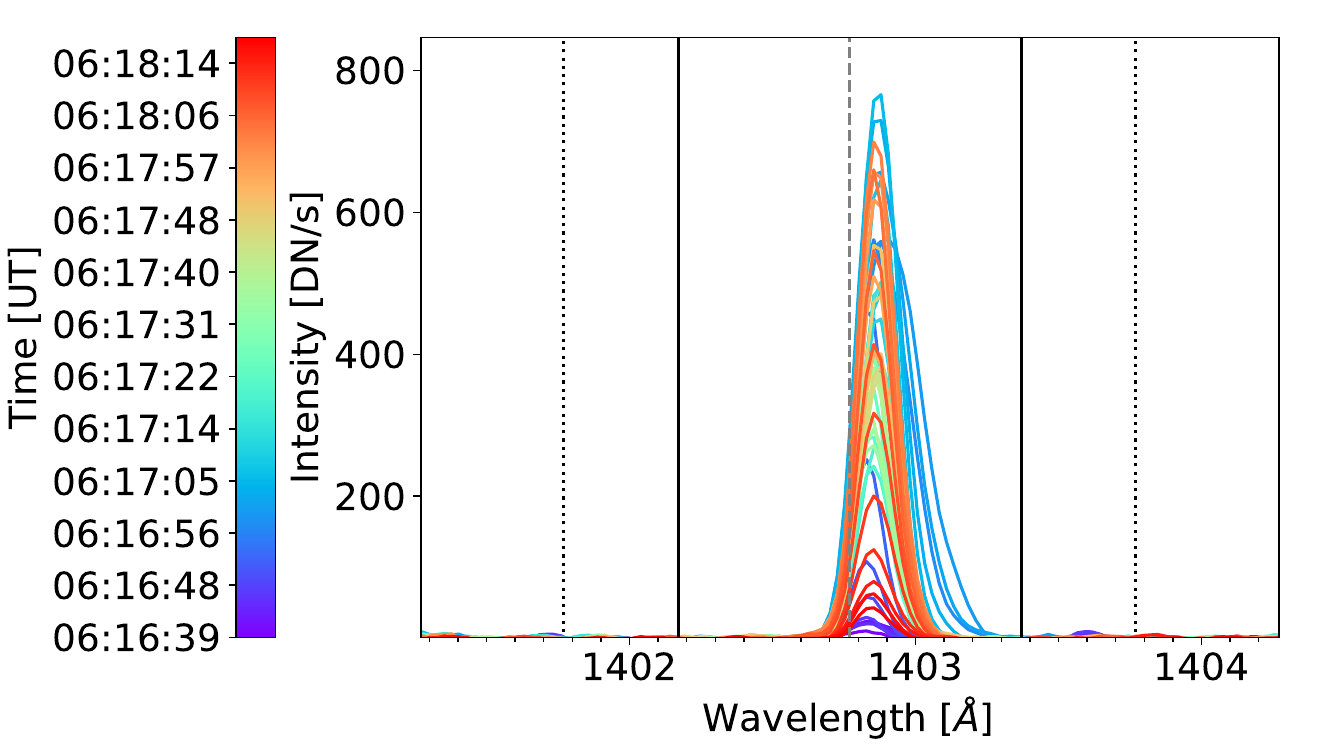}
\includegraphics[width=5.378cm, clip,   viewport=140 00 635 380]{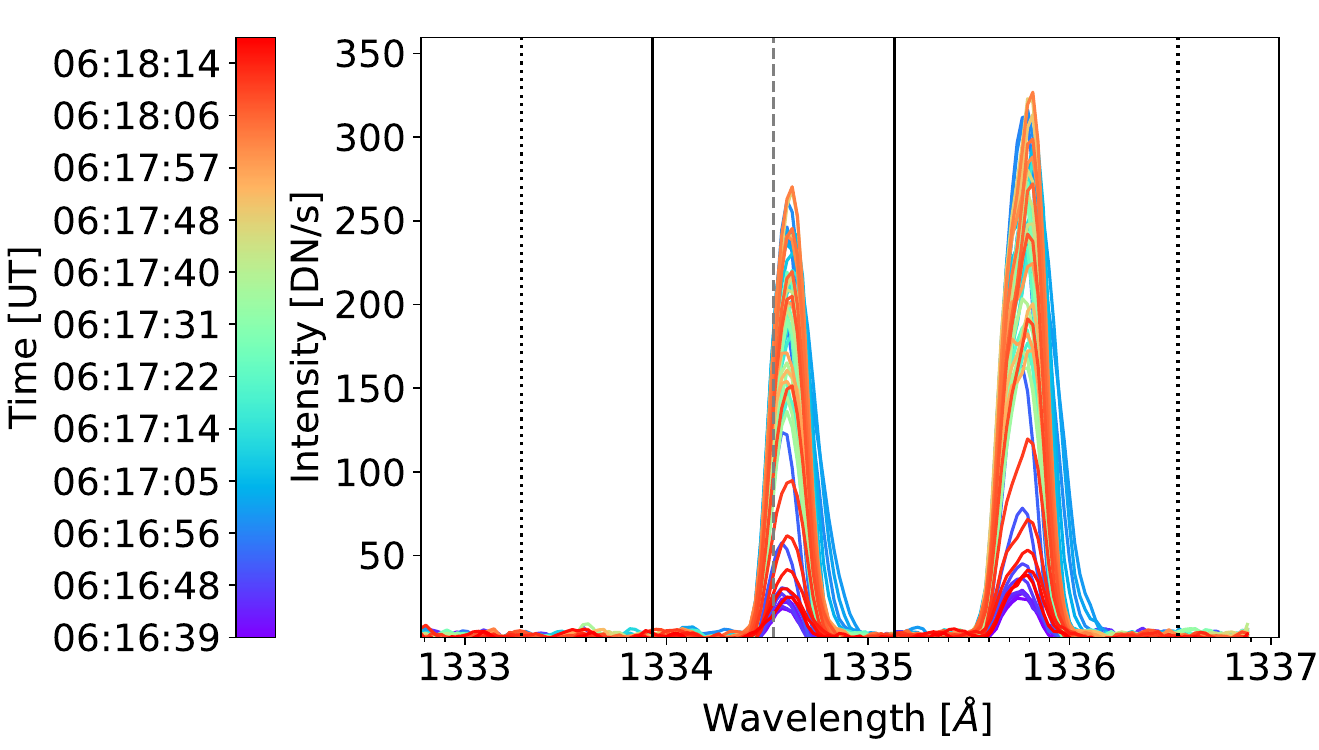}
\includegraphics[width=5.378cm, clip,   viewport=140 00 635 380]{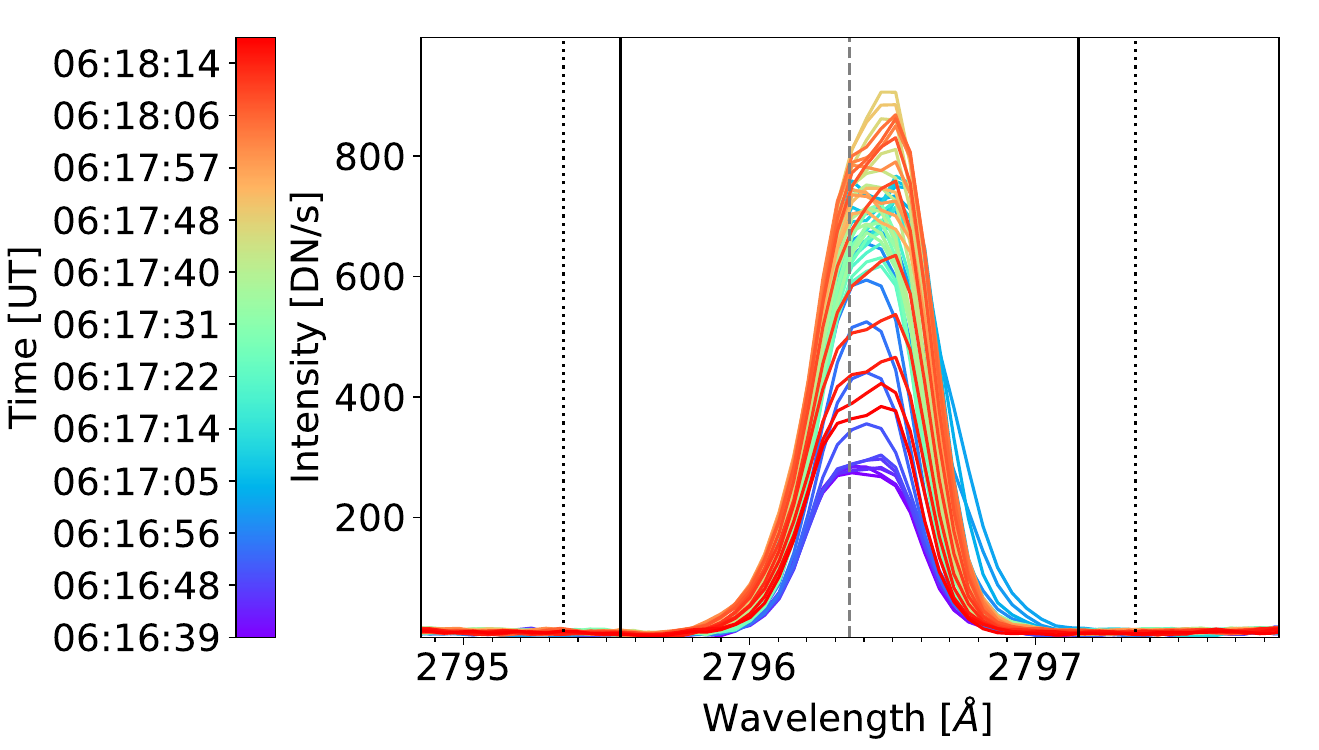}
\caption{\ion{Si}{4} 1402.77\,\AA~(left), \ion{C}{2} 1334.53\,\AA~(middle), and \ion{Mg}{2} 2796.35\,\AA~(right) line profiles during the brightening event B2. Dashed grey lines indicate the rest wavelengths of lines under study. Solid black lines delineate the spectral region used for the calculation of moments of each line, while the grey dotted lines border regions used for the subtraction of the FUV and NUV continua around the profiles. \label{fig:profiles_app}}
\end{figure*}

Example \ion{Si}{4} 1402.77\,\AA~(left), \ion{C}{2} 1334.53\,\AA~and 1335.66\,\AA~(middle), and \ion{Mg}{2} 2796.35\,\AA~(right) spectral line profiles observed during the B2 are plotted in Appendix Figure~\ref{fig:profiles_app}. In each panel, the rest wavelengths are indicated using the dashed gray line, while the solid black lines indicate the limits for calculating the moments (Section \ref{sec:obs_properties}). The dotted black lines designate the inner boundaries of spectral regions used for the FUV and NUV continuum estimation, with the outer boundary corresponding to the plotted edge of the spectral window. For typical values of the total intensities, Doppler velocities, and line broadening determined via the moment analysis see Section \ref{sec:obs_properties} and Figure~\ref{fig:inspection}.
\section{\ion{Mg}{2} k3 intensities}
\label{sec:appendix_b}

\begin{figure*}[h]
\centering
\includegraphics[width=18.00cm, clip,   viewport=00 00 510 350]{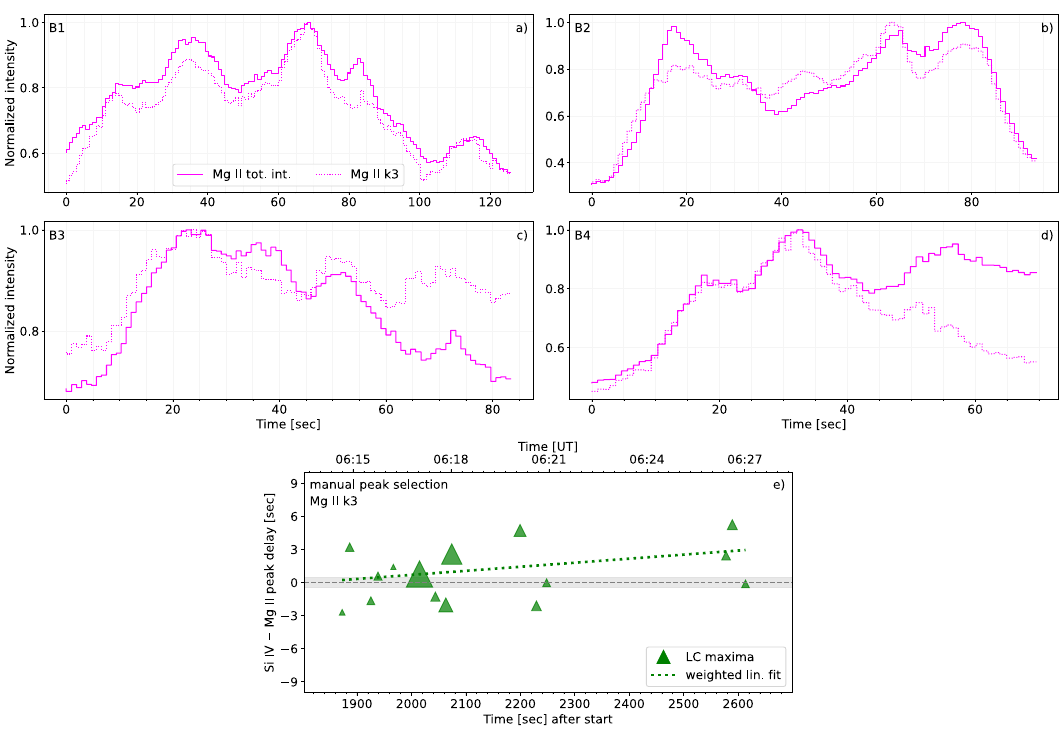} 
\caption{Comparison of normalized \ion{Mg}{2} 2796.35\,\AA~lightcurves obtained via computing the total intensities (0th moment, solid curves) and the k3 component (dotted curves) during the B1 -- B4 (panels a -- d). Panel e presents the delays between the maxima of \ion{Si}{4} and \ion{Mg}{2} line intensity peaks a function of time during the flare. The panel has the same design as Figure~\ref{fig:delays}, but the delays were computed using the \ion{Mg}{2} k3 component intensities instead of the total intensity. \label{fig:k3_analysis}} 
\end{figure*}

Unlike the typically single-peaked \ion{Si}{4} profiles, the \ion{Mg}{2} spectra often consist of multiple components under quiescent conditions, or at certain flare locations \citep{polito23,kerr24b}. The line core is commonly defined as k3 and the  blue and red peaks as k2v and k2r, respectively.  Only some of the analyzed profiles showed traces of these additional components, and most of the profiles were single-peaked (Appendix Figure~\ref{fig:profiles_app}), in agreement with previous analyses of \ion{Mg}{2} flare characteristics in bright ribbons \citep[e.g.,][]{kerr15, panos18, polito23, sainzdalda23}. Nevertheless, the \texttt{iris\_get\_mg\_features\_lev2} \citep{pereira13} was employed to probe the intensities and Doppler shifts of these components when they existed. Upon visual inspection of fits to the profiles we found that the line core or peak was typically well-fitted by the k3 component, and we focused on this component here. Appendix Figure~\ref{fig:k3_analysis}a -- d presents comparison of the \ion{Mg}{2} intensities corresponding to the total (solid line) and k3 (dotted line) intensities during the B1 -- B4. The k3 component lightcurves are generally similar to those of the total intensity, but they do not exhibit the same peaks and the peaks are in some cases less prominent, e.g. at $t = 85$\,s in panel a, or $t = 20$\,s in panel b. Further, the k3 lightcurve in panel c is noisier than the one corresponding to the total intensity.

The delays between the peak maxima along the \ion{Si}{4} total intensity lightcurves (Figure~\ref{fig:lightcurves}) and the \ion{Mg}{2} k3 lightcurves are plotted in Appendix Figure~\ref{fig:k3_analysis}e. This panel has the same style as Figure~\ref{fig:delays}. The linear fit to the observed delays (green dotted line) is consistent with that of the integrated intensity plotted in Figure~\ref{fig:delays}, albeit with more scatter around zero. The mean delay using the k3 intensities lowered to $\delta_{\mathrm{t}} = 0.7 \pm 0.6$\,s with $\sigma = 2.4$\,s, while the mean positive delay is $\delta_{\mathrm{t}} = 2.6 \pm 0.6$\,s with $\sigma = 1.6$\,s (see Appendix \ref{sec:appendix_d} for further details).

\section{Automatic detection of intensity peaks}
\label{sec:appendix_c}

\begin{figure*}[h]
\centering
\includegraphics[width=8.90cm, clip,   viewport=00 00 780 430]{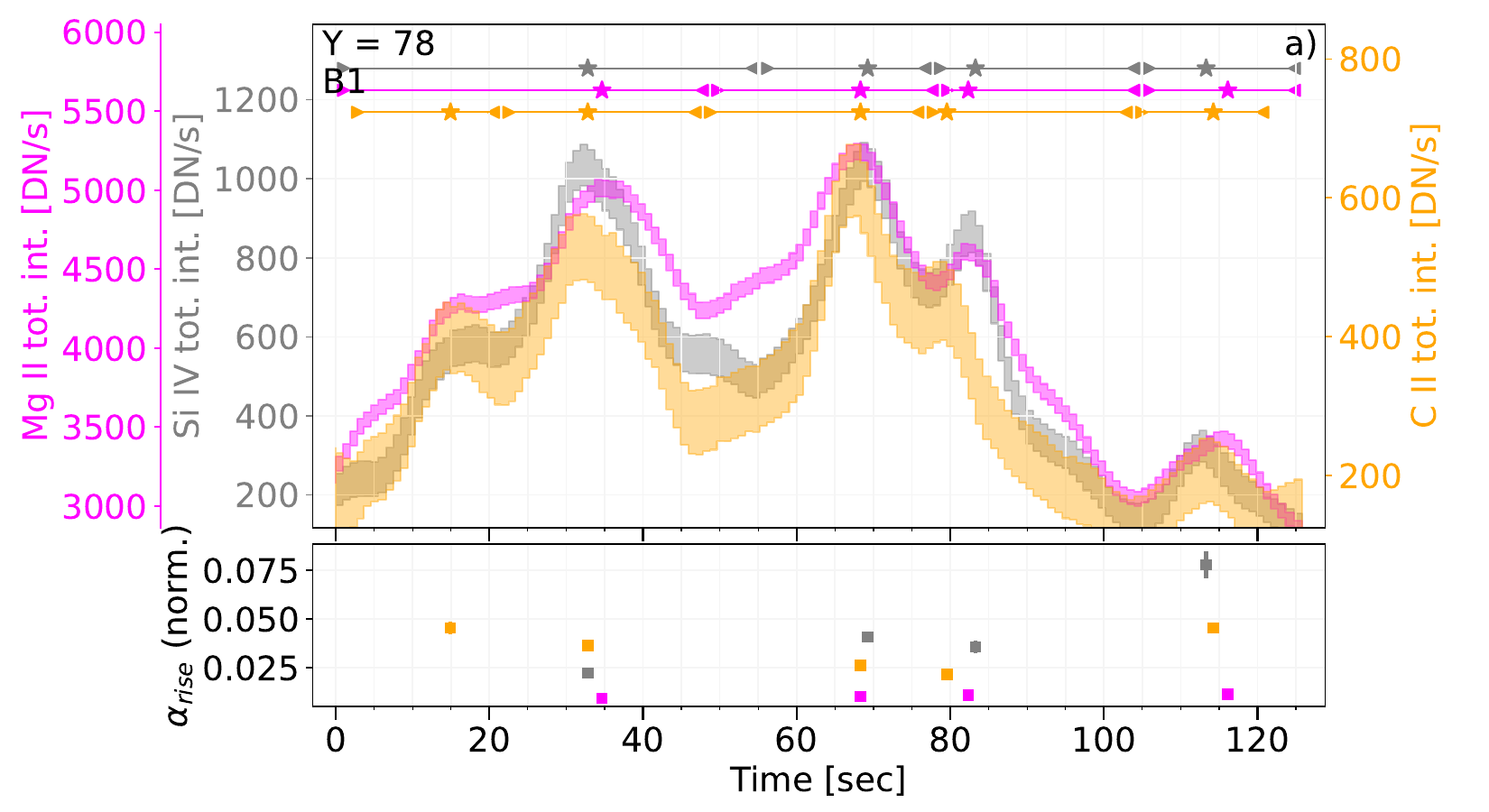} 
\includegraphics[width=8.90cm, clip,   viewport=00 00 780 430]{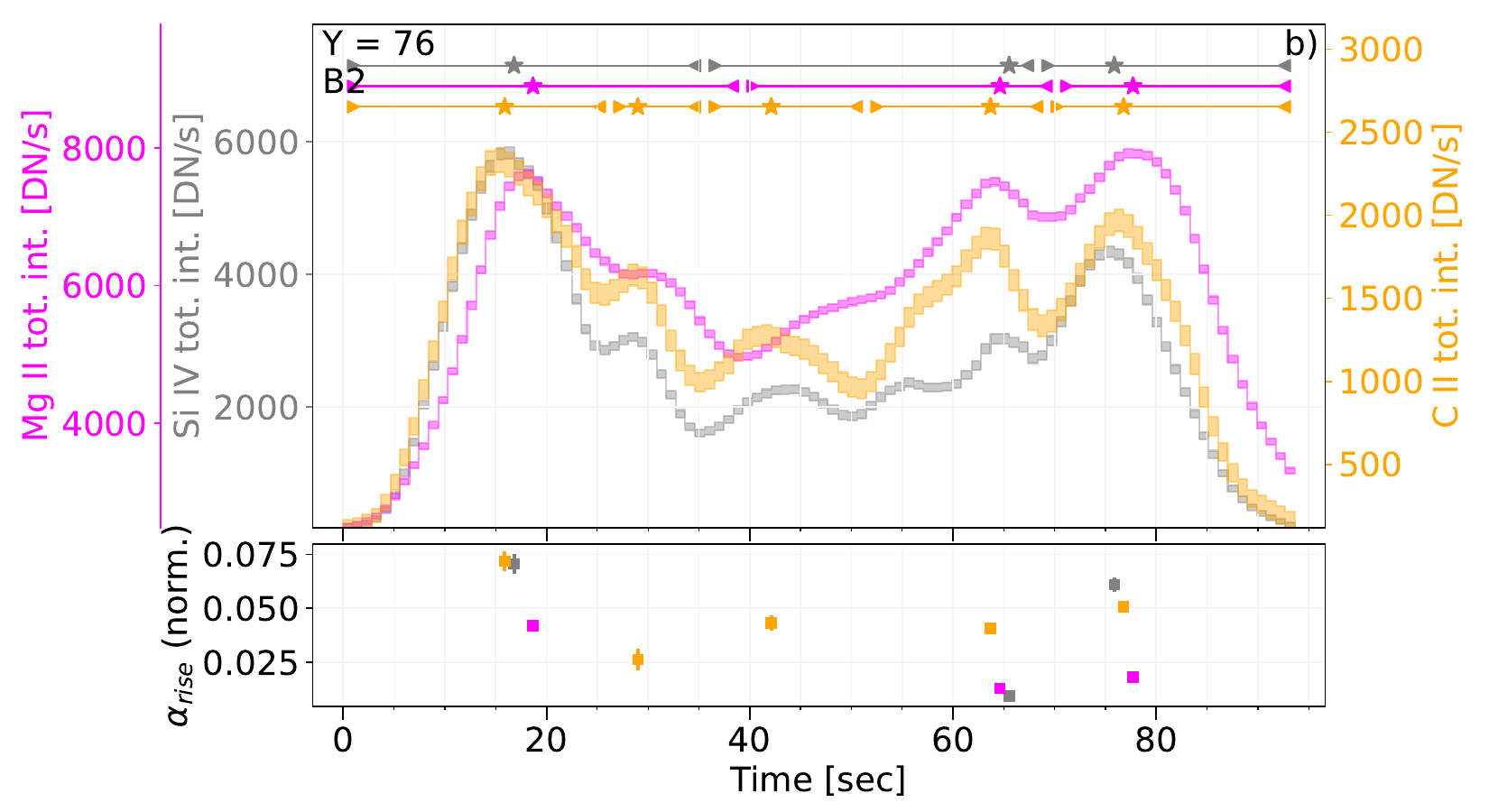} 
\\
\includegraphics[width=8.90cm, clip,   viewport=00 00 780 430]{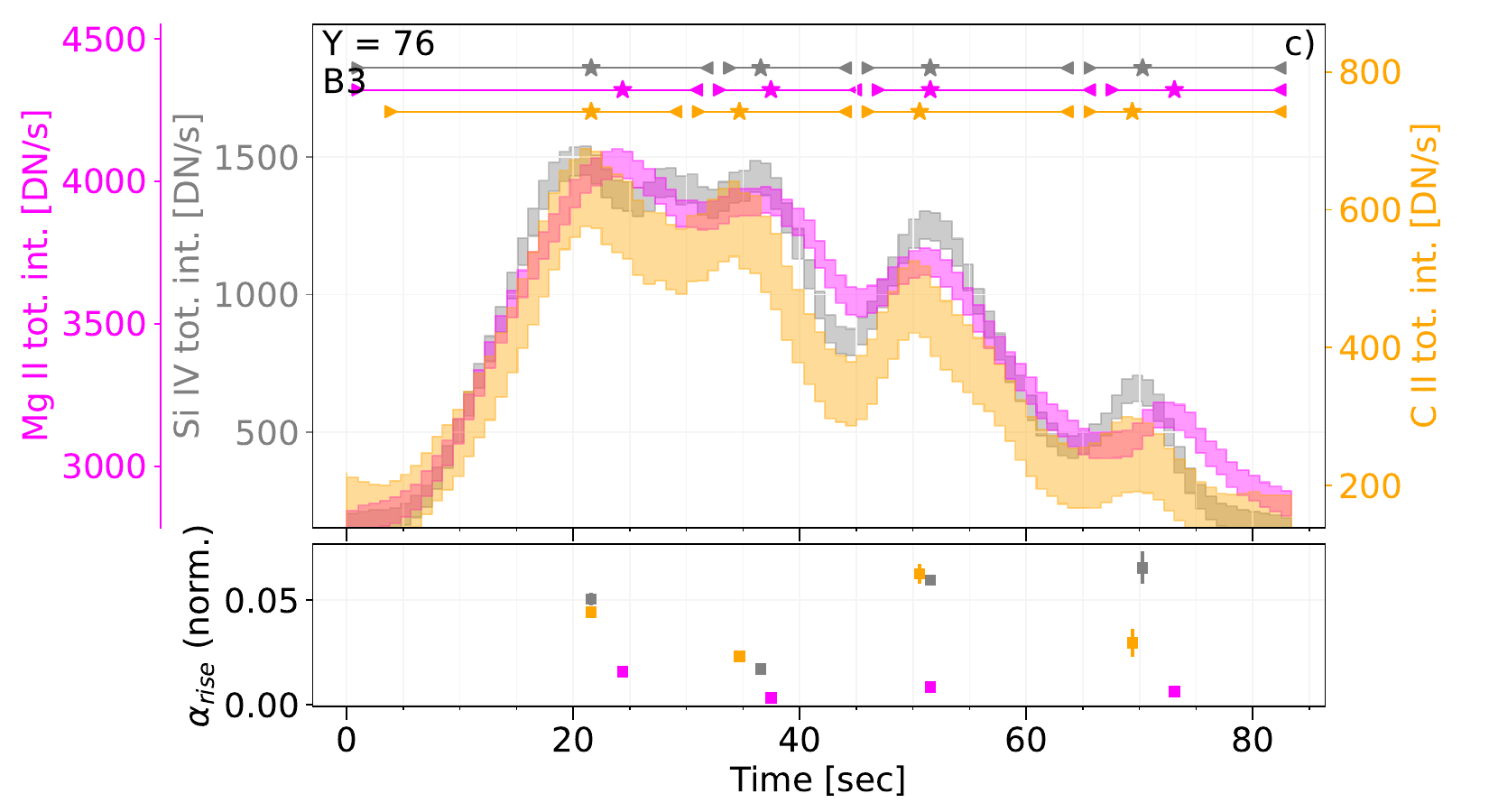} 
\includegraphics[width=8.90cm, clip,   viewport=00 00 780 430]{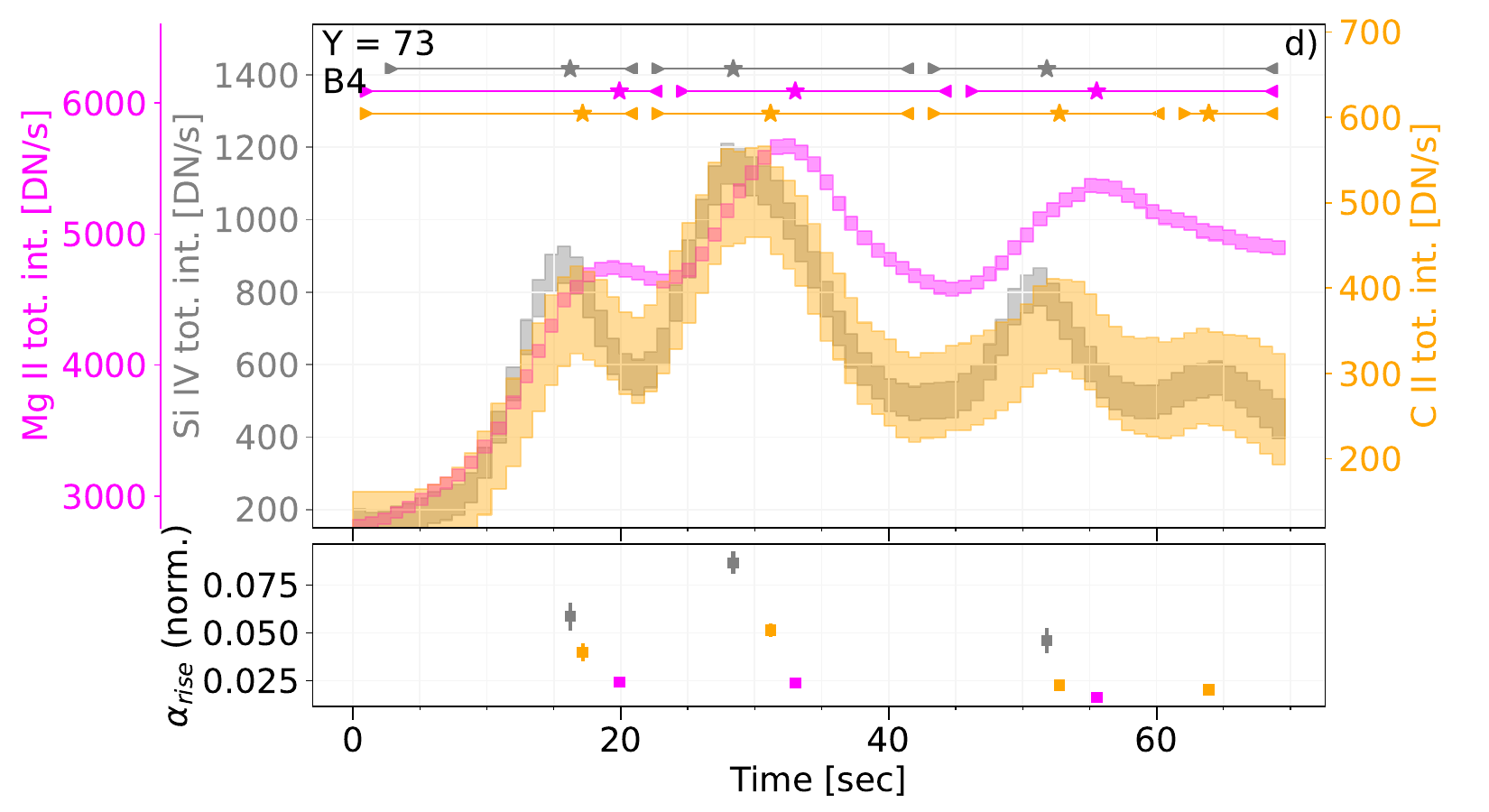} 
\caption{\ion{Si}{4} 1402\,\AA~(grey), \ion{Mg}{2} 2796.35\,\AA~(magenta), and \ion{C}{2} 1334.53\,\AA~(orange) line lightcurves during the B1 -- B4 brightening events in the ribbon observed by IRIS. The lightcurves were smoothed over additional 5 time pixels compared to those in Figure~\ref{fig:lightcurves}. The colored horizontal bars above the lightcurves indicate automatically-identified peak intervals. The triangle symbols are the left and right bases (peak onset and end times), while the stars designate the maximum of each peak. The bottom part of each panel indicates the slopes of each peak after normalization.   \label{fig:lightcurves_app}}
\end{figure*}

As stated in Section \ref{sec:obs_lightcurves}, the manual selection of the time intervals containing intensity peaks in different ions was in some cases difficult due to the characteristics of the peaks as well as presence of the noise. Peaks and peak intervals in the \ion{Si}{4} 1402.77\,\AA, \ion{Mg}{2} k 2796.35\,\AA, and \ion{C}{2} 1334.53\,\AA~lightcurves were therefore also determined automatically using the \texttt{find\_peaks} function of the SciPy library \citep{virtanen20}. Unlike our manual approach based on the identification of peak onset and end times (peak bases), this function directly returns positions and properties of peaks in an arbitrary signal. 

Input parameters of this function include several criteria guiding the detection of the peaks, including their height, width, amplitude, or the mutual distance between the peaks. We found the method to yield the most reliable results when the peak detection was constrained by the the minimal peak width of 2 time bins and the minimal separation between the peaks of at least 5 time bins. The performance of the method was negatively affected by the presence of the noise in the lightcurves, leading to false peak identifications. The lightcurves (intensities as well as their uncertainties) were therefore additionally smoothed using the Savitzy-Golay filter (\texttt{savgol\_filter} function in SciPy) with a boxcar size of 5 time pixels. Once the peaks in the lightcurves of the three lines have been identified, the peak bases were simply adapted from intensity minima in time intervals separating the detected peaks, or the peaks and the edges of the lightcurves. The peaks and peak intervals were subsequently used to measure the delays between the \ion{Si}{4} and \ion{Mg}{2} k line emission (Appendix \ref{sec:appendix_d}).

The number of the automatically-identified peaks in lightcurves of different ions varied in few instances. While the delays between peak maxima can easily be obtained by comparing the peaks closest in time, the inconsistent time intervals need to be accounted for in the computation of the delays inferred from the centroid and cross-correlation methods (Appendix \ref{sec:appendix_d}). This was achieved by merging peak intervals in the lightcurves exhibiting a larger number of peaks over the same time interval. To illustrate this in a simplified example, if for example one ion exhibited one peak in the interval [0, 6], while the other showed two peak intervals at [0, 3] and [3, 6], the two intervals would coalesce into one longer interval [0, 6] containing two peaks. The algorithm essentially mimics our approach from the main analysis where matching (or similar) peak intervals across different ions were performed manually (Section \ref{sec:obs_lightcurves}). 

Appendix Figure~\ref{fig:lightcurves_app} presents the smoothed lightcurves corresponding to the total line intensities, with peak intervals determined as disclosed above. The design and layout of the panels is identical to Figure~\ref{fig:lightcurves}; the upper portions of panels a -- d present the lightcurves during the four brightening events \mbox{B1 -- B4}, and their bottom parts show the slopes of the individual intensity peaks. Although the lightcurve smoothing resulted in a decrease of \ion{Si}{4} and \ion{Mg}{2} peak detections during the B1 and B2 (panels a, b), the method identified an additional peak during the B3 (panel c). {The ascending slopes of the \ion{Si}{4} and \ion{C}{2} peaks are consistently larger than those of \ion{Mg}{2}, what corresponds to the results described in Section \ref{sec:obs_lightcurves} where the peaks and peak bases were selected manually. This means that the noise level, occasionally hindering the identification of the peak bases, did not affect the slope calculation and the conclusion drawn from this analysis. An exception to this trend is again visible in the case of the single \ion{Mg}{2} peak at $t = 50 - 70$\,s during the B2 (Appendix Figure \ref{fig:lightcurves_app}b) where the method did not identify a low-amplitude \ion{Si}{4} peak at $t \sim 55$\,s (see Section \ref{sec:obs_lightcurves} for details)}. The delays between the automatically-detected \ion{Si}{4} and \ion{Mg}{2} peaks are plotted in Appendix Figure~\ref{fig:delays_app} and discussed in Appendix \ref{sec:appendix_d}. 

\section{Additional methods for inferring \ion{Si}{4} and \ion{Mg}{2} delays}
\label{sec:appendix_d}

\begin{figure*}[h]
\centering
\includegraphics[width=6.00cm, clip,   viewport=50 40 605 610]{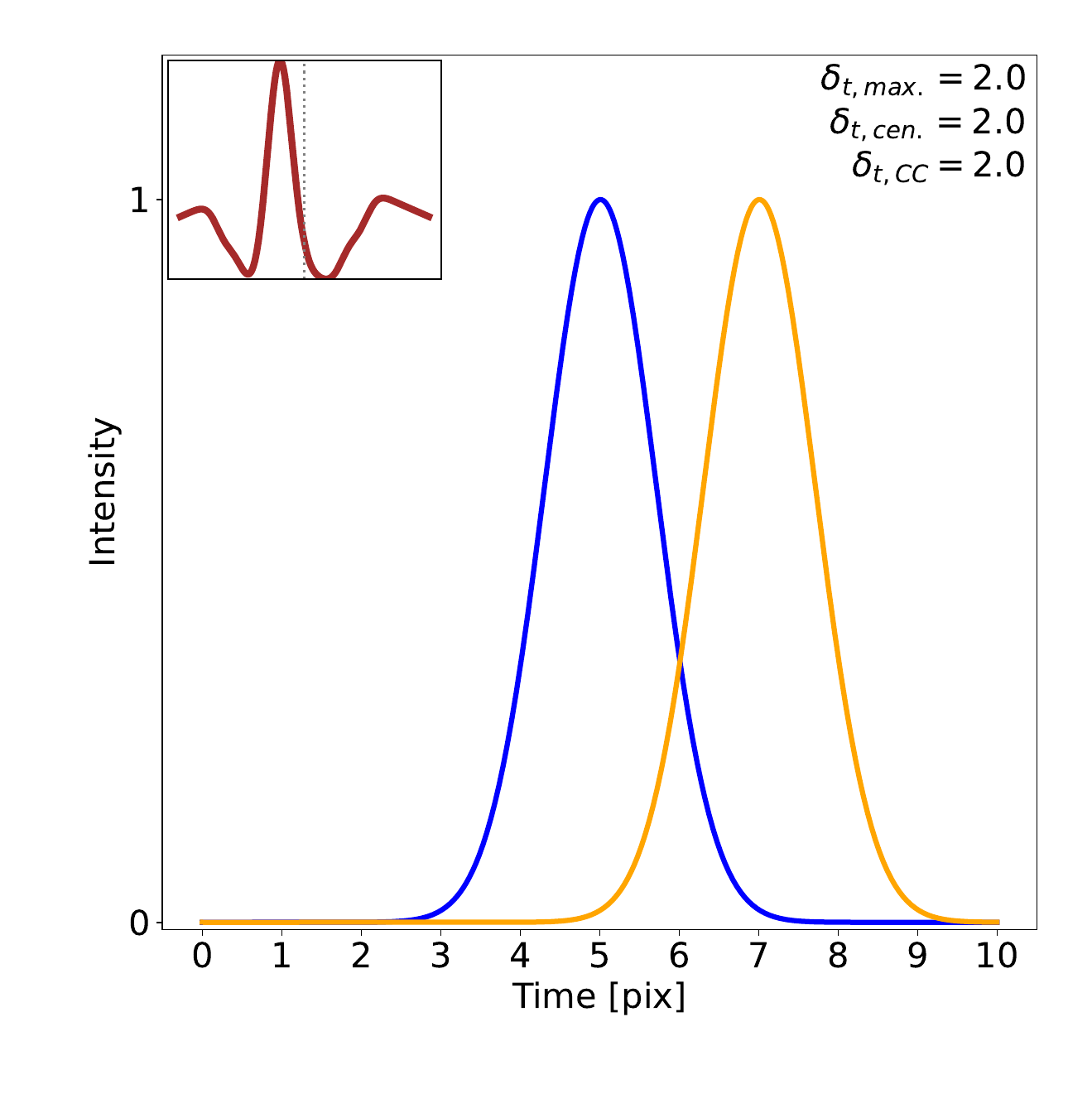}
\includegraphics[width=5.57cm, clip,   viewport=90 40 605 610]{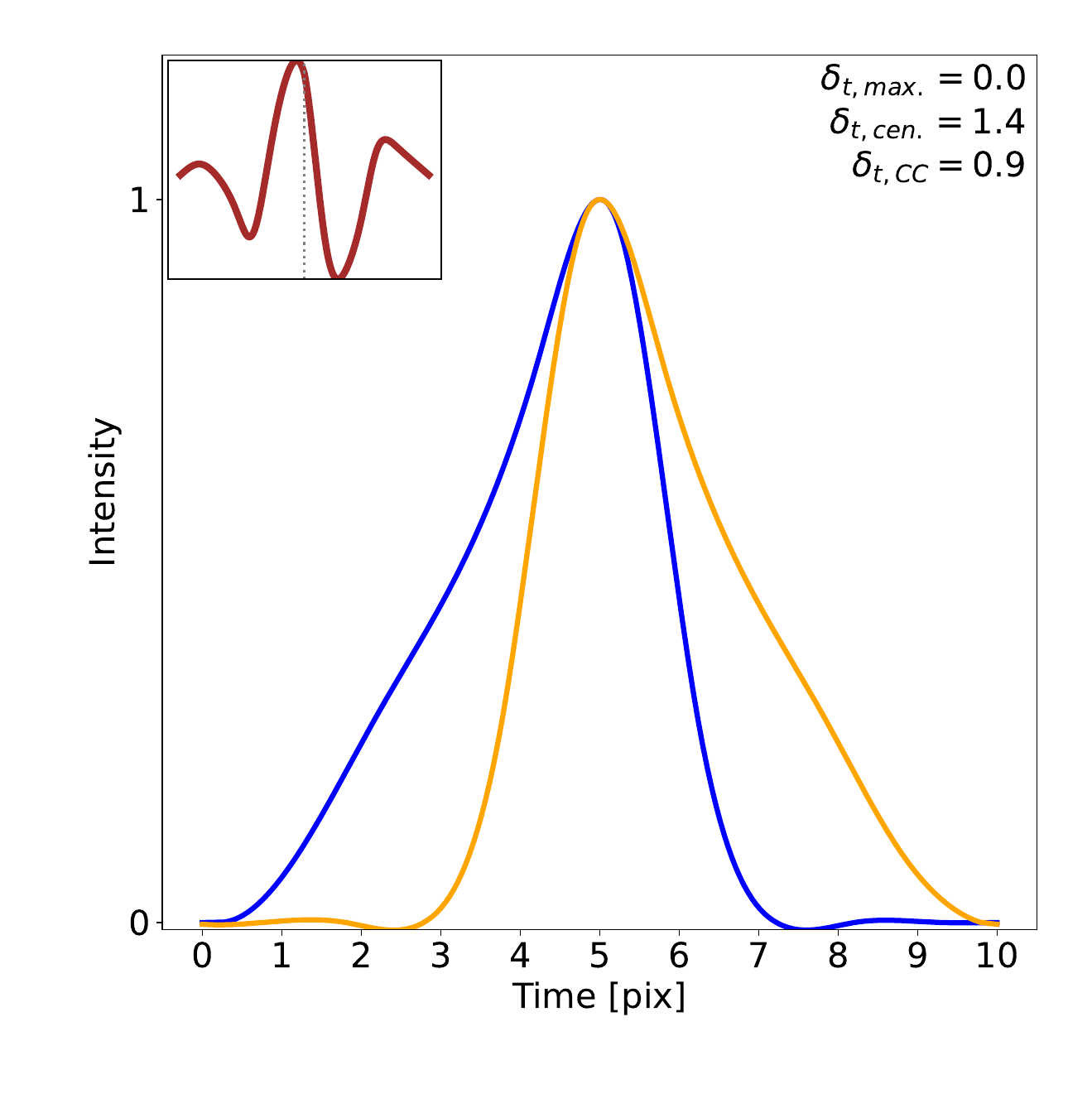}
\includegraphics[width=5.57cm, clip,   viewport=90 40 605 610]{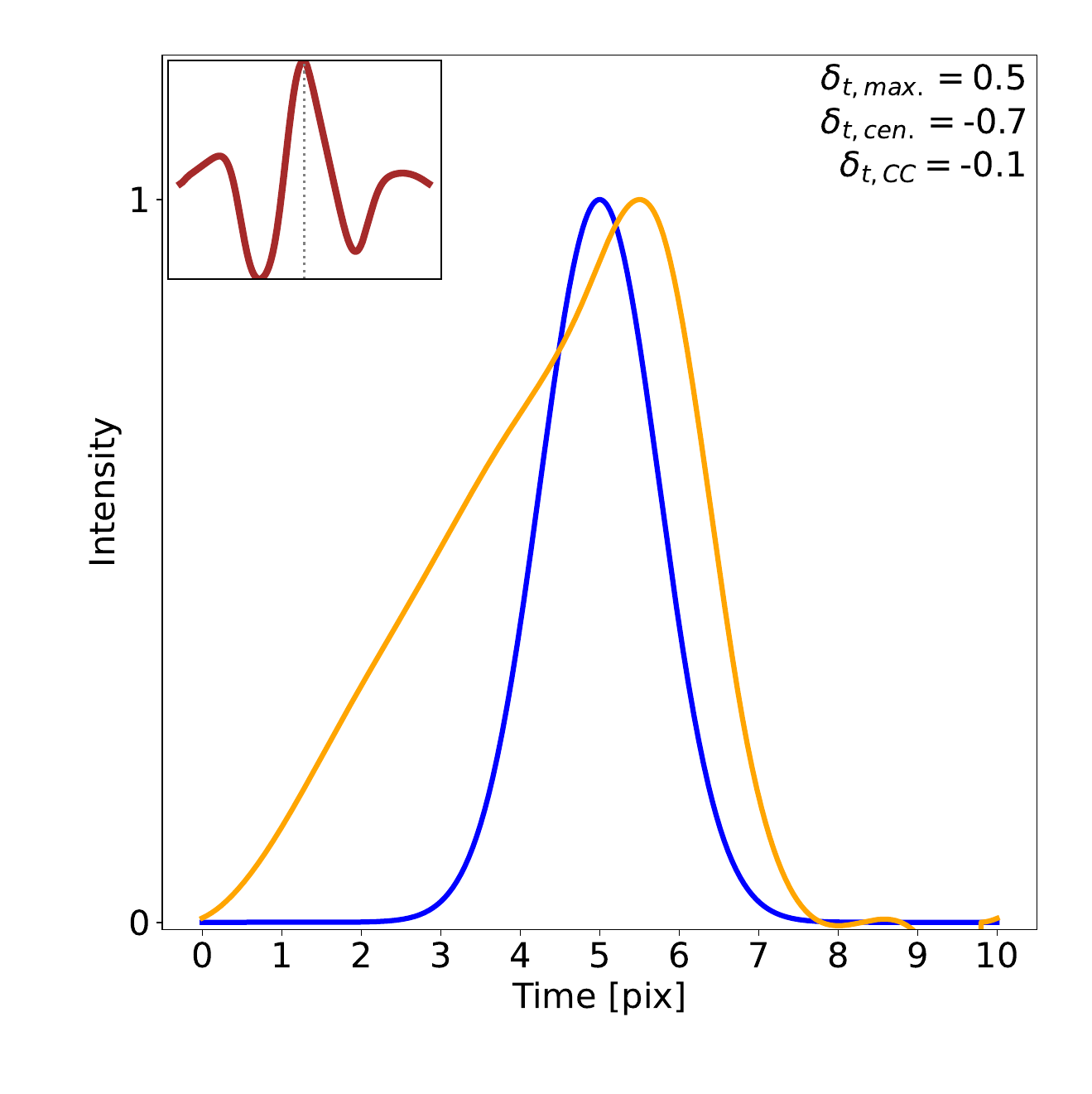}
\caption{Comparison of methods employed to infer the peak times between intensity peaks on selected pairs  of functions. See Appendix \ref{sec:appendix_d} for details. \label{fig:toymodel_app}}
\end{figure*}

Subtraction of peak times is likely the easiest and the most straightforward way to quantify the delays between the intensity peaks. This method does not account for sometimes complicated time evolution of the individual peaks (Section \ref{sec:obs_lightcurves}). We have therefore extended our study by calculating the delays between the centroids of the peaks determined from the 1st moment in each time interval limited by the peak bases. The analysis is also complemented by cross-correlating the individual peak intervals observed in the two lines. The cross-correlations were performed using functions available within the \texttt{sunkit\_image} library \citep{freij23}, see \citet[][]{barnes19} for the method. The time lag between the intensity peaks in different ions in the corresponding time interval were inferred from a single-Gaussian fit to the core of the cross-correlation function. In cases where the shape of the cross-correlation function could not be approximated by a Gaussian, the time lag was adapted from the maximum of the cross-correlation function. Differences between the peak delays inferred using the selected methods are demonstrated on three pairs of functions with distinct time evolution, plotted in Appendix Figure~\ref{fig:toymodel_app}. The insertion in the top-left corner of each panel represents the cross-correlation function of the two signals. 
\begin{itemize}
\item{The left panel depicts two Gaussians, blue one peaking at time $t = 5$\,s and the orange one peaking later at $t = 7$\,s. All three methods employed to study the delays between the intensity peaks show the same time delay $\delta_t = 2$\,s between the two signals.}
\item{The middle panel presents two asymmetric signals whose maxima correspond to $t = 5$\,s. The blue signal is characterized by a gentle increase to its peak followed by a rapid decrease. The orange signal is a mirrored representation of the orange one; it peaks rapidly and then slowly decreases. This time evolution results in positive delays (blue signal `peaking' first) inferred from the centroids $\delta_{t, \mathrm{cen.}}$ and the cross-correlation $\delta_{t, \mathrm{CC}}$.}
\item{The example presented in the right panel of Appendix Figure~\ref{fig:toymodel_app} shows a blue Gaussian and an orange asymmetric signal. Even though the blue Gaussian peaks before the orange signal ($\delta_{t, \mathrm{max}} = 0.5$\,s) the initial rise of the orange signal results in small but negative delays inferred from the centroids and cross-correlations.}
\end{itemize}

\begin{figure*}[h]
\centering
\includegraphics[width=8.80cm, clip,   viewport=00 00 710 350]{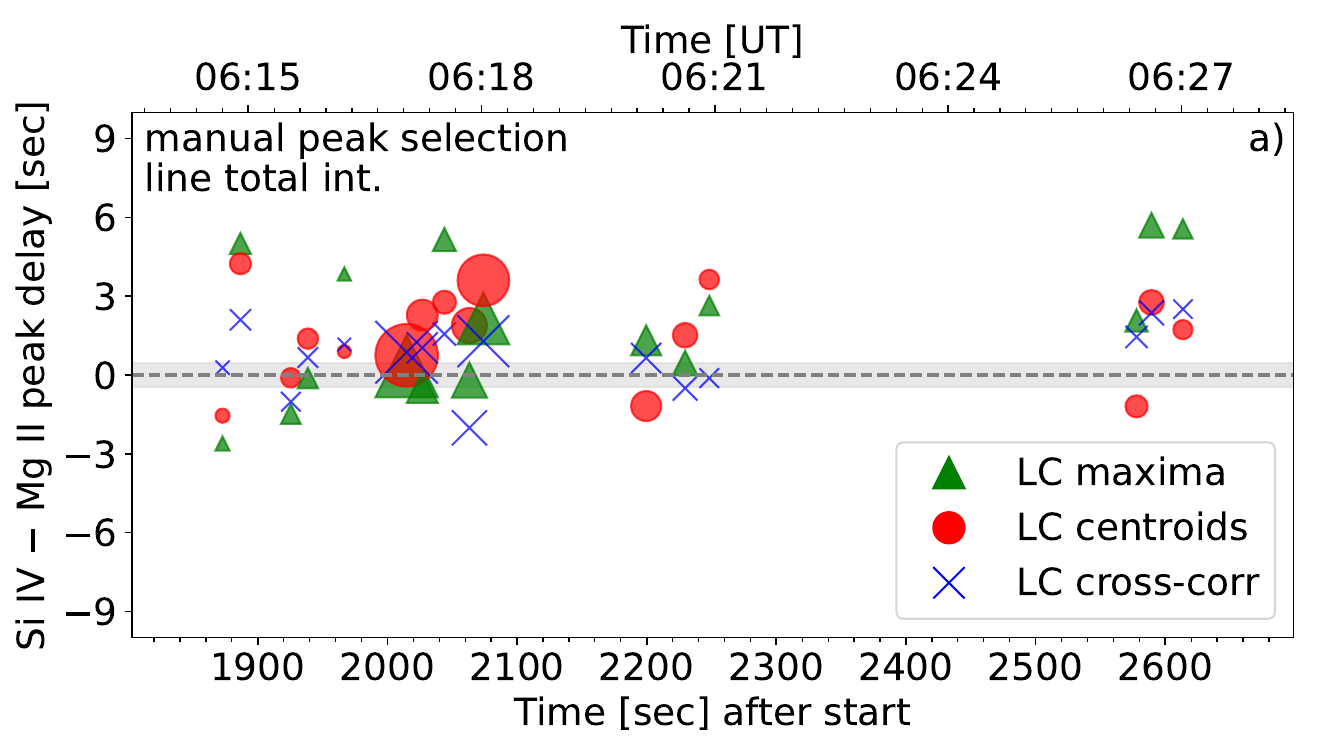}
\includegraphics[width=8.80cm, clip,   viewport=00 00 710 350]{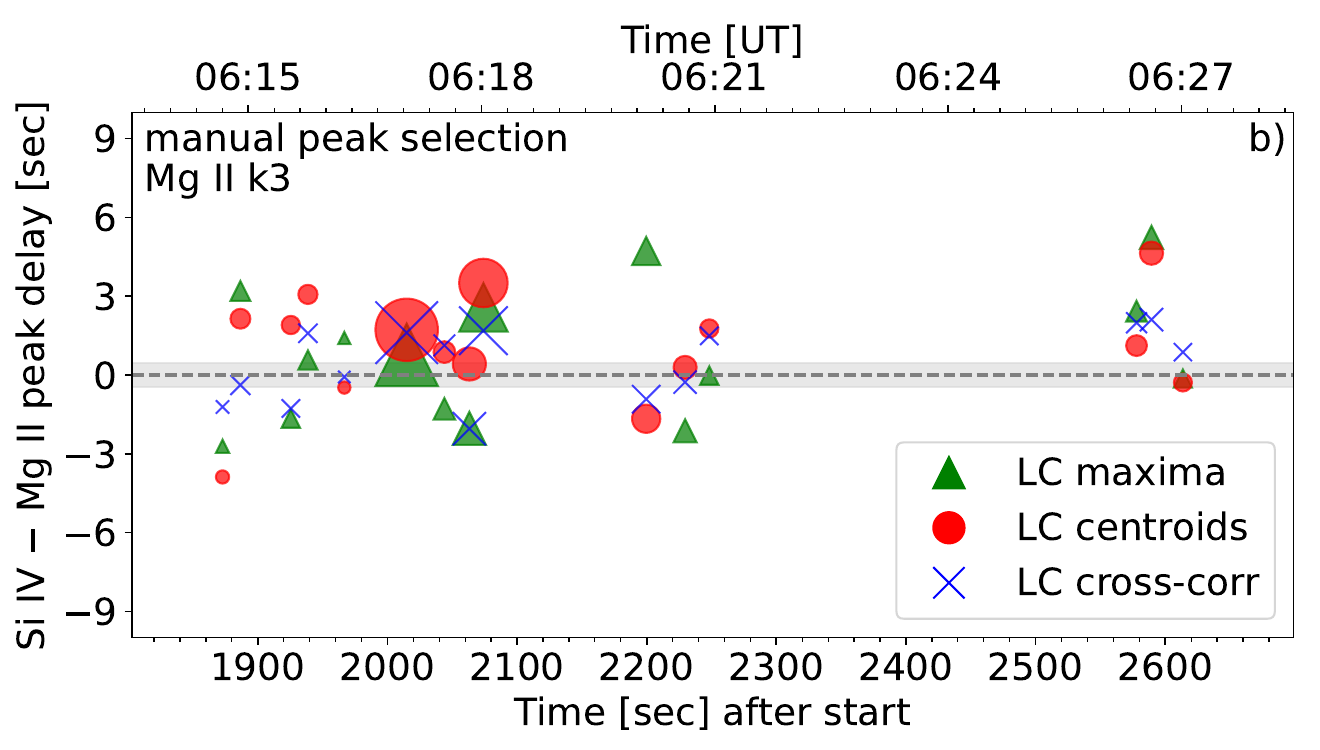}
\\
\includegraphics[width=8.80cm, clip,   viewport=00 00 710 350]{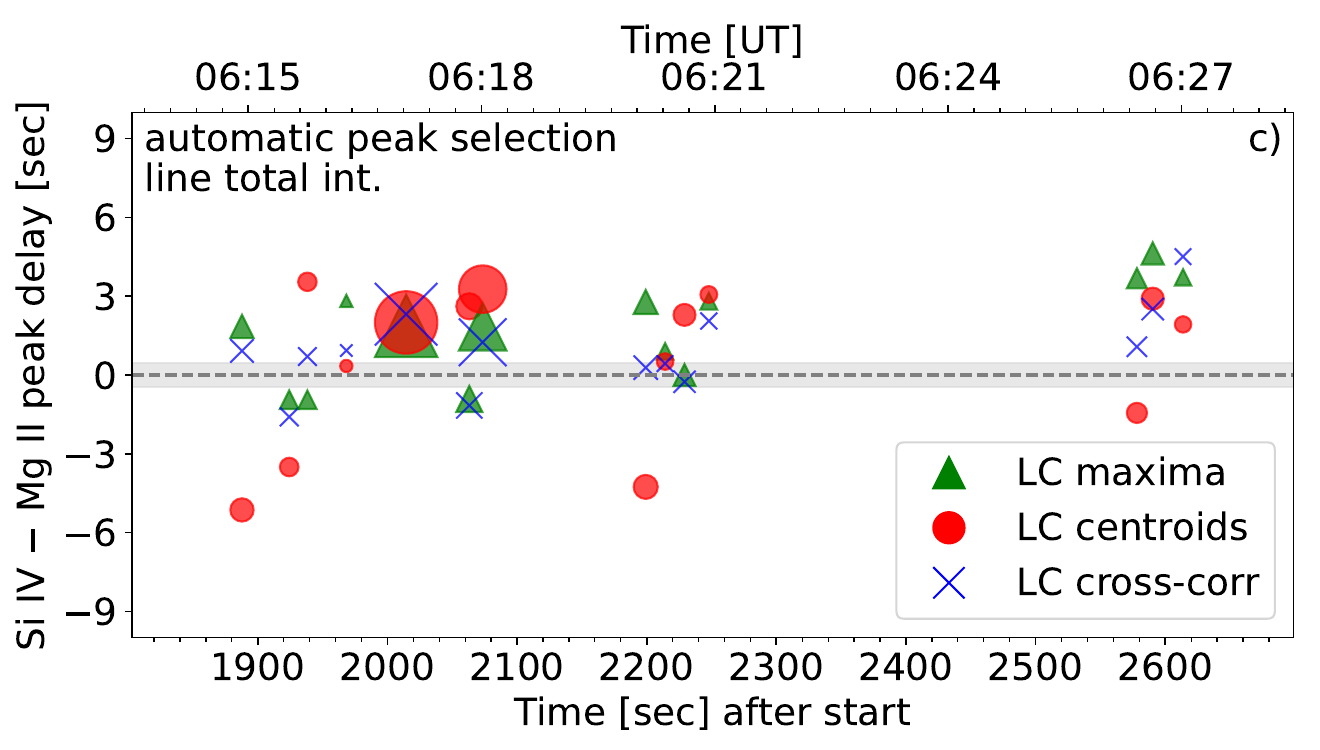}
\includegraphics[width=8.80cm, clip,   viewport=00 00 710 350]{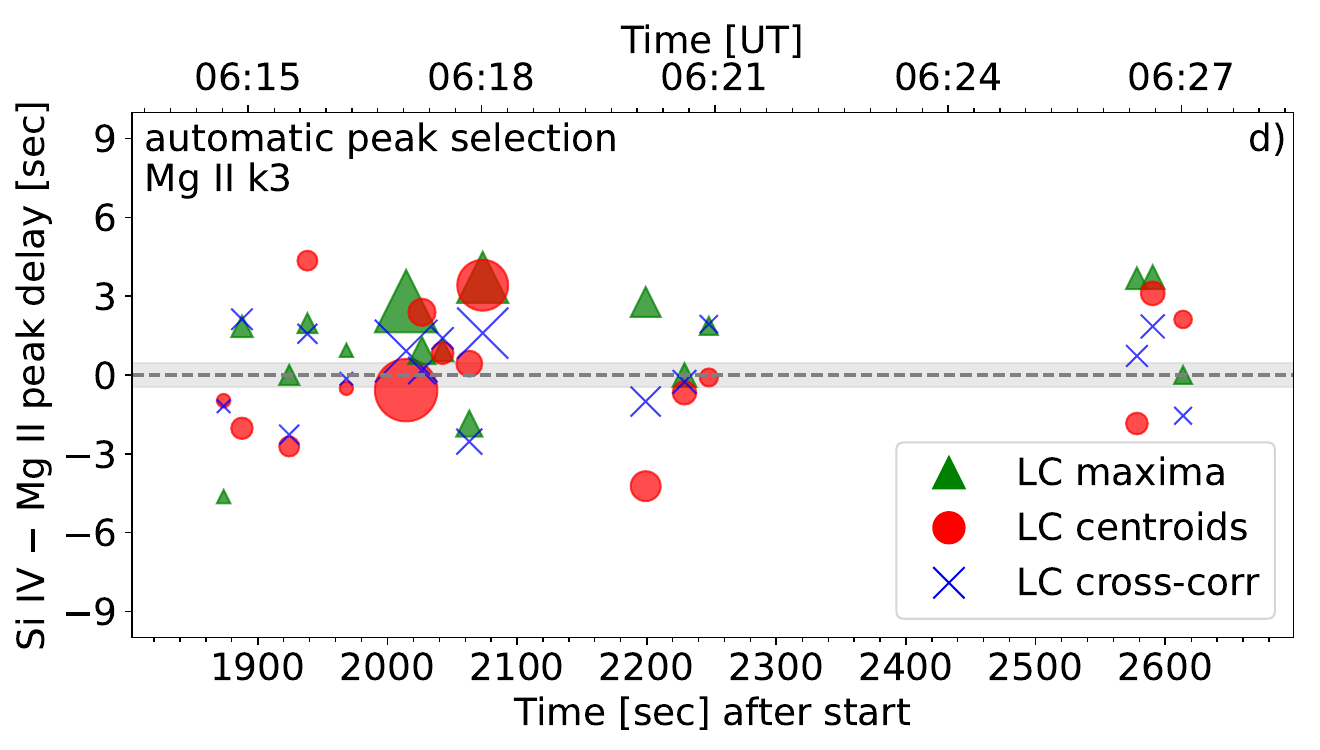}

\caption{Delays between \ion{Si}{4} and \ion{Mg}{2} k intensity peaks as a function of time during the flare. Panels a and b plot delays calculated in peak intervals determined manually. Automatic peak detection algorithm was used for the calculation of the delays plotted in panels c and d. Different symbols distinguish between the methods used for determining the peak times, and their size is given by the amplitude of \ion{Si}{4} peaks. Panels a and c present peak delays along lightcurves where both \ion{Si}{4} and \ion{Mg}{2} k line intensities were obtained via the moment analysis. \ion{Mg}{2} k3 component intensities were used for the calculation of delays plotted in panels b and d. The gray horizontal strip at $t = 0$\,s corresponds to $\pm \Delta_{t}/2$. \label{fig:delays_app}}
\end{figure*}

Appendix Figure~\ref{fig:toymodel_app} demonstrates the effects of slopes and symmetry of peaks on the delays inferred from the cross-correlations and peak centroids. The delays between the observed \ion{Si}{4} and \ion{Mg}{2} intensity peaks inferred via the three methods differ significantly for some peaks. Appendix Figure~\ref{fig:delays_app} presents the time evolution of the delays between \ion{Si}{4} and \ion{Mg}{2} line emission during the impulsive phase of the flare. The figure has the same layout as Figure~\ref{fig:delays}. The delays between the peak maxima (green triangles) are complemented by the delays inferred from peak centroids (red circles) and cross-correlations (blue crosses). The top row of Figure~\ref{fig:delays_app} a, b shows delays of peaks and peak intervals determined manually, while the automatic peak detection (Appendix \ref{sec:appendix_c}) was used to identify the peaks whose delays are plotted in panels c and d. Delays plotted in panels a, c were computed using the total line intensities, whereas the right column (panels b, d) shows the delays between the \ion{Si}{4} 1402.77\,\AA~total intensity and the k3 component of the \ion{Mg}{2} 2796.35\,\AA~line. 

The distribution of the datapoints in Appendix Figure~\ref{fig:delays_app} confirms that \ion{Si}{4} peaks typically occur before \ion{Mg}{2}. This result holds regardless of the method used to
\begin{enumerate}
\item{detect peaks and peak intervals (upper vs. bottom row),}
\item{measure \ion{Mg}{2} k line intensities (left vs. right column),}
\item{infer peak time and/or peak delay (symbols).}
\end{enumerate}
The delays between the centroids (circles) of the automatically-detected peaks of the \ion{Si}{4} total intensity and the \ion{Mg}{2} k3 component lightcurve, plotted in panel d, present an exception to this behavior. This is a result of additional smoothing and the fact that the k3 component intensity peaks were less prominent (Appendix \ref{sec:appendix_c}). The longest delays are typically found between peak maxima (triangles). Cross-correlating of the peak intervals, on the other hand, results in the lowest delays. $\sim 60 - 85$\% of the delays plotted in Appendix Figure~\ref{fig:delays_app} are lower than $3 \Delta_{t}$ ($\approx 2.8$\,s), depending on the method of \ion{Mg}{2} k line intensity measurement and the peak selection method.

\clearpage

\bibliographystyle{aasjournal}    %% Author (year)
\bibliography{bibliography}

\end{document}